\documentclass[aps,10pt,tightenlines,amsmath,amssymb]{revtex4}
\usepackage{graphicx}
\usepackage{dcolumn}
\usepackage{bm}

\begin{document}


\title{Extragalactic photon--axion-like particle oscillations up to 1000 TeV}

\author{Giorgio Galanti}
\email{gam.galanti@gmail.com}
\affiliation{INAF, Osservatorio Astronomico di Brera, Via Emilio Bianchi 46, I -- 23807 Merate, Italy}

\author{Marco Roncadelli}
\email{marco.roncadelli@pv.infn.it}
\affiliation{INFN, Sezione di Pavia, Via A. Bassi 6, I -- 27100 Pavia, Italy, and INAF}

\begin{abstract}
Axion-like particles (ALPs) are attracting increasing interest since, among other things, they are a prediction of many extensions of the standard model of elementary particles physics and in particular of superstrings and superbranes. ALPs are very light, neutral, pseudo-scalar bosons which are supposed to interact with two photons. For their mass $m_a \ll 1 \, {\rm eV}$ and two-photon coupling $g_{\gamma \gamma a}$ in a suitable range they can give rise to very interesting astrophysical effects taking place in the X- and $\gamma$-ray bands. Specifically, throughout the present paper we are concerned with photon-ALP oscillations in the very-high-energy band $\bigl({\rm VHE}, 100 \, {\rm GeV} \lesssim {\cal E} \lesssim 100 \, {\rm TeV} \bigr)$ and beyond, which ought to occur in the photon beam emitted by far-away blazars and are triggered by the domain-like random extragalactic magnetic field ${\bf B}_{\rm ext}$. Because of the presence of the extragalactic background light (EBL) -- which is the infrared/optical/ultraviolet radiation emitted by all galaxies during the cosmic evolution --  when a VHE photon scatters off an EBL photon an $e^+ e^-$ pair can be created, which causes a rather strong dimming of the source. In the presence of photon-ALP oscillation things are different, since a photon travels sometimes as a true photon and sometimes as an ALP. Since ALPs do not interact with the EBL, the effective optical depth is somewhat reduced. But -- as a consequence -- the photon survival probability gets {\it strongly enhanced} with respect to the prediction of conventional physics, thereby {\it greatly increasing} the photon transparency  in the VHE band so that the corresponding horizon gets enlarged to a considerable extent. While all this is well known and already studied in detail [Phys. Rev. D {\bf 84}, 105030 (2011), ibid. D {\bf 87}, 109903 (E) (2013)], the new effect of photon dispersion on the cosmic microwave background (CMB) becomes very important at high enough energies. The aim of the present paper is to take it systematically into account. Actually, two widely different energy scales are associated with it. One is ${\cal E}_H = {\cal O} (5 \, {\rm TeV)}$, above which the effect in question starts to become dominant and makes the {\it single} random realizations of the beam propagation process -- the only ones that are observable -- to exhibit small energy oscillations: this is a crucial {\it prediction} of our model. The other energy scale is ${\cal E}_{\rm eq}$ above which the oscillation length becomes smaller than the coherence length of ${\bf B}_{\rm ext}$: typically ${\cal E}_{\rm eq}  = {\cal O} (40 \, {\rm TeV})$ with a large uncertainty. Thus, previously used domain-like models of ${\bf B}_{\rm ext}$ would generally give wrong results above ${\cal E}_{\rm eq}$ and a more realistic model for ${\bf B}_{\rm ext}$ becomes {\it compelling}, like the one very recently developed by the authors. Remarkably, we have been able to derive the corresponding photon survival probability $P^{\rm ALP}_{\gamma \to \gamma} ({\cal E}_0, z)$ {\it analytically and exactly} up to observed energies ${\cal E}_0 = 1000 \, {\rm TeV}$ and redshift up to $z = 2$, a fact that drastically shortens the computation time in the derivation of the results presented in this paper. Specifically, for 7 simulated blazars we exhibit the plots of the $P^{\rm ALP}_{\gamma \to \gamma} ({\cal E}_0, z)$ along 1000 random realizations versus ${\cal E}_0$, for different values of $z$ and four values of the model parameters. Our predictions can be tested by the new generation of $\gamma$-ray observatories like CTA, HAWC, GAMMA-400, LHAASO, TAIGA-HiSCORE and HERD. Finally, for our guessed values of $m_a$ and $g_{\gamma \gamma a}$ our ALP can be detected in the upgrade of ALPS II at DESY, the planned experiments IAXO, STAX and ABRACADABRA as well as with other techniques developed by Avignone and collaborators.

\end{abstract}


\maketitle


\section{Introduction}

Different extensions of the standard model of elementary particles physics generally predict a different set of new particles, but a common requirement is that one of them should be the {\it axion}, namely the pseudo-Goldstone boson associated with the global Peccei-Quinn symmetry ${\rm U}(1)_{\rm PQ}$ proposed as a natural solution to the strong CP problem (for a review see~\cite{axionrev1,axionrev2,axionrev3,axionrev4}). Quite often -- especially within models arising from superstrings and superbranes -- the axion comes together with {\it axion-like particles} (ALPs), which are very light, neutral, pseudo-scalar bosons (for a review, see~\cite{alp1,alp2}). 

\begin{figure}[h]
\centering
\includegraphics[width=0.30\textwidth]{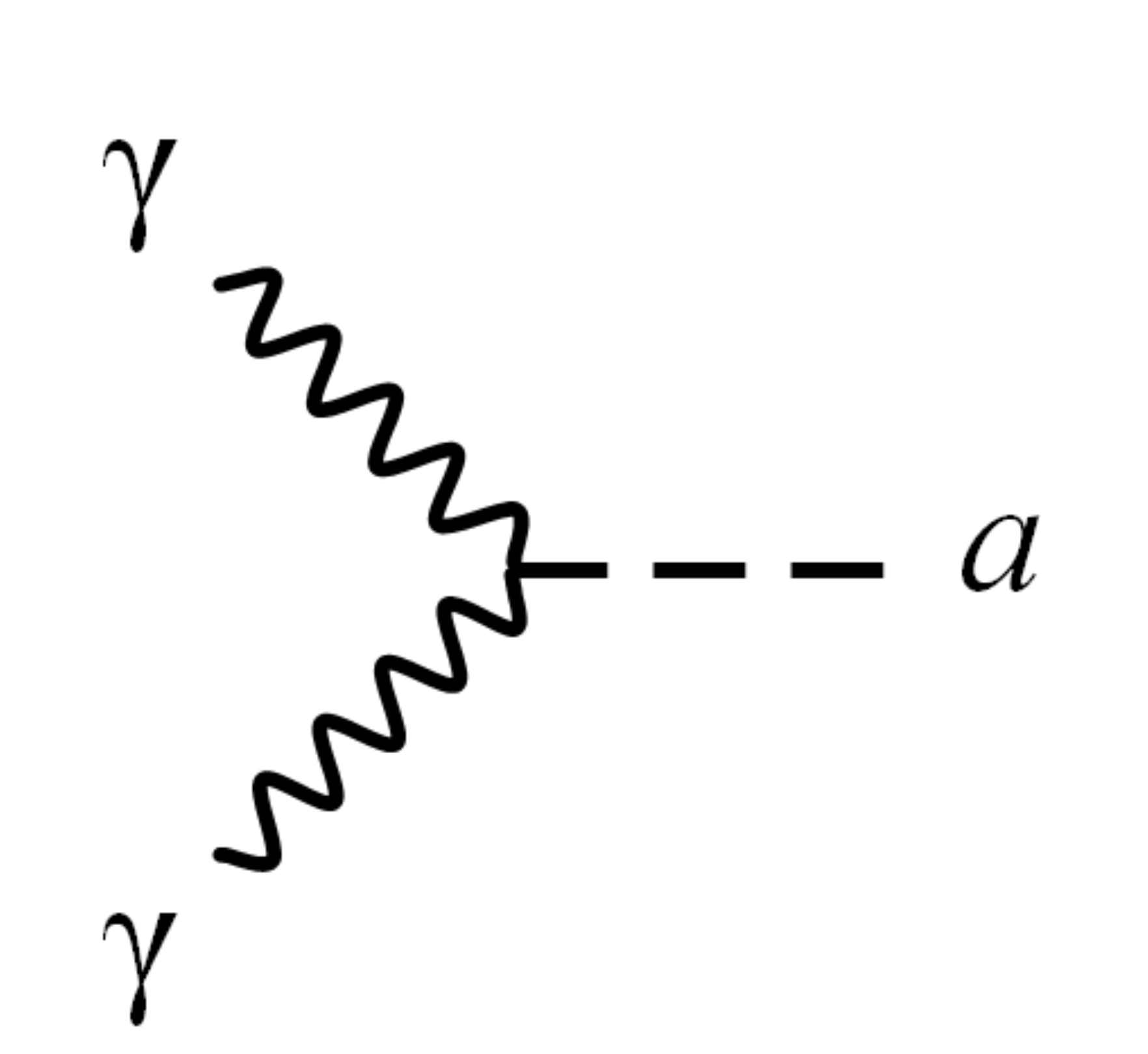}
\caption{\label{immagine3} Photon-photon-ALP vertex with coupling constant 
$g_{a \gamma \gamma}$.}
\end{figure}

While for the axion the mass $m$ is related to the two-photon coupling $g_{a \gamma \gamma}$ by $m = 0.7 \, k \, \bigl(g_{a \gamma \gamma} \,10^{10} \, {\rm GeV} \bigr) \, {\rm eV}$ with $k = {\cal O} (1)$~\cite{cgn1995}, in the case of ALPs their mass $m_a$ and  
two-photon coupling $g_{a \gamma \gamma}$ are {\it unrelated} quantities. Further, while the axion must have specific couplings to two gluons and to quarks in order for the Peccei-Quinn mechanism to work, what really matters for ALPs -- henceforth denoted by $a$ for simplicity -- is the two-photon coupling $g_{a \gamma \gamma}$: additional couplings can be present but do not play any significant role and since they are uninteresting for our purposes they will be discarded. So, the only piece of new physics arises solely from the interaction Lagrangian $g_{a \gamma \gamma} \, a \, {\bf E} \cdot {\bf B}$, which is represented by the Feynman diagram in Figure~\ref{immagine3} -- ${\bf E}$ and ${\bf B}$ are the electric and magnetic field, respectively.  

Actually, during the last fifteen years or so ALPs have attracted an ever growing interest. We can identify four reason behind for this fact. 

\begin{itemize}

\item One reason is certainly their ubiquity, in the sense that many extensions of the standard model of elementary particles physics -- even along diverging directions -- predict the existence of ALPs (for an incomplete sample of references, see~\cite{turok1996,string1,string2,string3,string4,string5,axiverse,abk2010,cicoli2012,cicoli2013,dias2014,cicoli2014a,conlon2014,cicoli2014b,conlon2015,cicoli2017,scott2017,
concon2017,cisterna1,cisterna2}).

\item Another reason is that in a suitable region of the parameter plane $(m_a, g_{a \gamma \gamma})$ ALPs are very good candidates for cold dark matter~\cite{preskill,abbott,dine,arias2012}. 

\item The third reason is that for another region of the parameter plane $(m_a, g_{a \gamma \gamma})$ -- which can overlap with the previous one -- ALPs give rise to very interesting astrophysical effects (for an incomplete sample of references, see~\cite{khlopov1992,khlopov1994,khlopov1996,raffelt1996,masso1996,csaki2002a,csaki2002b,cf2003,dupays,mr2005,fairbairn,mirizzi2007,drm2007,bischeri,sigl2007,dmr2008,shs2008,dmpr,mm2009,crpc2009,cg2009,bds2009,prada1,bronc2009,bmr2010,mrvn2011,prada2,gh2011,pch2011,dgr2011,frt2011,hornsmeyer2012,hmmr2012,wb2012,hmmmr2012,trgb2012,friedland2013,wp2013,cmarsh2013,mhr2013,hessbound,gr2013,hmpks,mmc2014,acmpw2014,rt2014,wb2014,hc2014,mc2014,bw2015,payez2015,trg2015,bgmm2016,fermi2016,day2016,sigl2016,sigl2017a,mcdmarsh2017,bnt2017,vlm2017,marshfabian2017,zung2018,pani2018,zhang2018,mnt2018}).

\item The last reason is that the region of the parameter plane $(m_a, g_{a \gamma \gamma})$ relevant for astrophysical effects can be probed in the laboratory within the next few years thanks to the upgrade of ALPS II at DESY~\cite{alps2}, the planned experiments IAXO~\cite{iaxo} and STAX~\cite{stax}, as well as with other strategies developed by Avignone and collaborators~\cite{avignone1,avignone2,avignone3}. Moreover, if the bulk of the dark matter is made of ALPs they can also be detected by the planned ABRACADABRA experiment~\cite{abracadabra}.

\end{itemize}

Coming back to our main line of development, any new physical process involving ALPs comes about by combining the Feynman diagram in Figure~\ref{immagine3} with those of the standard model (examples of this game will be shown in Section III). 

\begin{figure}[h]
\centering
\includegraphics[width=.40\textwidth]{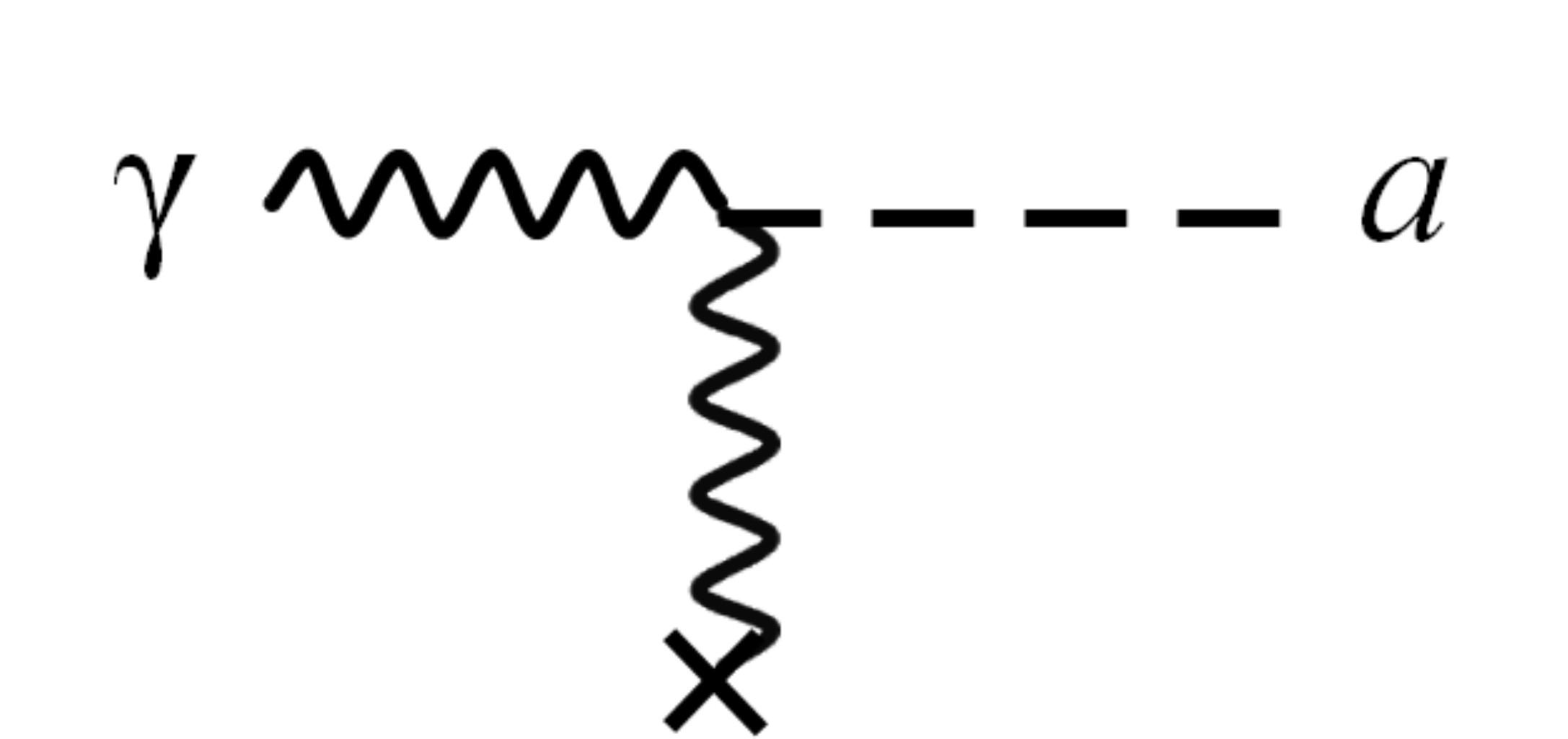}
\caption{\label{immagine4} $\gamma \to a$ conversion in the extragalactic magnetic field ${\bf B}_{\rm ext}$.}
\end{figure}

In the present paper we are merely concerned with the behavior of ALPs in the presence of the {\it extragalactic} magnetic field ${\bf B}_{\rm ext}$. In a sense, this is the simplest possible effect which is represented by the Feynman diagram shown in Figure~\ref{immagine4}. As a consequence, the mass matrix of the photon-ALP system becomes off-diagonal, thereby implying that the mass eigenstates differ from the interaction eigenstates. Because ${\bf B}_{\rm ext}$ is stationary (apart from cosmological effects which turn out to be irrelevant), it is evident from Figure~\ref{immagine4} that processes $\gamma \to a$ and $a \to \gamma$ are both allowed and energy conserving.

\begin{figure}[h]
\centering
\includegraphics[width=0.8\textwidth]{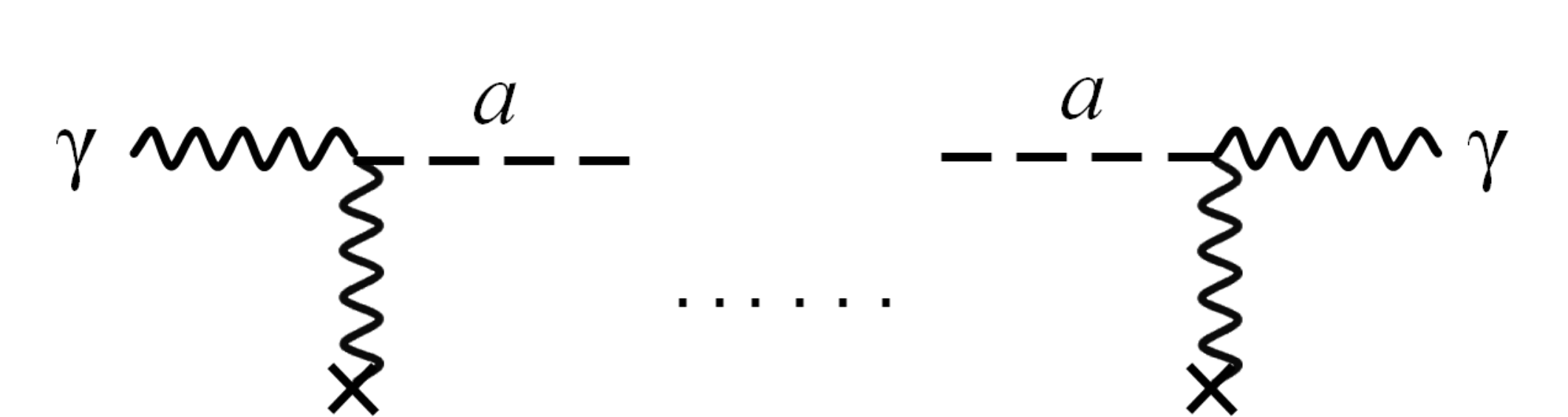}
\caption{\label{fey1} Schematic view of a photon-ALP oscillation in the extragalactic magnetic field ${\bf B}_{\rm ext}$.}
\end{figure}

Throughout this paper, we will focus our attention on the behavior of a monochromatic photon beam of energy ${\cal E}$ in the {\it very-high-energy} band $\bigl({\rm VHE}, 100 \, {\rm GeV} \lesssim {\cal E} \lesssim 100 \, {\rm TeV} \bigr)$ and beyond, emitted by a {\it blazar}, namely an Active Galactic Nucleus (AGN) with one jet pointing occasionally towards us. We suppose that the line of sight lies along the $y$-direction. As the beam propagates in ${\bf B}_{\rm ext}$, a succession of processes $\gamma \to a$ and $a \to \gamma$ takes place. But this means that photon-ALP {\it oscillations} $\gamma \leftrightarrow a$ -- as shown in Figure~\ref{fey1} -- occur inside the beam~\cite{sikivie,anselmo,rs1988}. They are very similar to flavor oscillations for massive neutrinos, apart from the need of the external magnetic field ${\bf B}_{\rm ext}$ in order to compensate for the spin mismatch, since photons have spin 1 while ALPs have spin 0. 

\

As a preliminary step, let us start to work within {\it conventional physics}. Accordingly, when a VHE beam photon $\gamma_{\rm VHE}$ of energy ${\cal E}$ scatters off a background photon $\gamma_{\rm BG}$ of energy $\epsilon$ an $e^+ e^-$ pair can be created according to the Breit-Wheeler process $\gamma_{\rm VHE} + \gamma_{\rm BG} \to e^+ + e^-$~\cite{breitwheeler,heitler}, thereby removing a photon from the beam and so causing a dimming of the observed blazar. It can be shown that the cross-section for the considered process is {\it maximal} for~\cite{gouldschreder} 
\begin{equation}
\label{0805201a}
\epsilon \simeq \left(\frac{900 \, {\rm GeV}}{{\cal E}} \right) {\rm eV}~.
\end{equation}  
Therefore, for $100 \, {\rm GeV} \lesssim {\cal E} < 100 \lesssim {\rm TeV}$ we get 
\begin{equation}
\label{11052018a}
0.009 \, {\rm eV}  \lesssim \epsilon  \lesssim 9 \, {\rm eV}~.
\end{equation}  
Hence $\epsilon$ ranges from the far-infrared to the ultraviolet. Unfortunately, within this band the sky is dominated by the Extragalactic Background Light (EBL), which is the total light emitted by galaxies during their whole evolution (for a review, see~\cite{dwek}). As a consequence, the VHE photon beam emitted by blazars undergoes a severe EBL-absorption before being detected (an updated quantitative estimate of this effect is contained in~\cite{dgr2013}). Such an opacity is quantified by the optical depth $\tau_{\rm CP} ({\cal E}_0, z)$  for a blazar at redshift $z$ and observed at energy ${\cal E}_0$. Correspondingly, the photon survival probability for the VHE photon beam propagation after its journey to us is given by 
\begin{equation}
\label{mr13112017bQ}
P_{\gamma \to \gamma}^{\rm CP} ({\cal E}_0, z) =  e^{- \, \tau_{\rm CP} ({\cal E}_0, z)}~. 
\end{equation}  

As first shown in 2007 by De Angelis, Roncadelli and Mansutti~\cite{drm2007}, photon-ALP oscillations in the beam -- occurring in extragalactic space -- can drastically reduce the EBL opacity. Basically, the gist is as follows. Owing to photon-ALP oscillations, a photon acquires a `split personality' as it travels towards us: sometimes it behaves as a {\it true photon} and undergoes EBL absorption, but sometimes it behaves as an ALP which is {\it insensitive} to the EBL (as we shall see in Section III). Thus, the effective optical depth $\tau_{\rm ALP} ({\cal E}_0, z)$ in extragalactic space becomes {\it smaller} than $\tau_{\rm CP} ({\cal E}_0,z)$, and Eq. (\ref{mr13112017bQ}) becomes
\begin{equation} 
\label{a012122010Wq}
P_{\gamma \to \gamma}^{\rm ALP} ({\cal E}_0,z) = e^{- \tau_{\rm ALP} ({\cal E}_0,z)}~.
\end{equation}
So, a look at Eqs. (\ref{mr13112017bQ}) and (\ref{a012122010Wq}) entails that even a {\it small decrease} of $\tau_{\rm ALP} ({\cal E}_0, z)$ with respect to $\tau_{\rm CP} ({\cal E}_0, z)$  implies a {\it large enhancement} of $P_{\gamma \to \gamma}^{\rm ALP} ({\cal E}_0,z)$ as compared to $P_{\gamma \to \gamma}^{\rm CP} ({\cal E}_0, z)$. In this way the EBL absorption can indeed be drastically reduced, thereby greatly enlarging the VHE transparency and the corresponding $\gamma$-ray horizon. A much more detailed development of this idea is reported in~\cite{dgr2011}. 

Manifestly, the quantitative aspects of the above scenario -- ultimately encoded in $P_{\gamma \to \gamma}^{\rm ALP} ({\cal E}_0,z)$ -- depend on $m_a$, $g_{a \gamma \gamma}$, and on the morphology and strength of ${\bf B_{\rm ext}}$. While the region of the parameter plane $(m_a, g_{a \gamma \gamma})$ for which extragalactic photon-ALP oscillations give rise to an observable effect will be discussed in Section III, the issue concerning ${\bf B}_{\rm ext}$ should be briefly addressed here because this is the main novelty of the present analysis (we shall come back to this point in more detail in Section IV). Hereafter, ${\bf B}_{\rm ext}$ will be denoted by ${\bf B}$ for notational simplicity.

Actually, this topic is highly nontrivial and has been discussed carefully in our previous paper to be referred to as GR2018a~\cite{gr2018a}, and so here we report only a short summary of its conclusions.

\

It has become customary to described ${\bf B}$ by means of a domain-like network with the following properties: all magnetic domains have the same size $L_{\rm dom}$ equal to the ${\bf B}$ coherence length, and in each domain ${\bf B}$ is assumed to be homogeneous and to have the same strength ${\bf B}$, but that its direction changes {\it randomly} and {\it discontinuously} passing from one domain to the next (for a review, see~\cite{kronberg1994,grassorubinstein}). Because of the latter fact, this model will be referred to as having {\it sharp edges}, and so we call it {\it domain-like sharp-edges} (DLSHE) model. 

As a matter of fact, the main physical motivation of {\it any} domain-like model devised to describe ${\bf B}$ relies upon the very realistic assumption that its seeds are galaxies, which magnetize extragalactic space through strong galactic outflows of ionized matter which get amplified by turbulence. Two well known possibilities -- which are not mutually exclusive -- are young dwarf galaxies~\cite{kronberg1999} and
quasars~\cite{furlanettoloeb}. In either case, it has been shown that the resulting ${\bf B}$ has a coherence length ${\cal O} (1 \, {\rm Mpc})$ and strength ${\cal O} (1 \, {\rm nG})$. Whence $L_{\rm dom} = {\cal O} (1 \, {\rm Mpc})$ and $B = {\cal O} (1 \, {\rm nG})$. Unfortunately, it is still impossible to determine the strength of ${\bf B}$ in every domain, but the overall picture seems to suggest that it is nearly the same in all domains, a fact captured by the DLSHE model.

Yet, the DLSHE model is a highly mathematical idealization, due to the jump of ${\bf B}$ from one domain to the next. While this fact makes it very easy to solve the beam propagation equation within a single domain, it leads to correct results only under an unstated assumption: the oscillation length $l_{\rm osc}$ must be {\it considerably larger} than $L_{\rm dom}$. Indeed, in such a situation only a small fraction of an oscillation is coherently affected by ${\bf B}$ (recall that coherence is preserved only inside single domains). As a consequence, the behaviour of ${\bf B}$ across the edges becomes irrelevant. 

As first pointed out in 2015 by Dobrynina, Kartavtsev and Raffelt~\cite{raffelt2015}, at very high energies photon dispersion on the CMB becomes the dominant effect, which causes $l_{\rm osc}$ to decrease. As a consequence, things change drastically whenever $l_{\rm osc} \lesssim L_{\rm dom}$, because in this case a whole oscillation -- or even several oscillations -- probe the whole domain, and if it is unphysical like in the DLSHE model then the results come out unphysical as well. 

The simplest way out of this problem is to smooth out the edges in such a way that the change of the ${\bf B}$ {\it direction} is continuous across the domain edges, even if it is still {\it random}. Therefore, in either case only a random {\it single realization} of the beam propagation process is {\it observable at once}. As a consequence, the photon survival probability can be denoted as $P^{\rm ALP}_{\gamma \to \gamma} \bigl({\cal E}_0,z; \phi (y), \theta (y) \bigr)$, where $\phi (y)$ and $\theta (y)$ are the two angles that fix the direction of ${\bf B} (y)$ in space at point $y$ along the beam. The price to pay in order to work with such a {\it domain-like smooth-edges} (DLSME) model is that the beam propagation equation within a single domain is three-dimensional and very difficult to solve analytically. As shown in GR2018a, such an equation turns out to become effectively {\it two-dimensional}. In order to get a feeling about such a dimensional reduction, two facts should be kept in mind.

\begin{itemize}

\item First -- as suggested by the above two models~\cite{kronberg1999,furlanettoloeb} -- the {\it strength} of ${\bf B}$ should vary rather little in different domains, and so we can average it over many domains and attribute in first approximation the resulting value to {\it each} domain, denoting it for simplicity again by $B$.  

\item We recall that the interaction term is $g_{a \gamma \gamma} \, a \, {\bf E} \cdot {\bf B}$, which entails that ALPs in the beam along the $y$-direction couple {\it only} to the component of ${\bf B}_T (y)$ which is transverse to the beam. Because of the previous point, we consistently take the {\it strength} of ${\bf B}_T (y)$ constant and equal to 
\begin{equation}
\label{19052018a}
B_T = \left(\frac{2}{3} \right)^{1/2} \, B~. 
\end{equation}

\end{itemize}

With some effort, the two-dimensional beam propagation equation can be solved exactly and analytically. Very remarkably, such a solution is {\it undistinguishable} from the numerical solution of the above three-dimensional exact equation (more about this in GR2018a). This means that the whole physics of the problem is confined inside the planes $\Pi (y)$ perpendicular to the beam rather than being spread out throughout the full 
three-dimensional space. Correspondingly $P^{\rm ALP}_{\gamma \to \gamma} \bigl({\cal E}_0, z; \phi (y), \theta (y) \bigr) \to P^{\rm ALP}_{\gamma \to \gamma} \bigl({\cal E}_0, z; \phi (y) \bigr)$, where $\phi (y)$ is the angle between ${\bf B}_T (y)$ and a fixed fiducial $x$-direction {\it equal} in all domains (namely in all planes $\Pi (y)$).

\

So far, blazars have been observed in the VHE band with the presently operating Imaging Atmospheric Cherenkov Telescopes (IACTs) H.E.S.S. (High Energy Stereoscopic System)~\cite{hess}, MAGIC (Major Atmospheric Gamma Imaging Cherenkov Telescopes)~\cite{magic} and VERITAS (Very Energetic Radiation Imaging Telescope Array System)~\cite{veritas}, which reach energies up to ${\cal O} (10 \,  {\rm TeV})$. Accordingly, the corresponding $\gamma \leftrightarrow a$ oscillation length $l_{\rm osc}$ turns out to be much larger than $L_{\rm dom}$, and so the standard DLSHE models can be -- and has been -- correctly used. 

But with the advent of the new generation of VHE observatories like CTA (Cherenkov Telescope Array)~\cite{cta}, HAWC (High-Altitude Water Cherenkov Observatory)~\cite{hawc}, GAMMA 400 (Gamma-Astronomy Multifunction Modules Apparatus)~\cite{g400}, LHAASO (Large High Altitude Air Shower Observatory)~\cite{lhaaso}, TAIGA-HiSCORE (Tunka Advanced Instrument for Gamma-ray and Cosmic ray Astrophysics-Hundred Square km Cosmic ORigin Explorer)~\cite{desy} and HERD (High Energy cosmic-Radiation Detection)~\cite{herd} the situation changes dramatically. Because they explore the whole VHE band end even beyond, the phenomenon of photon dispersion on the CMB becomes crucial~\cite{raffelt2015}. As a consequence, it turns out that as ${\cal E}$ becomes larger and larger $l_{\rm osc}$ gets smaller and smaller. Defining the energy 
${\cal E}_{\rm eq}$ such that $l_{\rm osc} ({\cal E}_{\rm eq}) \equiv  L_{\rm dom}$, the 
need to employ the DLSME model for ${\cal E} \gtrsim {\cal E}_{\rm eq}$ becomes {\it compelling}. Specifically, we will see that ${\cal E}_{\rm eq} = {\cal O} (40 \, {\rm TeV})$ with a large uncertainty
\

The paper is organized as follows. In Section II we cursorily sketch the main properties of ALPs, we recall the propagation equation of a photon/ALP beam in extragalactic space and we show how to obtain the photon-ALP conversion probability for a pure state as well as the photon survival probability for mixed states. In addition, two important regimes -- depending on the beam energy -- are addressed. In Section III we take into account all available bounds on the model parameters, along with the allowed effects. We also consider some realistic benchmark values of these parameters. In Section IV we briefly recall our domain-like smooth-edges (DLSME) model (GR2018a). In Section V we explain how the DLSME model can be framed within the cosmological context, and we present the solution of the beam propagation equation (the explicit form is too cumbersome to be reported here and the reader can find it in GR2018a). In Section VI we exhibits our results, in particular the Figures where the photon survival probability from a blazar to us is plotted versus the detected energy ${\cal E}_0$. We do that for 7 blazars at different redshifts $z$, and -- for each of them -- for different values of our benchmark values. The physical meaning of our results is discussed in Section VII. The relation with previous work is the subject of Section VIII. Finally, in Section IX we offer our conclusions. The Appendix shortly summarizes an argument that explains why -- contrary to conventional physics -- flat spectrum radio quasars (FSRQs) emit up to $400 \, {\rm GeV}$. Such a mechanism consists of photon-ALP oscillations as framed within standard blazar emission models, and works for values of the parameters that are in full agreement with those leading to the enhanced transparency in the VHE band. 
 
\section{Main properties of axion-like particles}

In the present Section we recall very briefly the main properties of ALPs that will be needed in the subsequent discussion (a more thorough account can be found in GR2018a).

According to the previous discussion, the ALP Lagrangian is
\begin{equation}
\label{lagr}
{\cal L}_{\rm ALP} =  \frac{1}{2} \, \partial^{\mu} a \, \partial_{\mu} a - \, \frac{1}{2} \, m_a^2 \, a^2 - \, \frac{1}{4} g_{a \gamma \gamma} \, F_{\mu\nu} \tilde{F}^{\mu\nu} a = \frac{1}{2} \, \partial^{\mu} a \, \partial_{\mu} a - \, \frac{1}{2} \, m_a^2 \, a^2 + g_{a \gamma \gamma} \, {\bf E} \cdot {\bf B} \, a~,
\end{equation}     
where ${\bf E}$ and ${\bf B}$ are the electric and magnetic components of the electromagnetic tensor $F_{\mu\nu}$ ($\tilde{F}^{\mu\nu}$ is its dual).

Clearly, in Eq. (\ref{lagr}) ${\bf E}$ is the electric field of a beam photon while ${\bf B}$ is the extragalactic magnetic field. 

Within this Section we confine our attention to a single generic $n$-th magnetic domain ($1 \leq n \leq N$), and for notational simplicity drop the sub-index $n$.

Because we are working in the regime ${\cal E} \gg m_a$, the short-wavelength approximation (WKB) can safely be employed, which implies that the photon/ALP beam propagation equation becomes a Schr\"odinger-like equation with $t$ replaced by the coordinate $y$ along the beam. It reads
\begin{equation}
\label{eqprop}
\left( i \frac{d}{d y} + {\cal E} + {\cal M} ({\cal E}, y) \right) \, \psi (y) = 0~,
\end{equation}
with
\begin{equation}
\label{smvett}
\psi ( y ) \equiv \left(
\begin{array}{c}
\gamma_1 ( y ) \\
\gamma_2 ( y ) \\
a( y ) \\
\end{array}
\right)~,
\end{equation}
where $\gamma_1 ( y )$ and $\gamma_2 ( y )$ are the photon amplitudes with polarization along the $x$- and $z$-axis, respectively, while $a ( y )$ is the ALP amplitude. This achievement due to Raffelt and Stodolsky~\cite{rs1988} is of great importance, since it allows the beam to be handled by means of the {\it formalism of non-relativistic quantum mechanics}.

Hence, denoting by $U \bigl({\cal E}; y, y_0; \phi (y) \bigr)$ the transfer matrix -- which is the solution of Eq. (\ref{eqprop}) with initial condition $U \bigl(y_0, y_0; \phi (y) \bigr) = 1$ -- we can write the propagation of a generic wave function as
\begin{equation}
\label{trmatr}
\psi (y) = U \bigl({\cal E}; y, y_0; \phi (y) \bigr) \, \psi(y_0)~.
\end{equation}
A simplification comes about by setting 
\begin{equation}
\label{expon}
U \bigl({\cal E}; y, y_0; \phi (y) \bigr) = e^{ i {\cal E} \left( y - y_0 \right)} \, {\cal U} \bigl({\cal E}; y, y_0; \phi (y)  \bigr)~,
\end{equation}
where ${\cal U} \bigl({\cal E}; y, y_0; \phi (y) \bigr)$ is the transfer matrix corresponding to  the reduced Sch\"odinger-like equation
\begin{equation}
\label{redeqprop}
\left(i \frac{d}{d y} + {\cal M} ({\cal E}, y) \right) \psi( y ) = 0~.
\end{equation}
As a matter of fact, the wave function describes a linearly polarized beam, but since in the VHE band the polarization cannot be measured we have to employ the density matrix $\rho (y)$ which satisfies the Von Neumann-like equation associated with Eq. (\ref{eqprop}) -- equivalently with Eq. (\ref{redeqprop}) -- namely 
\begin{equation}
\label{vneum}
i \frac{d \rho (y)}{d y} = \rho (y) \, {\cal M}^{\dag} ({\cal E}, y) - {\cal M} ({\cal E}, y) \, \rho (y)~,
\end{equation}
whose solutions can be represented in terms of ${\cal U} \bigl(y , y_0; \phi (y) \bigr)$ as
\begin{equation}
\label{unptrmatr}
\rho \bigl( y; \phi (y) \bigr) = {\cal U} \bigl({\cal E}; y , y_0; \phi (y) \bigr) \, \rho (y_0) \, {\cal U}^{\dag} \bigl({\cal E}; y, y_0; \phi (y) \bigr)~.
\end{equation}
Therefore, the probability that the beam in the initial state $\rho_0$ at $y_0$ will be found in the final state $\rho (y)$ at $y$ reads
\begin{equation}
\label{unpprob}
P^{\rm ALP}_{\gamma \to \gamma} \bigl({\cal E}; y, \rho; y_0, \rho_0; \phi (y) \bigr) = {\rm Tr} \Bigl[\rho \, {\cal U} \bigl({\cal E}; y, y_0; \phi (y) \bigr) \, \rho_0 \, {\cal U}^{\dag} \bigl({\cal E}; y, y_0; \phi (y) \bigr) \Bigr]
\end{equation}
with ${\rm Tr} \, \rho (y_0) = {\rm Tr} \, \rho (y) =1$, as shown in GR2018a. 

Let us now come back to the mixing matrix ${\cal M} ({\cal E}, y)$ entering Eq. (\ref{redeqprop}). In the present situation it has the following form
\begin{equation}
\label{matr}
{\mathcal M} ({\cal E}, y) \equiv
\left(
\begin{array}{ccc}
\Delta_{\rm CMB} ({\cal E}) + \Delta_{\rm abs}  ({\cal E}) + \Delta_{\rm pl}  ({\cal E}) & 0 & \Delta_{\rm a \gamma} \, 
{\rm sin} \,\phi (y) \\[8pt]
0 & \Delta_{\rm CMB} ({\cal E}) + \Delta_{\rm abs} ({\cal E}) + \Delta_{\rm pl} ({\cal E}) & \Delta_{\rm a \gamma} \, 
{\rm cos} \,\phi (y) \\[8pt]
\Delta_{\rm a \gamma} \, {\rm sin} \,\phi (y) & \Delta_{\rm a \gamma} \, {\rm cos} \, \phi (y) & \Delta_{aa}  ({\cal E})
\end{array}
 \right)~.
\end{equation}
The physical meaning of the terms of ${\cal M} ({\cal E}, y)$ is as follows. The contribution from photon dispersion on the CMB is $\Delta_{\rm CMB} ({\cal E})= 0.522 \cdot 10^{-42} \, {\cal E}$~\cite{raffelt2015}, the contribution from the EBL absorption is $\Delta_{\rm abs}  ({\cal E}) = i/\bigl(2 \lambda_{\gamma} ({\cal E}) \bigr)$ where $\lambda_{\gamma} ({\cal E})$ denotes the corresponding photon mean free path inside a single domain (more about this, in Subsection V.B), the contribution from the plasma frequency of the ionized intergalactic medium is $\Delta_{\rm pl} ({\cal E}) = - \, \omega^2_{\rm pl}/(2 {\cal E})$ while the remaining terms are $\Delta_{\rm a \gamma} = g_{a \gamma \gamma} \, B_T/2$ and $\Delta_{a a} ({\cal E}) = - \, m_a^2/(2 {\cal E})$. Finally, we recall that $ \omega_{\rm pl}$ is related to the electron number density $n_e$ by~\cite{rs1988} 
\begin{equation}
\label{SI90708a}
\omega_{\rm pl} = 3.69 \cdot 10^{- 11} \left(\frac{n_e}{{\rm cm}^{- 3}} \right)^{1/2} \, {\rm eV}~.
\end{equation}
So, the considered photon/ALP beam can formally be regarded as a {\it non-relativistic, 
three-level, unstable quantum system}~\cite{dgr2011}.

\

Strictly speaking, the equations to be reported below should be computed by using the exact results expressed by Eqs. (54) and (91) of GR2018a. However, they would lead to unacceptably cumbersome expressions which shed no light on what is going on. For this reason, until the end of this Section we work within the DLSHE model -- which amounts to set $\phi (y) = 0$ into Eq. (\ref{matr}) -- whose validity extends up to ${\cal E} = {\cal O} (40 \, {\rm TeV})$ for a realistic choice of $\langle L_{\rm dom} \rangle$ of the magnetic domains (as we shall see in Subsection V.A). In a sense, we are following a perturbative approach.

Thanks to the fact that the EBL absorption is independent of the $\gamma \leftrightarrow a$ oscillation length $l_{\rm osc} ({\cal E})$ (as we shall see in Subsection V.B), it can be momentarily neglected in the forthcoming analysis. Actually, we have~\cite{rs1988}
\begin{equation}
\label{mr16112017a}
l_{\rm osc} ({\cal E}) = 2 \pi \left[\left(\frac{m_a^2  - \omega_{\rm pl}^2 }{2 {\cal E}} + 0.522 \cdot 10^{-42} \, {\cal E} \right)^2  + \bigl(g_{a \gamma \gamma} \, B_T \bigr)^2 \right]^{-1/2}~,
\end{equation}
and so it is a simple exercise to show that the photon-ALP conversion probability takes the form
\begin{equation}
\label{SI9tris} 
P_{\gamma \to a} ({\cal E}, y) = \left(\frac{g_{a \gamma \gamma} \, B_T \, l_{\rm osc} 
({\cal E})}{2 \pi} \right)^2 {\rm sin}^2 \left(\frac{\pi y}{l_{\rm osc} ({\cal E})} \right)~, \ \ \ \ \ \ \ \ \ \ \ \ y \leq L_{\rm dom}~.
\end{equation}

Because ${\bf B}$ and $g_{a\gamma\gamma}$ enter ${\cal L}_{\rm ALP}$ in the combination $g_{a\gamma\gamma} \, B_T$, it is quite useful to define  
\begin{equation}
\label{09032018a}
\xi \equiv \left(\frac{B_T}{{\rm nG}} \right) \bigl(g_{a\gamma\gamma} \, 10^{11} \, {\rm GeV}   \bigr)~. 
\end{equation}

In the rest of this Section we shall try to express all quantities in terms of $\xi$.

\

Let us first define the {\it low-energy threshold} ${\cal E}_L$ and the {\it high-energy threshold} ${\cal E}_H$ as 
\begin{equation}
\label{eqprop1q}
{\cal E}_L \equiv \frac{|m_a^2 - \omega^2_{\rm pl}|}{2 g_{a \gamma \gamma} \, B_T} \simeq \frac{2.56}{\xi} \left| \left(\frac{m_a}{{\rm neV}} \right)^2 - \, \left(\frac{{\omega}_{\rm pl}}{{\rm neV}} \right)^2 \right| \, {\rm TeV}~,  
\end{equation}
and 
\begin{equation}
\label{18042018e}
{\cal E}_H \equiv 1.92 \cdot 10^{42} \, g_{a \gamma \gamma} \, B_T \simeq 3.74 \cdot 10^2 \, \xi \, {\rm GeV}~, 
\end{equation} 
respectively, where the second equalities in both equations arise from Eq. (\ref{09032018a}). 

\

Then three different regimes are allowed, depending on which of the three terms in the square brackets in Eq. (\ref{mr16112017a}) dominates over the others  (see GR2018a).

\begin{itemize}

\item ${\cal E} < \, {\cal E}_L$ -- This is the {\it low-energy weak mixing regime}, in which the term $\propto {\cal E}^{- 1}$ dominates. Accordingly, we have
\begin{equation}
l_{\rm osc} ({\cal E}) \simeq \frac{4 \pi \, {\cal E}}{|m_a^2  - \omega_{\rm pl}^2|}~,
\label{18042018c} 
\end{equation}
and
\begin{equation}
P_{\gamma \to a} ({\cal E}, L_{\rm dom}) \simeq \left(\frac{2 \, g_{a \gamma \gamma} \, B_T   \, {\cal E}}{|m_a^2  - \omega_{\rm pl}^2| } \right)^2 {\rm sin}^2 \left(\frac{|m_a^2  - \omega_{\rm pl}^2| \, L_{\rm dom}}{4 \, {\cal E}} \right)~.
\end{equation}
However, since we will not commit ourselves with this case throughout the paper there is no need to discuss its physical properties any further.

\item ${\cal E}_L < {\cal E} < {\cal E}_H$ -- This is the intermediate-energy or {\it strong mixing regime} where the ${\cal E} = {\rm constant}$ term dominates. Correspondingly, we get
\begin{equation}
\label{18042018d}
l_{\rm osc} \simeq \frac{2 \pi}{g_{a \gamma \gamma} \, B_T} \simeq 2.05 \cdot 10^2 \, 
\xi^{- 1} \, {\rm Mpc}~,
\end{equation}
\begin{equation}
\label{18042018cc} 
P_{\gamma \to a} (L_{\rm dom}) \simeq {\rm sin}^2 \left(\frac{g_{a \gamma \gamma} \, B_T  \, L_{\rm dom}}{2} \right) \simeq {\rm sin}^2 \left[1.54 \cdot 10^{- 2} \, \xi \left(\frac{L_{\rm dom}}{{\rm Mpc}} \right) \right]~.
\end{equation}
Manifestly, $l_{\rm osc}$ and $P_{\gamma \to a} (L_{\rm dom})$ turn out to be {\it independent} of both $m_a$ and ${\cal E}$, and $P_{\gamma \to a} (L_{\rm dom})$ becomes {\it maximal}: note that $m_a$ enters ${\cal E}_L$ and nowhere else.

\item ${\cal E} > {\cal E}_H$ -- This is the {\it high-energy weak mixing regime}, which is in a sense a sort of reversed low-energy weak mixing regime where however the term $0.522 \cdot 10^{-42} \, {\cal E}$ dominates over $g_{a \gamma \gamma} \, B_T$ (here the first term in $l_{\rm osc} ({\cal E})$ can safely be neglected). Accordingly, we find 
\begin{equation}
\label{18042018ee}
l_{\rm osc} ({\cal E}) \simeq \frac{1.20 \cdot 10^{43}}{\cal E} \simeq 76.15 \left(\frac{{\rm TeV}}{{\cal E}} \right) {\rm Mpc}~,
\end{equation}
\begin{equation}
\label{18042018f}
P_{\gamma \to a} ({\cal E}, L_{\rm dom}) \simeq 1.39 \cdot 10^{- 1} \, \xi^2 \left(\frac{{\rm TeV}}{{\cal E}} \right)^2 \, {\rm sin}^2 \left[4.12 \cdot 10^{- 2} \left(\frac{L_{\rm dom}}{{\rm Mpc}} \right) \left(\frac{{\cal E}}{{\rm TeV}} \right) \right]~.
\end{equation}
Evidently, $l_{\rm osc} ({\cal E})$ decreases with increasing ${\cal E}$ -- as already anticipated -- and $P_{\gamma \to a} ({\cal E}, L_{\rm dom}) $ exhibits oscillation in ${\cal E}$: this reflects the fact that the individual realizations of the beam propagation are also oscillating functions of ${\cal E}$. Moreover -- since  $P_{\gamma \to a} ({\cal E}, L_{\rm dom}) \propto {\cal E}^{-2}$ -- as ${\cal E}$ increases the photon-ALP oscillations become unobservable at some point. 

\end{itemize}

\section{Parameters, bounds and resulting effects}

We are now in position to take into account the various observational bounds on our model parameters. In order to make progress, we use the equations derived in the last part of the previous Section within the DLSHE model, and we will need to assume some benchmark values for the parameters in order to figure out the magnitude of the various effects. 

\begin{itemize}   

\item Exactly the same upper bound on $g_{a \gamma \gamma}$ has been derived by the CAST experiment at CERN~\cite{cast} and by the analysis of some particular stars in globular clusters~\cite{straniero}. Both read $g_{a \gamma \gamma} < 0.66 \cdot 10^{- 10} \, {\rm GeV}^{- 1}$ for $m_a < 0.02 \, {\rm eV}$ at the $2 \sigma$ level.

\item The most stringent upper bound on the strength of the extragalactic magnetic field is 
$B < 1.7 \, {\rm nG}$ on the Mpc scale at the $2 \sigma$ level arises from a global analysis of rotation measures~\cite{pshirkov2016}. 

\item Therefore, owing to Eqs. (\ref{19052018a}) and (\ref{09032018a}) the combination of these two bounds yields $\xi < 9.20$. Below, we will take $\xi = 0.5$, $\xi = 1.0$, $\xi = 2.0$, $\xi = 5.0$ as benchmark values.

\end{itemize}

\

Before addressing other bounds, we want to prove our previous statement that ALP are insensitive to the EBL, and more generally that for all practical purposes in the present situation ALPs interact {\it neither with matter nor with radiation} (in spite of their two-photon coupling). 

\begin{figure}[h]
\centering
\includegraphics[width=.40\textwidth]{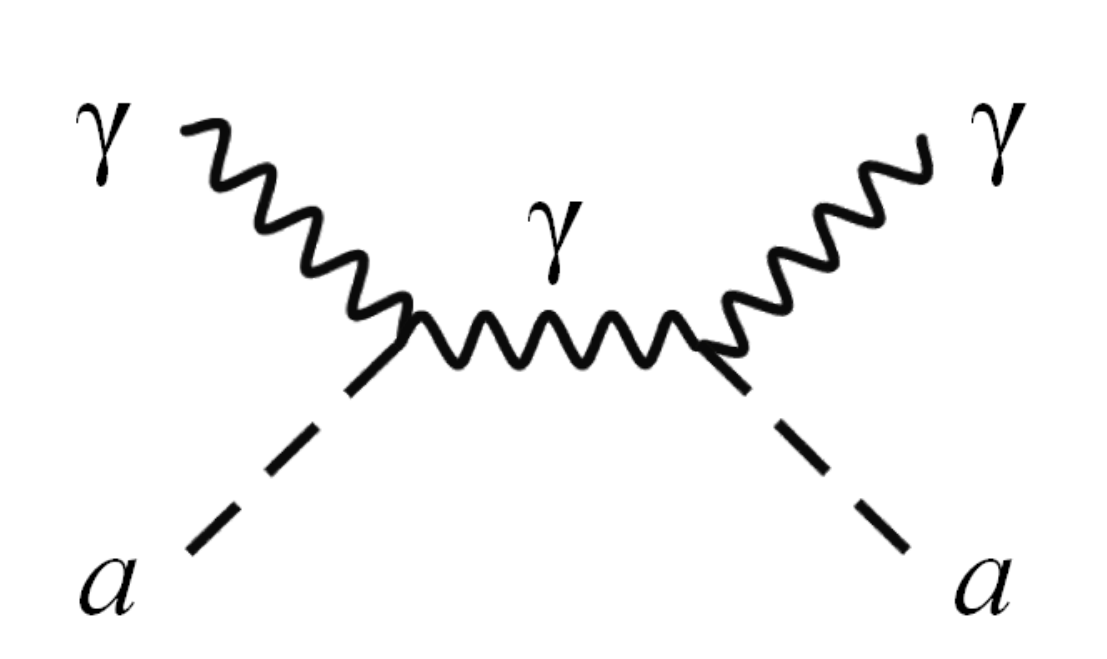}
\caption{\label{ALPph} Feynman diagram for the $a \gamma \to a \gamma$ scattering.}
\end{figure}

\begin{figure}[h]
\centering
\includegraphics[width=0.40\textwidth]{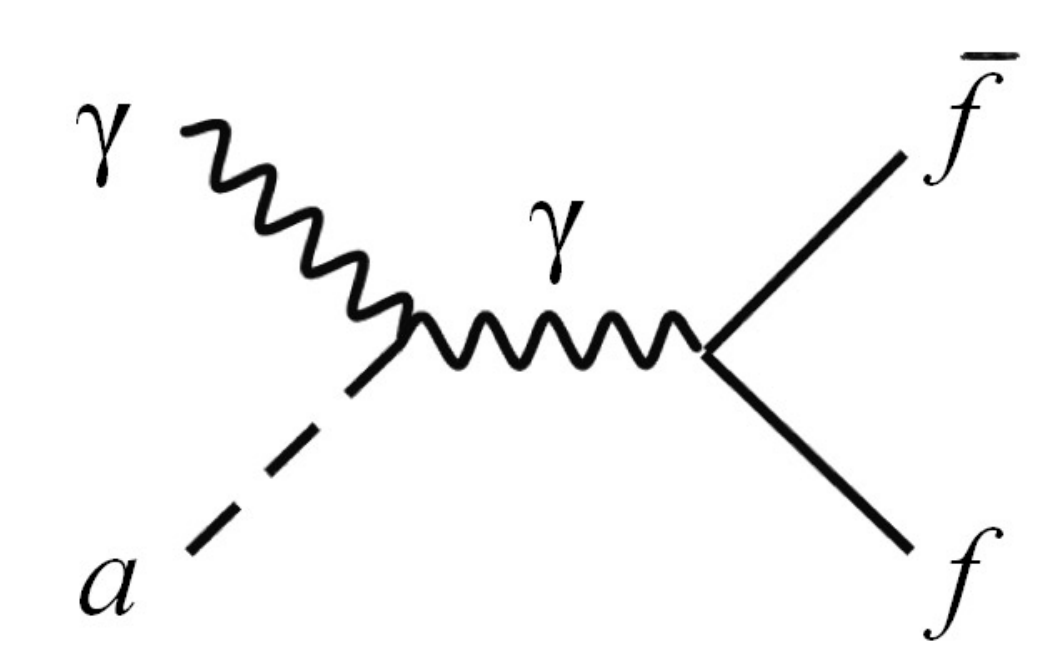}
\caption{\label{fey5Q} Feynman diagram for the $a + \gamma \to f + 
{\overline f}$ scattering in the $s$-channel, and for the $a + f \to \gamma + f$ scattering in the $u$-channel.}
\end{figure}

ALPs might interact with the EBL only through two processes: $a + \gamma \to a + \gamma$ and $a + \gamma \to f + {\overline f}$, where $f$ denotes a generic charged fermion. Consider first the process $a + \gamma \to a + \gamma$, which is represented by the $s$-channel of the Feynman diagram in Figure~\ref{ALPph}. A simple estimate gives 
$\sigma (a + \gamma \to a + \gamma) = {\cal O} \bigl(s \, g^4_{a \gamma \gamma} \bigr)$, and enforcing the CAST bound we get $\sigma (a + \gamma \to a + \gamma) < {\cal O} (10^{- 68}) \, \bigl({\cal E}_a/{\rm GeV} \bigr) \, \bigl(\epsilon/{\rm GeV} \bigr) \, {\rm cm}^2$, where ${\cal E}_a$ is the ALP energy and $\epsilon$ is the energy of an EBL photon. Now, since we consider VHE photons and $\gamma \leftrightarrow a$ oscillations are energy conserving, it follow that ${\cal E}_a = {\cal E}$.  On the other hand, from Eq. (\ref{11052018a}) we obtain $\sigma (a + \gamma \to a + \gamma) < {\cal O} (10^{- 76}) \, \bigl({\cal E}/{\rm GeV} \bigr) \, {\rm cm}^2$, which even for ${\cal E} = 1000 \, {\rm TeV}$ becomes $\sigma (a + \gamma \to a + \gamma) < {\cal O} (10^{- 70}) \, {\rm cm}^2$. The second process is $a + \gamma \to f + {\overline f}$ in the $s$-channel of the Feynman diagram in Figure~\ref{fey5Q}. Manifestly, $\sigma (a + f \to \gamma + f)  = {\cal O} \bigl(\alpha \, g^2_{a \gamma \gamma} \bigr)$, which -- thanks to the CAST bound --  becomes $\sigma (a + f \to \gamma + f) < {\cal O} (10^{- 50}) \, {\rm cm}^2$. Finally, the process $a + f \to \gamma + f$ is represented by the $u$-channel of the Feynman diagram in Figure~\ref{fey5Q}. So, apart from kinematic factors ${\cal O} (1)$ we still end up with 
$\sigma (a + f \to \gamma + f)  < {\cal O} (10^{- 50}) \, {\rm cm}^2$. Because such cross-sections are extremely small and the density of extragalactic space is so tiny, our statement is proved.

\

Coming back to our main line of development, what about ${\cal E}_L$ and ${\cal E}_H$? 

Let us start from ${\cal E}_L$. We clearly want to maximize the $\gamma \leftrightarrow a$ oscillation effect in extragalactic space and so we suppose that the strong mixing regime 
sets in just at the lower end of the VHE band, which means that ${\cal E}_L  \simeq  100 \, {\rm GeV}$ (a more general situation wherein this assumption will be relaxed shall be addressed in a future publication). As a consequence, Eq. (\ref{eqprop1q}) entails 
\begin{equation} 
\label{mr16112017d}  
\left| \left(\frac{m_a}{{\rm neV}} \right)^2 - \, \left(\frac{{\omega}_{\rm pl}}{{\rm neV}} \right)^2 \right| \simeq 3.91 \cdot 10^{- 2} \, \xi \lesssim 0.36~, 
\end{equation} 
where again the bound $\xi < 9.20$ has been used. It is well known that the Universe at large redshift is ionized owing to the lack of the Gunn-Peterson effect~\cite{peeblesbook}. But also the local Universe ($z < 1$) is ionized, even if the determination of the number density of free electrons $n_e$ is a tricky business. Because this point turns out to be fairly unimportant for us, we refer the reader to~\cite{csaki2002b} for a detailed discussion, and we simply quote the result: $n_e < 6 \cdot 10^{- 9} \, {\rm cm}^{- 3}$ for $\Omega_B = 0.045$ and $h_0 = 0.65$. Hence, normalizing this value to the most recent one obtained by PLANCK for $\Omega_B = 0.049$ and $h_0 = 0.67$~\cite{planck}, Eq. (\ref{SI90708a}) yields $\omega_{\rm pl} < 3 \cdot 10^{- 15} \, {\rm eV}$. Therefore, the second term in the l.h.s. of Eq. (\ref{mr16112017d})  can be discarded (as it will become apparent in a moment). Hence, from condition (\ref{mr16112017d}) we find that the resulting ALP upper mass bound is 
\begin{equation} 
\label{17052018a}  
m_a \simeq 1.98 \cdot 10^{- 10} \, \xi^{1/2} \, {\rm eV} \lesssim 6.0 \cdot 10^{- 10} \, {\rm eV}~,
\end{equation} 
which for our benchmark values becomes
\begin{equation}
m_a \simeq  \left\{ \begin{array}{llll}
1.40 \cdot 10^{- 10} \, {\rm eV}~, \ \ \ \ \ \ \ {\rm for} \ \ \ \ \ \ \ \xi = 0.5~; \\
1.98  \cdot 10^{- 10} \, {\rm eV}~, \ \ \ \ \ \ \ {\rm for} \ \ \ \ \ \ \ \xi = 1.0~; \\
2.79 \cdot 10^{- 10} \, {\rm eV}~, \ \ \ \ \ \ \ {\rm for} \ \ \ \ \ \ \ \xi = 2.0~; \\
4.43 \cdot 10^{- 10} \, {\rm eV}~, \ \ \ \ \ \ \ {\rm for} \ \ \ \ \ \ \ \xi = 5.0~. \\
\end{array} \right.
\label{18052018a}
\end{equation}
Note that since Eq. (\ref{mr16112017d}) entails that ${\cal E}_L \propto m_a^2$, an increase of $m_a$ leads to a larger increase of ${\cal E}_L$. Moreover, Eq. (\ref{18042018d}) allows us to compute $l_{\rm osc}$ in the strong mixing regime for our benchmark values. The result is
\begin{equation}
l_{\rm osc} \simeq  \left\{ \begin{array}{llll}
4.10 \cdot 10^2 \, {\rm Mpc}~, \ \ \ \ \ \ \ {\rm for} \ \ \ \ \ \ \ \xi = 0.5~; \\
2.05 \cdot 10^2 \, {\rm Mpc}~, \ \ \ \ \ \ \ {\rm for} \ \ \ \ \ \ \ \xi = 1.0~; \\
1.05 \cdot 10^2 \,  {\rm Mpc}~, \ \ \ \ \ \ \ {\rm for} \ \ \ \ \ \ \ \xi = 2.0~; \\
0.41 \cdot 10^2 \,  {\rm Mpc}~, \ \ \ \ \ \ \ {\rm for} \ \ \ \ \ \ \ \xi = 5.0~, \\
\end{array} \right.
\label{18052018b}
\end{equation}
thereby confirming our expectation that in this case $l_{\rm osc} \gg {\cal O} (1 \, {\rm Mpc})$.

\

Let us finally address ${\cal E}_H$. As before, only an upper bound can be derived. In fact, by combining Eq. (\ref{18042018e}) with the bound $\xi < 9.20$ we find that 
${\cal E}_H < 3.44 \, {\rm TeV}$, which shows that the strong mixing regime covers only the lower end of  the VHE band. According again to our benchmark values, Eq. (\ref{18042018e}) gives 
\begin{equation}
{\cal E}_H \simeq  \left\{ \begin{array}{llll}
1.87 \cdot 10^2 \, {\rm GeV}~, \ \ \ \ \ \ \ {\rm for} \ \ \ \ \ \ \ \xi = 0.5~; \\
3.74 \cdot  10^2 \, {\rm GeV}~, \ \ \ \ \ \ \ {\rm for} \ \ \ \ \ \ \ \xi = 1.0~; \\
7.48 \cdot  10^2 \,  {\rm GeV}~, \ \ \ \ \ \ \ {\rm for} \ \ \ \ \ \ \ \xi = 2.0~; \\
1.87 \, {\rm TeV}~, \ \ \ \ \ \ \ \ \ \ \ \ \ \ {\rm for} \ \ \ \ \ \ \ \xi = 5.0~. \\
\end{array} \right.
\label{19052018q}
\end{equation}

Still -- as stated in Section I -- the dominance of the photon dispersion on the CMB reduces $l_{\rm osc} ({\cal E})$ according to Eq. (\ref{18042018ee}) and concomitantly -- thanks to Eq. (\ref{18042018f}) -- reduces the photon survival probability. At first sight, we would be led to evaluate $l_{\rm osc} ({\cal E}_H)$ for our benchmark values as ${\cal E} > {\cal E}_H$. However, this would be an obsolete exercize. The real question is: at which energies do we have $l_{\rm osc} ({\cal E})  \lesssim L_{\rm dom}$? Because we shall take $L_{\rm dom}$ inside a range (as explained in Section V.A) a clear-cut answer is impossible. The best we can do is to ask at what energy $ {\cal E}_{\rm eq}$ the following condition $l_{\rm osc} ({\cal E}) = \langle L_{\rm dom} \rangle$ holds true. Since we shall see in Section V.A that our probability density for $L_{\rm dom}$ has $\langle L_{\rm dom} \rangle = 2 \, {\rm Mpc}$, then from Eq. (\ref{18042018ee}) it follows that this happens at ${\cal E}_{\rm eq} = {\cal O} (40 \, {\rm TeV})$. Needless to say, this value is affected by a rather large uncertainty. Still, by systematically employing the DLSME model no uncertainty shows up.

The overall emerging picture is therefore as follows. Just above ${\cal E} \simeq 100 \, {\rm GeV}$ we are in the strong mixing regime, where $l_{\rm osc}$ and $P_{\gamma \to a} (L_{\rm dom})$ turn out to be independent of both $m_a$ and ${\cal E}$, and $P_{\gamma \to a} (L_{\rm dom})$ becomes maximal. In addition, $l_{\rm osc} \gg L_{\rm dom}$, and so the DLSHE model is perfectly viable, thanks to Eq. (\ref{18052018b}). As the energy increases and reaches one of values reported in Eq. (\ref{19052018q}) -- or at most ${\cal E}_H < 3.44 \, {\rm TeV}$ -- the high-energy weak mixing regime takes over. Accordingly, the photon survival probability starts to decrease, but we still have $l_{\rm osc} ({\cal E}) \gg L_{\rm dom}$. Only when the energy reaches a value ${\cal E}_{\rm eq} = {\cal O} (40 \, {\rm TeV})$ starts the oscillation length $l_{\rm osc} ({\cal E})$ to become equal to $L_{\rm dom}$, and becomes smaller than that for ${\cal E}_{\rm eq} \gtrsim {\cal O} (40 \, {\rm TeV})$. Hence, for ${\cal E} \gtrsim {\cal O} (40 \, {\rm TeV})$ the use of our DLSME model developed in GR2018a is {\it compelling}.

\ 

Let us next turn to the bounds on $g_{a \gamma \gamma}$ which depend on $m_a$. Basically, use is made of the observed absence of the characteristic oscillating behavior of the individual realizations of the beam propagation around ${\cal E}_L$ (as explained in Section IV of GR2018a). Several bounds have been derived~\cite{wb2012,hessbound,gr2013,marshfabian2017,fermi2016,zhang2018,mnt2018}, but the only one which is very marginally relevant for the present work has been obtained by applying such a strategy by the Fermi/LAT collaboration -- observing the central galaxy NGC 1275 in the Perseus cluster -- obtaining $g_{a \gamma \gamma} < 5 \cdot 10^{- 12} \, {\rm GeV}^{- 1}$ at the $2 \sigma$ level for $5 \cdot 10^{- 10} \, {\rm eV} < m_a  < 500 \cdot 10^{- 10} \, {\rm eV}$~\cite{fermi2016}. Nevertheless, it does not affect at all our benchmark values of $m_a$ reported in Eq. (\ref{18052018a}).               

A quite different story concerns the bound derived from the lack of detection of ALPs 
supposedly emitted by the supernova SN1987A. They were assumed to convert into 
gamma-rays of the same energy in the Galactic magnetic field, and so they ought to have been observed by the Gamma-Ray Spectrometer (GRS) aboard the Solar Maximum Mission (SMM) satellite, even though its line of sight was orthogonal to the direction to supernova SN1987A. Generally speaking, it is always risky to draw reliable conclusions from such old data because of the unknown systematics. More precisely, in 1989 three members of the SMM satellite team Chupp, Westrand and Reppin published a paper about the possibility to detect the radiative decay of a heavy neutrino~\cite{chupp1989}. Because they had actually built the GRS, they were well aware of its limitations. And quite correctly, right at the beginning of their paper they put the following statement: `{\it In using the GRS data, it is important to recognize that a gamma-ray spectrometer in a satellite orbit experiences highly variable and sometimes unpredictable background variations. Unfortunately, some authors have assumed that the background of the SMM GRS referred to in International Astronomical Union Circular No. 4365 is the diffuse cosmic gamma-ray background. In fact, the GRS background is the sum of a number of components. Normally, the dominant component is generated by activation of the spacecraft and the scintillation detectors by primary cosmic rays (modulated by geomagnetic rigidity) and also trapped protons in the South Atlantic radiation anomaly for about 3 of the 15-16 satellite orbits per day. Unpredictable transient events include solar flares, cosmic gamma-ray bursts, particle precipitation events, and manmade events. For these reasons extreme care should be taken in reaching conclusions using the SMM GRS data}'~\cite{chupp1989}. 

Unfortunately, all these cautionary remarks have been ignored in the subsequent theoretical papers. In 1996 two papers almost simultaneously appeared which analyzed the ALP emission from supernova SN1987A~\cite{raffelt1996,masso1996}. From the lack of detection a bound on $g_{a \gamma \gamma}$ for nearly massless ALP was derived. In order to figure out the weight of this missed detection, we recall that GRS has detected in 10 years 177 Gamma Ray Bursts (GRBs), namely less than $5 \%$ of GRBs impinging on the satellite in that period~\cite{grs-grb}. Therefore, GRS detects a very limited subsample of GRBs selected by energy, position in the field of view, orbital and geomagnetic parameters (including delayed effects of activations). Thus, we conclude that the no-detection of photons coming from ALPs cannot be regarded as a proof that ALPs are not emitted by SN1987A.

Coming back to the two above mentioned papers, basically both of them address the protoneutron star (PNS) just before the bounce, but are actually framed within the vacuum plus the condition that electrons are fully degenerate, which are correctly supposed to play no dynamical role because of Pauli blocking. Apart from some obvious mistakes -- like the Debye screening of protons by other protons -- several important issues are neglected, like the existence of strong interactions, the fast rotation of the PNS and its huge magnetic field. 
Obviously, this is an unacceptable oversimplification. In 2015 a much more detailed follow-up analysis of the same problem appeared, which takes into account strong interactions {\it only} to the extent that they produce a small reduction of the nucleon mass and a slight proton degeneracy~\cite{payez2015}. Yet, such an analysis does not start from first principles, but {\it relies upon} the conclusions of ~\cite{raffelt1996,masso1996}, which therefore reverberate on the results. Among the mistakes are the following ones. 1) Protons are Debye screened by other protons!  2) The Primakoff ALP production mechanism is computed (with massless photons) by means of  the Minkowski metric, in spite of the fact that it must be treated with a metric which takes into account the properties of the medium, namely nuclear matter at twice the nuclear saturation density and $T \simeq 40 \, {\rm K}$, which is obviously very different from the Minkowski one of ordinary vacuum (this point is explained in great detail in Raffelt's book~\cite{raffeltbook}). 3) Moreover, the PNS typically has a rotation period of one millisecond, hence it is a non-inertial reference frame, so that the use of Minkowski metric is again wrong. 4) Magnetic fields as strong as $B = \bigl(10^{12} - 10^{16} \bigr) \, {\rm G}$ are simply discarded. A thorough analysis of this issue is reported in~\cite{bgr2018}, and in conclusion the claimed bound $g_{a \gamma \gamma} \lesssim 5.3 \cdot 10^{- 12} \, {\rm GeV}^{- 1}$ for $m_a \lesssim 4.4 \cdot 10^{- 10} \, {\rm eV}$~\cite{payez2015} is incorrect and cannot be used. 

\section{Sketch of the domain-like smooth-edges (DLSME) model}

As explained in Section I, our aim is to apply the DLSME domain model to a monochromatic photon/ALP beam of energy ${\cal E}$ emitted by a far-away blazar, propagating through extragalactic space  and reaching us.

Let us therefore briefly summarize this model (see GR2018a for a more detailed account). We suppose that there are $N$ domains between the blazar and us, and we number them in such a way that domain $1$ is the one closest to the blazar while domain $N$ is the one closest to us. Note that the same convention has been used in GR2018a but {\it not} in~\cite{dgr2011}. Momentarily, we suppose that all domains have the same length. We denote by $\{y_{D,n}\}_{0 \leq n \leq N}$ the set of coordinates which defines the beginning ($y_{D,n-1}$) and the end ($y_{D,n}$) of the $n$-th domain ($1 \leq n \leq N$) towards the blazar. As pointed out in Section I, we start from the three-dimensional case. 

Unfortunately, given our ignorance of the strength of ${\bf B}$ in every domain and the previous suggestion that it should vary rather little in different domains we imagine to average $B$ over many domains, and next we attribute the resulting value to each of them, so that $B$ -- but not ${\bf B}$ -- will henceforth be regarded as constant in first approximation. 

As already pointed out, in GR2018a we have found that the three-dimensional problem becomes fully equivalent to a two-dimensional one, since what matters is the direction of transverse component of the extragalactic magnetic field ${\bf B}_T ( y )$ in the planes $\Pi ( y )$ perpendicular to the beam. Therefore -- by considering a fixed fiducial $x$-direction  inside the planes $\Pi (y)$ which is the {\it same} for all domains -- we denote by $\{\phi_n\}_{1 \leq n \leq N}$ the set of angles that ${\bf B}_T ( y )$ forms with the $x$-direction in the middle of every domain. Owing to the previous assumption and Eq. (\ref{19052018a}) also $B_T$ but not ${\bf B}_T$ can be taken as constant in all domains.

Actually, since ${\bf B}_T ( y ) $ changes {\it randomly} from one domain to the next, in order for ${\bf B}_T ( y )$ to be continuous all along the beam it is necessary that it has {\it equal values} on both sides of every edge, e.g. the one between the $n$-th and the $(n + 1)$-th domain. So, we suppose that in the $n$-th domain ${\bf B}_T ( y )$ is homogeneous in the central part, but as the distance from the edge with the $(n + 1)$-th domain decreases we assume that ${\bf B}_T ( y )$ linearly changes in such a way to become equal to ${\bf B}_T ( y )$ on the same edge but as evaluated in the $(n + 1)$-th domain. In this way, the continuity of the components of ${\bf B}_T ( y )$ along the whole beam is ensured. 

\begin{figure}[h]       
\begin{center}
\includegraphics[width=.90\textwidth]{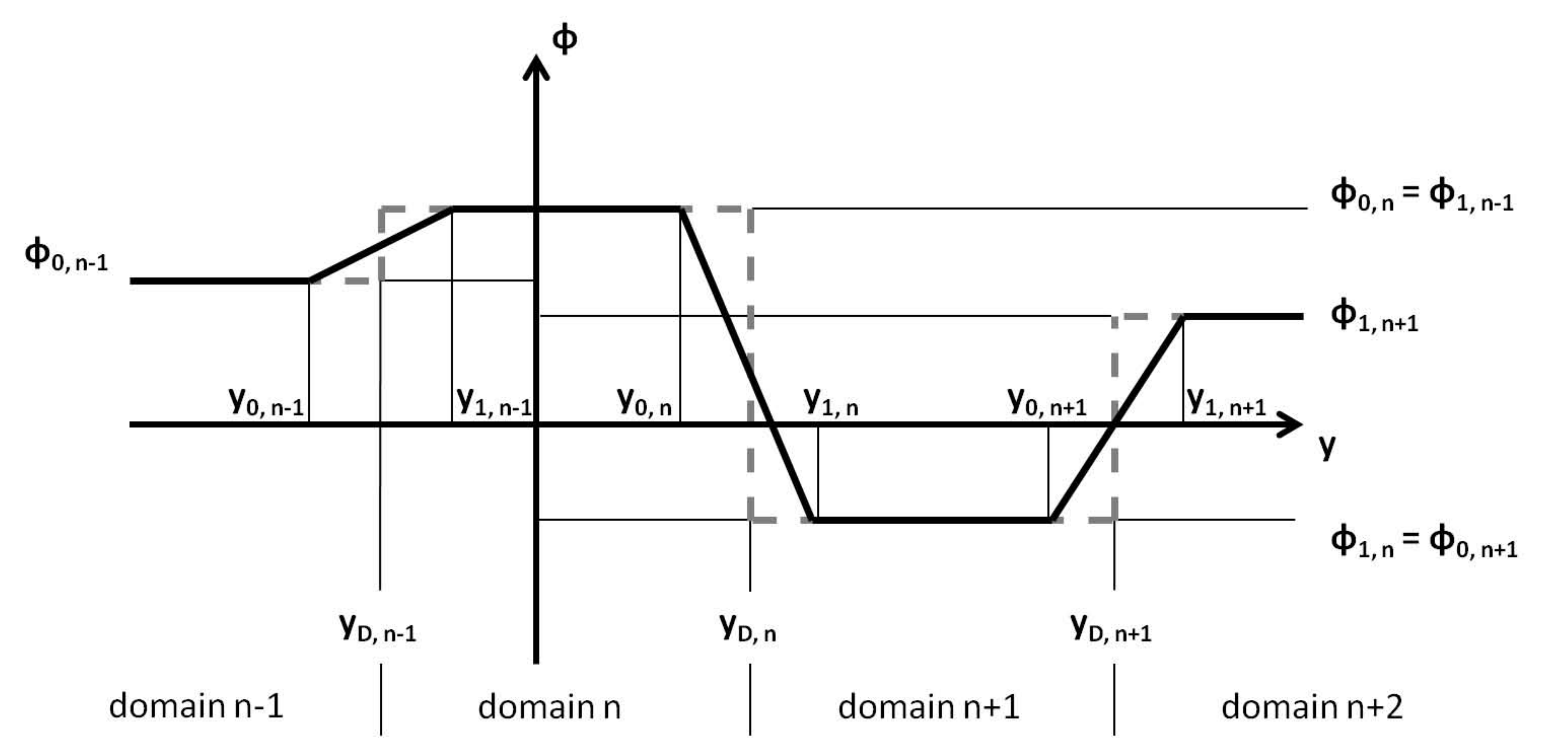}
\end{center}
\caption{\label{linear} 
DLSME model -- Behavior of the angle $\phi$ between ${\bf B}_T (y)$ and the fixed fiducial $x$-direction (equal for all domains) inside $\Pi (y)$. The solid black line is the new smooth version, while the broken gray line represents the usual jump of ${\bf B}_T (y)$ from one domain to the next in the DLSHE model. The horizontal solid and broken lines partially overlap. For illustrative simplicity, we have taken the same length for all domains. Note that the blazar is on the extreme left of the figure while the observer is on the extreme right.}
\end{figure}

A schematic view of this construction is shown in Figure~\ref{linear}.

In more mathematical terms, it is convenient to define the two quantities $y_{0,n}$ and $y_{1,n}$ as
\begin{equation}
\label{x0}
y_{0,n} \equiv y_{D,n} - \frac{\sigma}{2} \bigl(y_{D,n} - y_{D,n-1} \bigr)~, \,\,\,\,\,\,\,\,\,\, (1 \leq n \leq N - 1)~;
\end{equation}
\begin{equation}
\label{x1}
y_{1,n} \equiv y_{D,n} + \frac{\sigma}{2} \bigl(y_{D,n+1} - y_{D,n} \bigr)~, \,\,\,\,\,\,\,\,\,\, (1 \leq n \leq N - 1)~;
\end{equation}
where $\sigma \in [0,1]$ is the {\it smoothing parameter}. The interval $[y_{0,n},y_{1,n}]$ is the region where we apply the smoothing procedure, namely where the angle $\phi (y)$ changes smoothly from the value $\phi_{0,n} \equiv \phi_n$ in the $n$-th domain to the value $\phi_{0,n+1} \equiv \phi_{n+1}$ in the ($n+1$)-th domain. Clearly, for $\sigma = 0$ we have $y_{0,n} = y_{1,n}$, the smoothing region vanishes and we recover the DLSHE model. On the other hand, for $\sigma = 1$ then $y_{0,n}$ becomes the midpoint of the $n$-th domain, and likewise $y_{1,n}$ becomes the midpoint of the $(n+1)$-th domain: in this case the smoothing is maximal, because we never have a constant value of $\phi$ in any domain. The general case is of course intermediate -- represented by a value of $0 < \sigma < 1$ -- so that in the central part of a domain the angle is constant ($\phi_{0,n}$) and then it linearly joins the value of the constant angle in the next domain ($\phi_{1,n}$). Therefore, in a generic interval $[y_{1,n-1},y_{1,n}] \, (1 \leq n \leq N-1)$ we have  
\begin{equation}
\label{phi}
\phi (y) = \begin{cases}
\phi_{0,n} = {\rm constant}~, & y \in [y_{1,n-1},y_{0,n}]~;\\[8pt]
\phi_{0,n} + \dfrac{\phi_{0,n+1} - \phi_{0,n}}{y_{1,n} - y_{0,n}} \, (y - y_{0,n})~, & y \in [y_{0,n},y_{1,n}]~.
\end{cases} 
\end{equation}
We stress that within our convention the blazar redshift is $z \equiv z_0$, the points $y_{D,n-1}$ and $y_{D,n}$ defining the $n$-th domain have redshift $z_{n - 1}$ and $z_n$ ($z_n < z_{n - 1}$), respectively, and we set ${\overline z}_n \equiv \bigl(z_{n - 1} + z_n \bigr)/2$ for the average redshift of the $n$-th domain. Likewise, the emitted beam has energy ${\cal E}_0$, whereas the beam at points $y_{D,n-1}$ and $y_{D,n}$ has energy ${\cal E}_{n -1}$ and ${\cal E}_n$ (${\cal E}_{n -1} > {\cal E}_n$), respectively. Finally, we define the average energy of the $n$-th domain as ${\overline {\cal E}}_n \equiv \bigl({\cal E}_{n -1} + {\cal E}_n \bigr)/2$, and the observer has energy ${\cal E}_N$. As usual, 
${\cal E}_n = (1 + z_n) {\cal E}_N$ (${\cal E}_0 = (1 + z_0) {\cal E}_N)$.

\section{General strategy}

In order to make our approach fully defined, a few topics are still to be addressed: the probability density for the domains length, the magnetic field in a generic domain, and the EBL absorption in each domain. 

\subsection{Probability density for the domain lengths}

\begin{figure}[h]       
\begin{center}
\includegraphics[width=.44\textwidth]{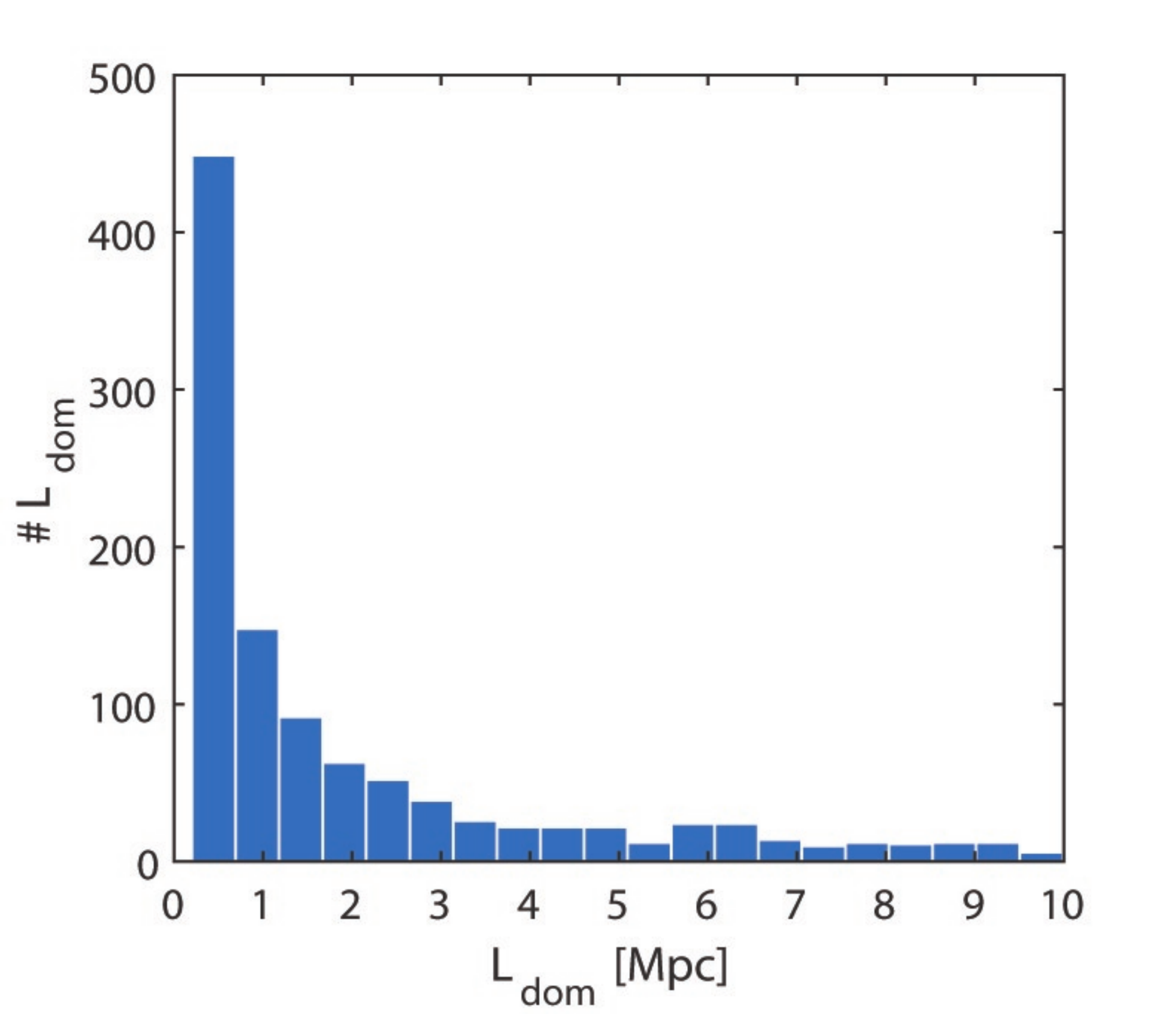}
\end{center}
\caption{\label{histLdom} 
Number of domains as a function of their size in the case $z = 0.5$.}
\end{figure}

Realistically, we expect the domains to be similar but of course not identical. Therefore, we contemplate a spread of the domain length $\{L_{\rm dom,n} \}_{1 \leq n \leq N}$ within a fixed range. Recalling the properties of the extragalactic magnetic field mentioned in Section I, at redshift $z = 0$ we take for the probability density of the domains length the power law $\propto L_{\rm dom}^{- 1.2}$ inside the range $0.2 \, {\rm Mpc} - 10 \, {\rm Mpc}$, which entails that $\langle L_{\rm dom} \rangle = 2 \, {\rm Mpc}$, indeed allowed by present bounds~\cite{durrerneronov}. It goes without saying that our choice is largely arbitrary. 
What about its $z$-dependence? Answering this question is an impossible task: because turbulence plays a crucial role in the amplification of the extragalactic magnetic fields ${\bf B}$, one cannot simply scale $L_{\rm dom,n} \propto1/(1 + {\overline z}_n)$. Strictly speaking, it should be kept in mind that $L_{\rm dom}$ is not just a simple length but the coherence length of ${\bf B}$. So, in order to avoid the risk of making wrong assumptions, we prefer to take the probability density of the domains length as $z$-independent. As an illustration, the corresponding histogram is shown in Figure~\ref{histLdom}.

\subsection{EBL absorption and magnetic field within a single domain}

What has yet to be done is to take EBL absorption into account and to determine  the magnetic field strength $B_{T,n}$ in the generic $n$-th domain of size $L_{{\rm dom},n}$. 

\begin{itemize}

\item The first goal can be achieved as follows. We employ the very recent EBL model of Franceschini and Rodighiero~\cite{franceschinirodighiero}. They have tabulated the values of the optical depth $\tau \bigl({\cal E}, z)$ for all values of ${\cal E}$ from $10 \, {\rm GeV}$ to $1000 \, {\rm TeV}$ and $z$ up to 2. Because the domain size is so small as compared to the cosmological standards, we can safely drop cosmological evolutionary effects within a single domain. Then -- as far as absorption is concerned -- what matters is the mean free path $\lambda_{\gamma,n} \equiv \lambda_{\gamma} ({\overline {\cal E}}_n)$ for the Breit-Wheeler scattering $\gamma_{\rm VHE} + \gamma_{\rm EBL} \to e^+ + e^-$, and so the term $i/2 \lambda_{\gamma,n}$ should be inserted into the 11 and 22 entries of the matrix (\ref{matr}). In order to evaluate $\lambda_{\gamma, n}$ we proceed as follows. We start by noting that in general -- recalling Eq. (\ref{mr13112017bQ}) -- by switching to the present notation the observed flux $\Phi_{\rm obs} ({\cal E}_N)$ is related to the emitted one $\Phi_{\rm em} \bigl({\cal E}_N (1 + z_0) \bigr)$ by
\begin{equation}
\label{11052018q}
\Phi_{\rm obs} ({\cal E}_N) = P_{\gamma \to \gamma}^{\rm CP} ({\cal E}_N, z) \, \Phi_{\rm em}  \bigl({\cal E}_N (1 +  z) \bigr) = e^{- \tau_{\rm CP} ({\cal E}_N, z)} \, \Phi_{\rm em}  \bigl({\cal E}_N (1 +  z) \bigr)~.
\end{equation}
We stress that we have to use Eq. (\ref{mr13112017bQ}) rather than Eq. (\ref{a012122010Wq}) because ALPs are {\it not} EBL absorbed. Next, in order to evaluate $\lambda_{\gamma, n}$ we use a trick. We imagine that two hypothetical, identical blazars $A$ and $B$ are located at both edges of the $n$-th domain along the line of sight and are observed by us, with $B$ farther than $A$. Moreover, we suppose that both $A$ and $B$ can be switched on and off at will. Now, when $A$ is on and $B$ is off, Eq. (\ref{11052018q}) takes the form
\begin{equation}
\label{11052018w}
\Phi_{A, {\rm obs}} \bigl({\cal E}_N \bigr) = e^{- \tau_{\rm CP} ({\cal E}_N, z_n)} \, \Phi_{A, {\rm em}}  \bigl({\cal E}_N (1 +  z_n) \bigr)~,
\end{equation}
whereas in the opposite case Eq. (\ref{11052018q}) becomes
\begin{equation}
\label{11052018e}
\Phi_{B, {\rm obs}} ({\cal E}_N) = e^{- \tau_{\rm CP} ({\cal E}_N, z_{n - 1})} \, \Phi_{B, {\rm em}}  \bigl({\cal E}_N (1 +  z_{n - 1}) \bigr)~.
\end{equation}
Obviously we have $\Phi_{A, {\rm obs}} ({\cal E}_N) = \Phi_{B, {\rm obs}} ({\cal E}_N)$ just by construction, and so by combining Eqs. (\ref{11052018w}) and (\ref{11052018e}) we see that the flux change across the $n$-th domain is
\begin{equation}
\label{MR3}
\Phi_A \bigl({\cal E}_n  \bigr) = e^{- \bigl[\tau_{\rm CP} ({\cal E}_N, z_{n - 1}) - \tau_{\rm CP} ({\cal E}_N, z_n) \bigr]} \, \Phi_B \bigl({\cal E}_{n - 1} \bigr)~, 
\end{equation}
since cosmological evolutionary effects have been discarded. Correspondingly, Eq. (\ref{MR3}) should take the usual non-cosmological form 
\begin{equation}
\label{MR4} 
\Phi_A \bigl({\overline {\cal E}}_n \bigr) = {\rm exp} \, \left(- \, \frac{L_{\rm dom,n}}{\lambda_{\gamma,n}} \right) \Phi_B \bigl({\overline{\cal E}}_n \bigr)~,
\end{equation}
and the comparison with Eq. (\ref{MR3}) ultimately yields
\begin{equation}
\label{MR5}
\lambda_{\gamma, n} = \frac{L_{\rm dom,n}}{\tau_{\rm CP} ({\cal E}_N, z_{n - 1}) - \tau_{\rm CP} ({\cal E}_N, z_n)}~, 
\end{equation}
which is the desired result.

\item In order to accomplish the second task, we note that because of the high conductivity of the IGM the magnetic flux lines can be thought as frozen inside it~\cite{grassorubinstein}. Therefore, the flux conservation during the cosmic expansion implies $B \propto (1 + z)^2$. Therefore -- owing to Eq. (\ref{19052018a}) -- in the $n$-th magnetic domain we have on average $B_{T,n} (y) = \bigl(B_{T,N}  (y) \bigr) \, \bigl(1 + {\overline z}_n \bigr)^2$, where of course $B_{T,N} (y)$ is the strength of ${\bf B}_T  (y)$ in the local Universe, namely in the domain closest to the observer ($z = 0$).

\end{itemize}

At this point, the mixing matrix (\ref{matr}) in a single $n$-th domain entering the reduced Schr\"odinger equation (\ref{redeqprop}) is just Eq. (\ref{matr}) in which ${\cal M} ({\overline {\cal E}}_n, y)$ has all terms replaced by those evaluated above with the sub-index $n$, and of course ${\cal E}_n = {\cal E}_N \, \bigl(1 + {\overline z}_n \bigr)$.

\subsection{Solving the beam propagation equation}

As a matter of fact, this task has been the main accomplishment of GR2018a, and consists of two steps. 

\begin{itemize}

\item Solving the beam propagation equation inside a {\it single} $n$-th domain. The solution turns out to be
\begin{equation}
\label{20052018q}
{\cal U}_n \bigl({\overline {\cal E}}_n; z_n, z_{n - 1}; \phi_{1,n}, \phi_{1,n-1} \bigr) = {\cal U}_{{\rm var},n} \bigl({\overline {\cal E}}_n ; z_n, z_{n - 1}; \phi_{1,n}, \phi_{0,n} \bigr) \, 
{\cal U}_{{\rm const}, n} \bigl({\overline {\cal E}}_n; z_n, z_{n - 1}; \phi_{0, n} \bigr)~, 
\end{equation}
for an arbitrary choice of the angle $\phi_{0, n}$. Unfortunately, the explicit forms of the two transfer matrices is much too cumbersome to be reported here, and the reader can found them in GR2018a: see its Eqs. (54) and (91) with ${\cal E} \to {\overline {\cal E}}_n$ and the appropriate conversions in order to go over from physical space to redshift space. 

\item We should obtain the {\it whole} transfer matrix from the blazar to us, namely along a single arbitrary realization of the whole beam propagation process. Starting from Eq. (\ref{20052018q}) it is a trivial implication of quantum mechanics that the equation we are looking for has presently the form
\begin{eqnarray}
&\displaystyle  {\cal U}_T \bigl({\cal E}_N; z; \{ \phi_n \}_{1 \le n \le N} \bigr) = {\cal U}_{{\rm const}, N} \bigl({\cal E}_N; z_N, z_{N - 1}; \phi_{0, N} \bigr) \times   \label{20052018w} \\
&\displaystyle \prod_{n=1}^{N-1} {\cal U}_{{\rm var}, n} \bigl({\overline {\cal E}}_n; z_n, 
z_{n - 1}; \phi_{1,n}, \phi_{0,n} \bigr) \ {\cal U}_{{\rm const}, n} \bigl({\overline {\cal E}}_n; z_n, z_{n - 1}; \phi_{0, n} \bigr)~. \nonumber
\end{eqnarray}
We emphasize that this product must be ordered in such a way that the transfer matrixes with smaller and smaller $n$ must be closer and closer to the source.

\end{itemize}

\section{Results}

Our final step consists in evaluating the photon survival probability from the blazar to us along an arbitrary realization of the whole beam propagation process. This goal is again trivially achieved thanks to the analogy with non-relativistic quantum mechanics, namely by extending Eq. (\ref{unpprob}) to $N$ domains. Because the photon polarization cannot be measured at the considered energies, we have to start with the unpolarized beam state and sum the result over the two final polarization states. So, for the reader's convenience we revert to the same, common notation used in Section I, namely ${\cal E}_N \to {\cal E}_0$, $z_N \to 0$, $z_0 \to z$. Accordingly, Eq. (\ref{unpprob}) takes the form
\begin{eqnarray}
&\displaystyle P^{\rm ALP}_{\gamma \to \gamma, {\rm unp}} \bigl({\cal E}_0; \rho_x, \rho_z; z, \rho_{\rm unp}; \, \{\phi_n \}_{1 \le n \le N} \bigr) = \label{20052018r} \\
&\displaystyle \sum_{i = x,z} {\rm Tr} \left[\rho_i \, {\cal U}_T \bigl({\cal E}_0; z; \{ \phi_n \}_{1 \le n \le N} \bigr) \, \rho_{\rm unp} \, {\cal U}_T \bigl({\cal E}_0; z; \{ \phi_n \}_{1 \le n \le N} \bigr)  \right] \nonumber 
\end{eqnarray}
with 
\begin{equation}
\label{sm3a}
\rho_x \equiv
\left(
\begin{array}{ccc}
1 & 0 & 0 \\
0 & 0 & 0 \\
0 & 0 & 0
\end{array}
 \right)~,\,\,\,\,\,\,\,\,\,\,
\rho_z \equiv
\left(
\begin{array}{ccc}
0 & 0 & 0 \\
0 & 1 & 0 \\
0 & 0 & 0
\end{array}
 \right)~,\,\,\,\,\,\,\,\,\,\,
\rho_{\rm unp} \equiv
\frac{1}{2}\left(
\begin{array}{ccc}
1 & 0 & 0 \\
0 & 1 & 0 \\
0 & 0 & 0
\end{array}
\right)~.
\end{equation}

\

Below, we plot the photon survival probability $P^{\rm ALP}_{\gamma \to \gamma, {\rm unp}} \bigl({\cal E}_0; \rho_x, \rho_z; z, \rho_{\rm unp}; \, \{\phi_n \}_{1 \le n \le N} \bigr)$ versus the observed energy ${\cal E}_0$ for 7 simulated blazars at $z = 0.02, 0.05, 0.1, 0.2, 0.5, 1, 2$, taking for each one our benchmark values $\xi = 0.5, 1, 2, 5$. For notational simplicity, we will denote $P^{\rm ALP}_{\gamma \to \gamma, {\rm unp}} \bigl({\cal E}_0; \rho_x, \rho_z; z, \rho_{\rm unp}; \, \{\phi_n \}_{1 \le n \le N} \bigr)$ simply as $P_{\gamma \to \gamma}^{\rm ALP} ({\cal E}_0, z)$. We have considered 1000 random realizations of the propagation process for each choice of $z$, $\xi$, ${\cal E}_0$. In all figures we have taken a random distribution of the domain length $L_{\rm dom}$: we have chosen a power law distribution function with exponent $\alpha = -1.2$ and domain length in the interval between the minimal value $L_{\rm dom}^{\rm min}=0.2 \, \rm Mpc$ and the maximal value $L_{\rm dom}^{\rm max}=10 \, \rm Mpc$. The resulting average domain length is $\langle L_{\rm dom} \rangle = 2 \, \rm Mpc$. We take the smoothing parameter 
$\sigma=0.2$ for the transition from one magnetic domain to the next. The dotted-dashed black line corresponds to conventional physics, the solid light-gray line to the median of all the realizations of the propagation process and the solid yellow line to a single realization with a random distribution of the domain lengths and of the orientation angles of the magnetic field inside the domains. The filled area is the envelope of the results on the percentile of all the possible realizations of the propagation process at 68 $\%$ (dark blue), 90 $\%$ (blue) and 99 $\%$ (light blue), respectively. In the upper-left panel we have chosen $\xi=0.5$, in the upper-right panel $\xi=1$, in the lower-left panel $\xi=2$ and in the lower-right panel $\xi=5$.

\newpage

\section*{Figures for ${\bf z = 0.02}$}

\begin{figure}[h]       
\begin{center}
\includegraphics[width=.53\textwidth]{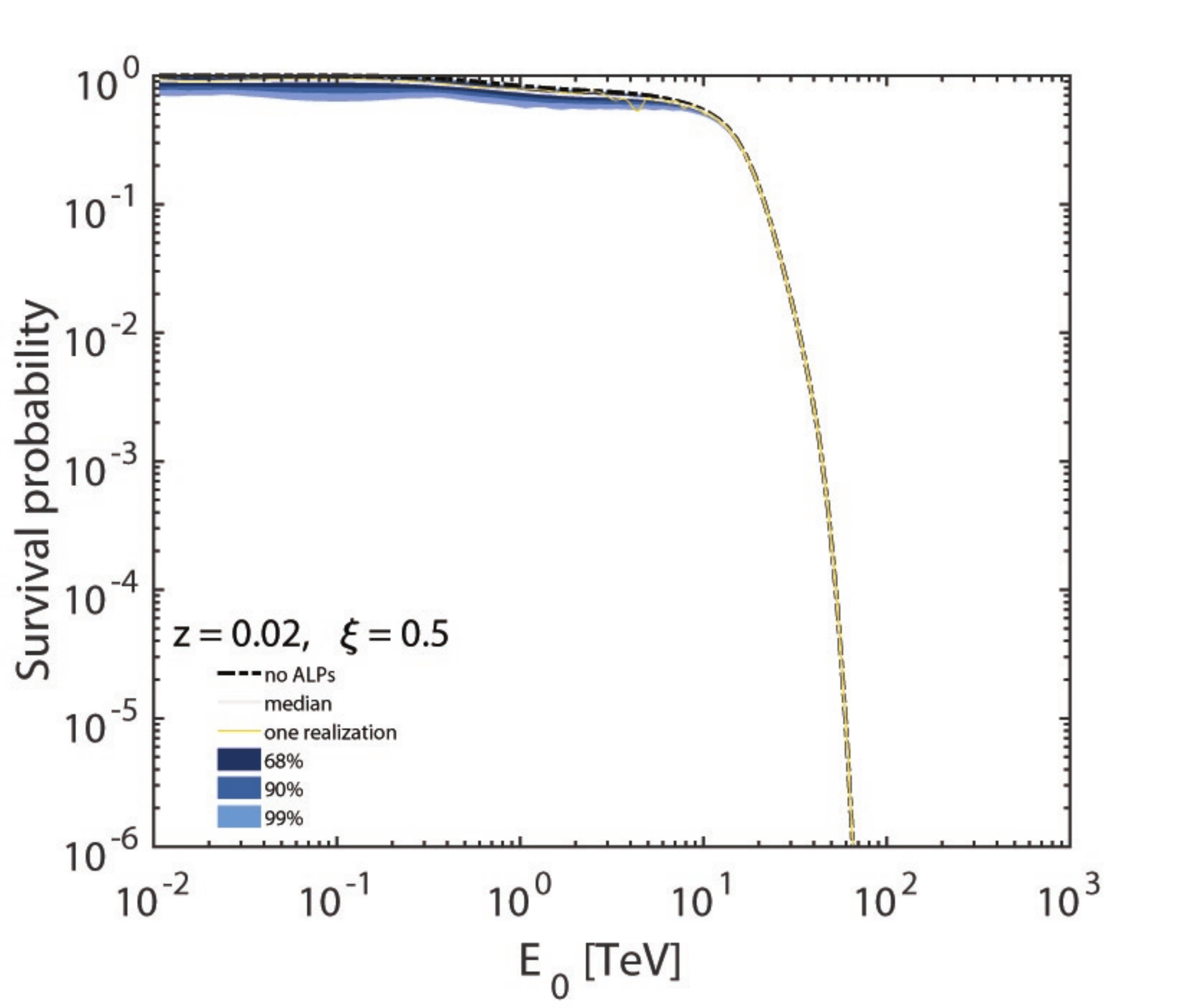}\includegraphics[width=.53\textwidth]{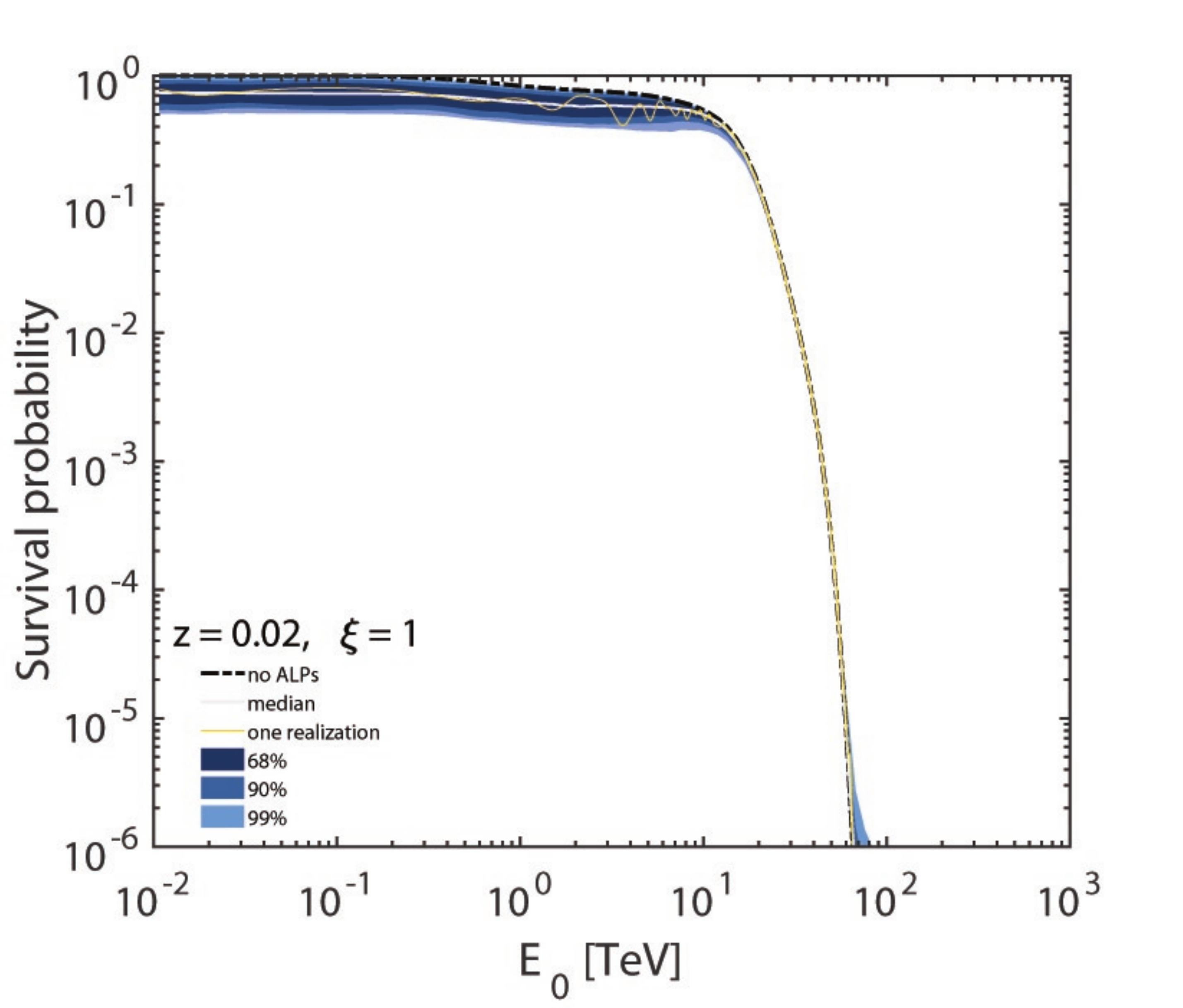}
\includegraphics[width=.53\textwidth]{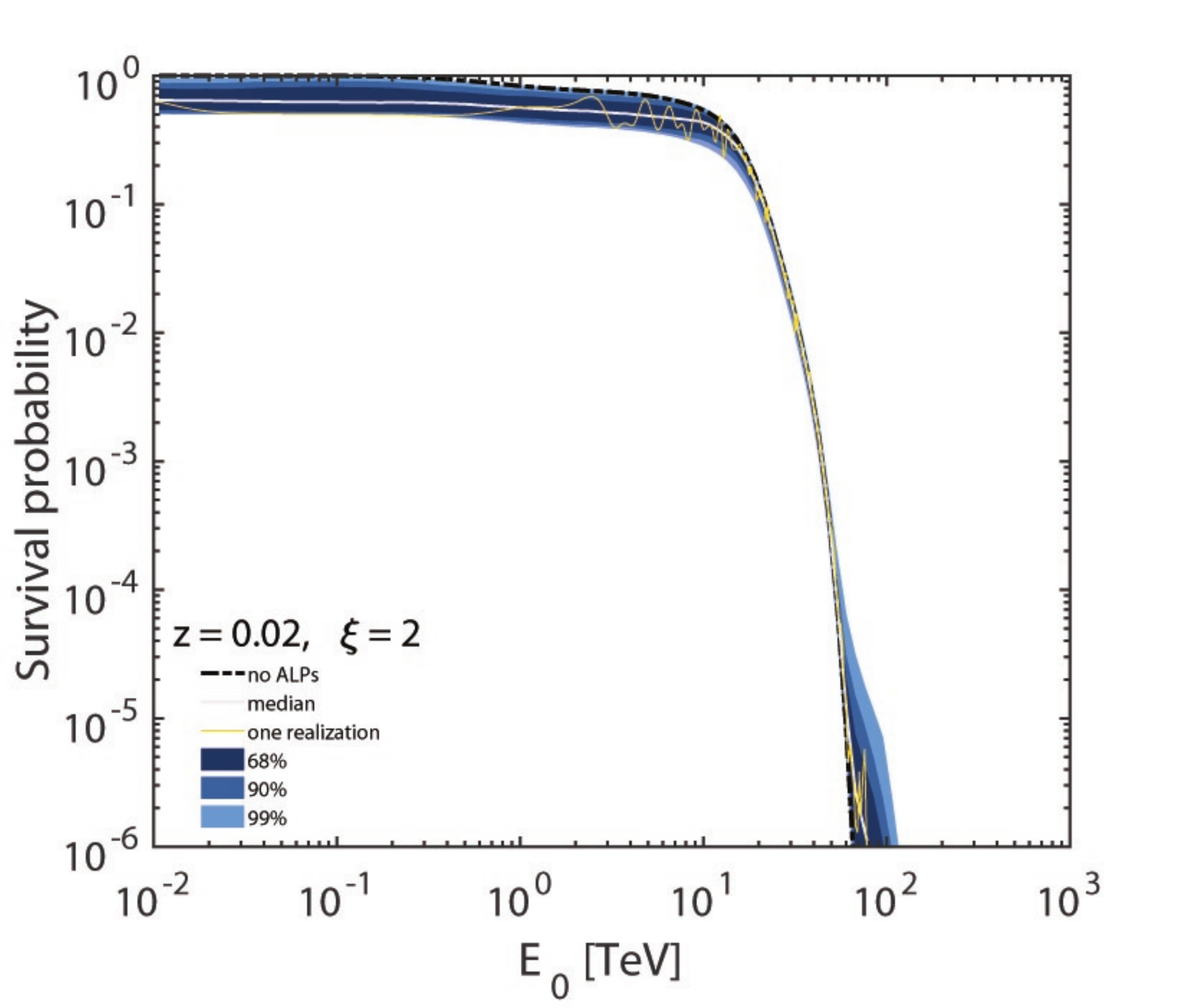}\includegraphics[width=.53\textwidth]{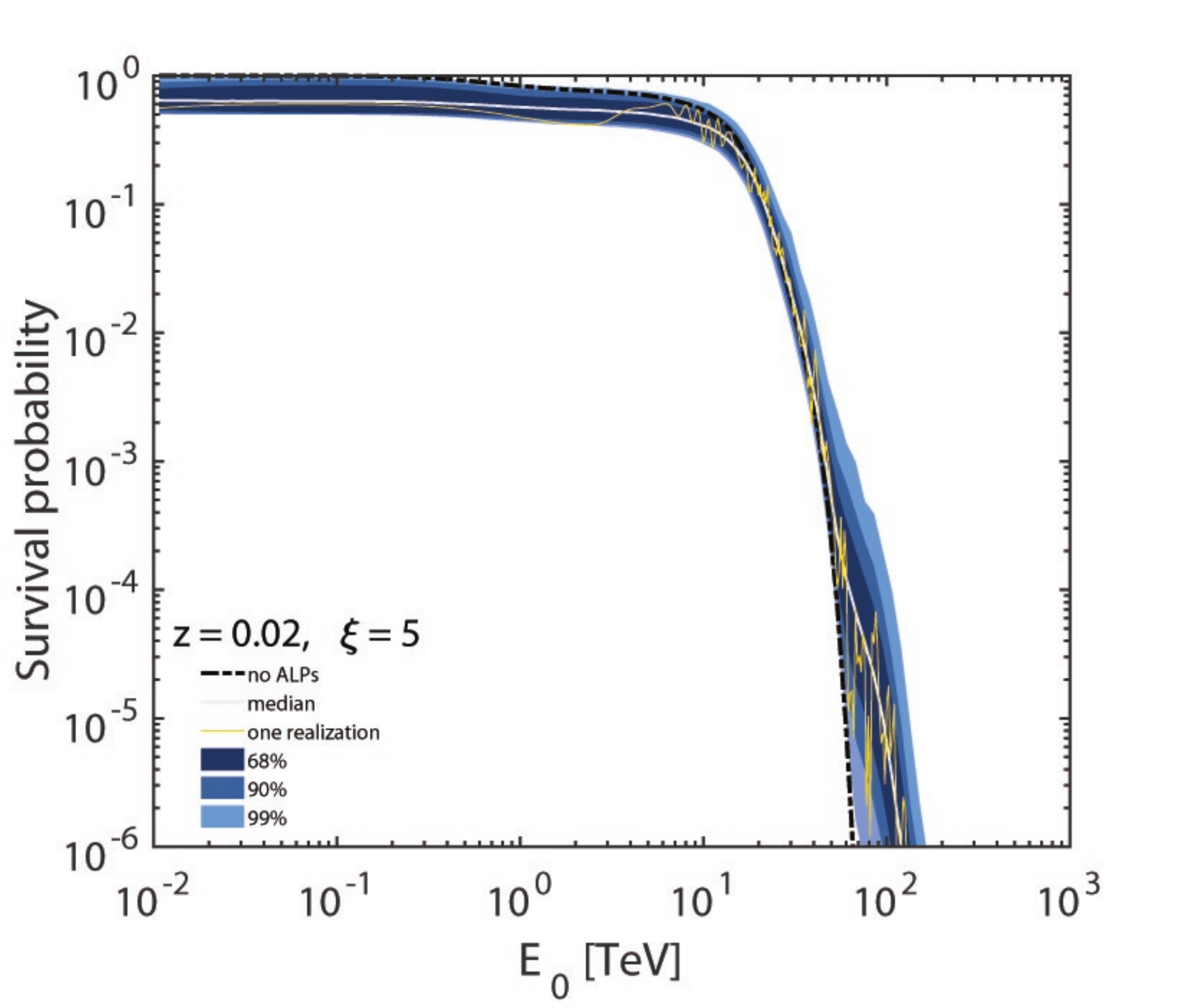}
\end{center}
\caption{\label{z002} 
Behaviour of $P_{\gamma \to \gamma}^{\rm ALP} ({\cal E}_0, z)$ versus the observed energy ${\cal E}_0$ for $z=0.02$. In all the figures we have taken a random distribution of the domain length $L_{\rm dom}$: we have chosen a power law distribution function with exponent $\alpha = -1.2$ and domain length in the interval between the minimal value $L_{\rm dom}^{\rm min}=0.2 \, \rm Mpc$ and the maximal value $L_{\rm dom}^{\rm max}=10 \, \rm Mpc$. The resulting average domain length is $\langle L_{\rm dom} \rangle = 2 \, \rm Mpc$. We consider a smoothing parameter with $\sigma=0.2$ for the transition from one magnetic field domain to the following one. The dotted-dashed black line corresponds to conventional physics, the solid light-gray line to the median of all the realizations of the propagation process and the solid yellow line to a single realization with a random distribution of the domain lengths and of the orientation angles of the magnetic field inside the domains. The filled area is the envelope of the results on the percentile of all the possible realizations of the propagation process at 68 $\%$ (dark blue), 90 $\%$ (blue) and 99 $\%$ (light blue), respectively. In the upper-left panel we have chosen $\xi=0.5$, in the upper-right panel $\xi=1$, in the lower-left panel $\xi=2$ and in the lower-right panel $\xi=5$.}
\end{figure}

\newpage

\section*{Figures for ${\bf z = 0.05}$}

\begin{figure}[h]       
\begin{center}
\includegraphics[width=.53\textwidth]{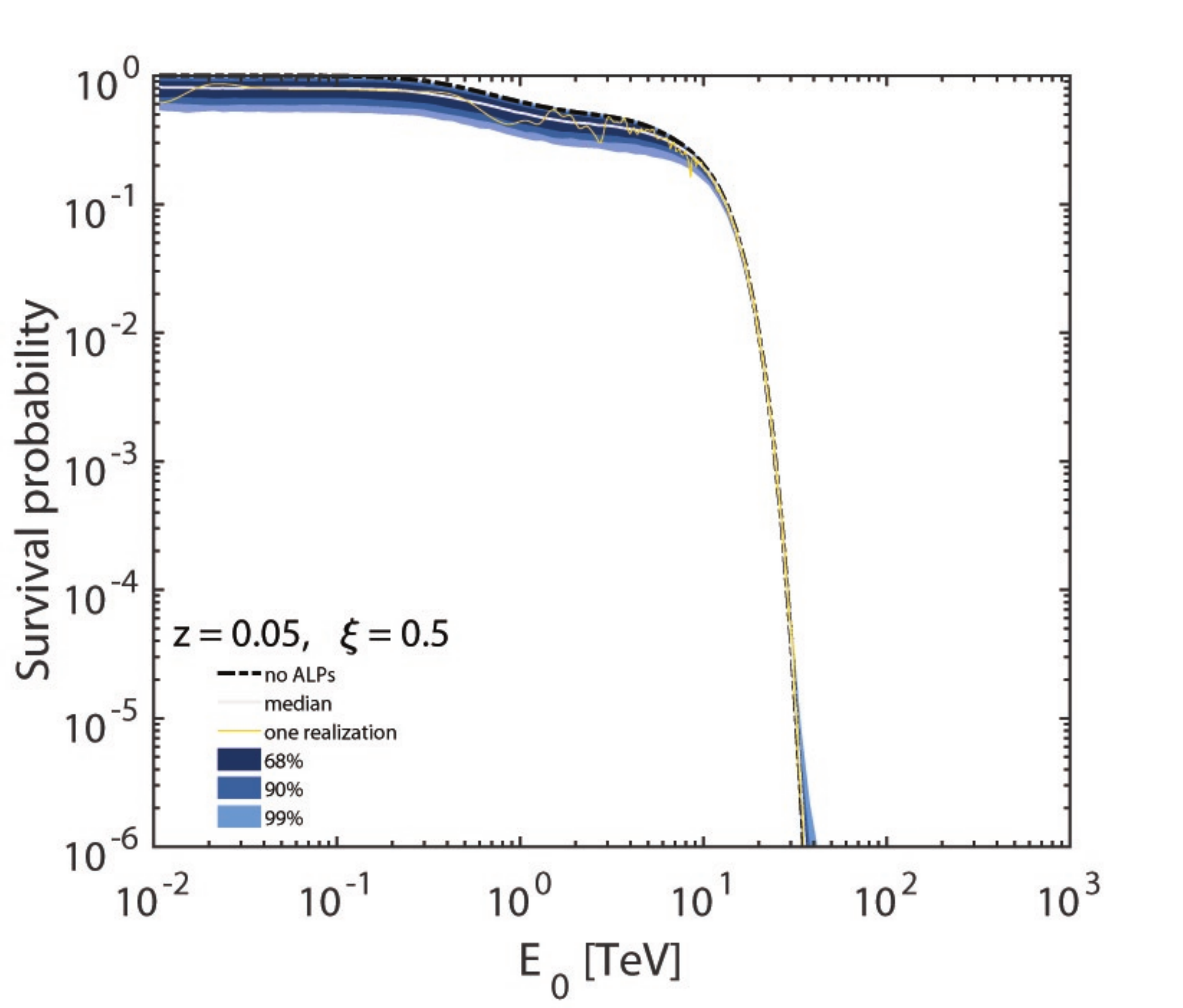}\includegraphics[width=.53\textwidth]{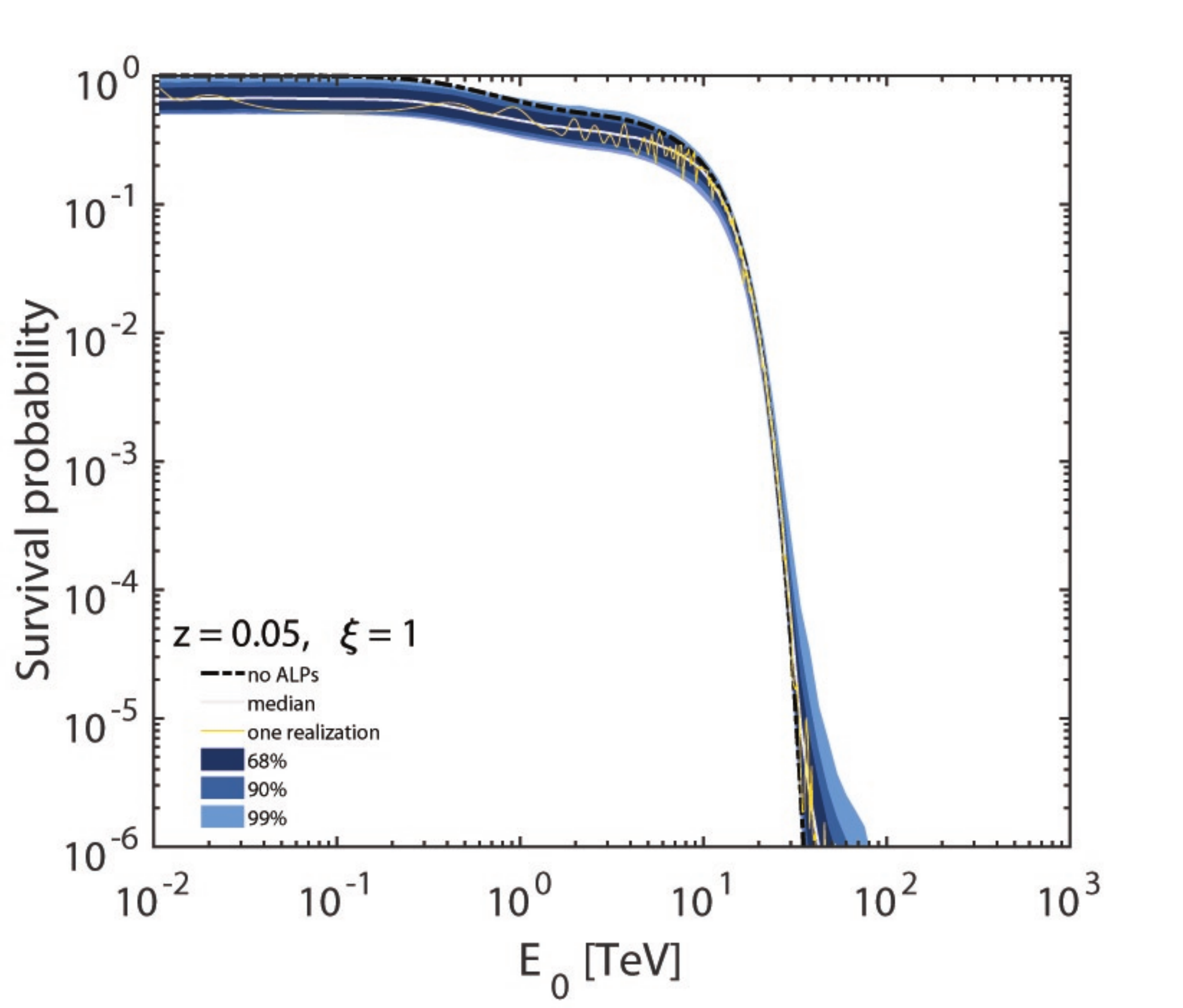}
\includegraphics[width=.53\textwidth]{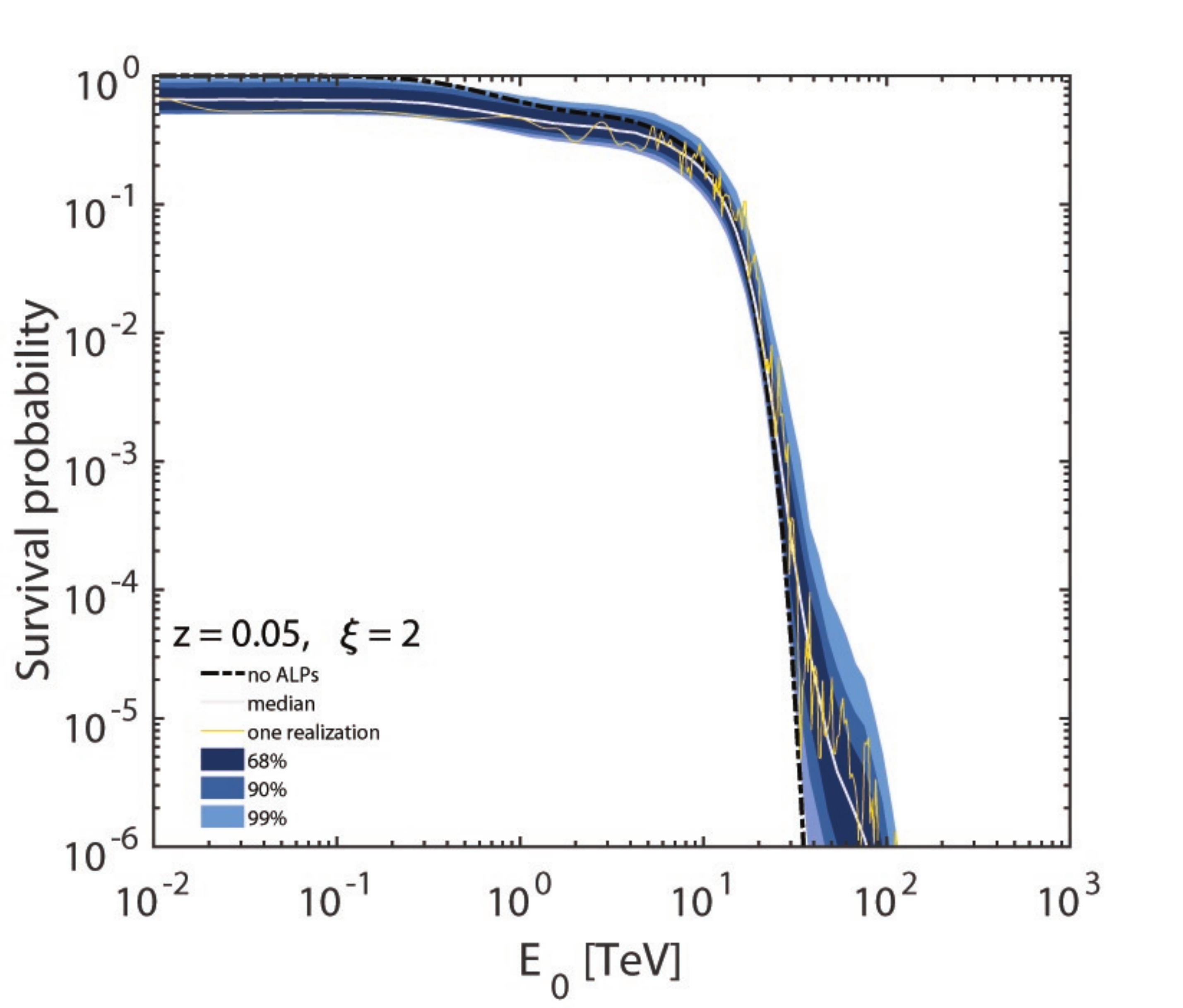}\includegraphics[width=.53\textwidth]{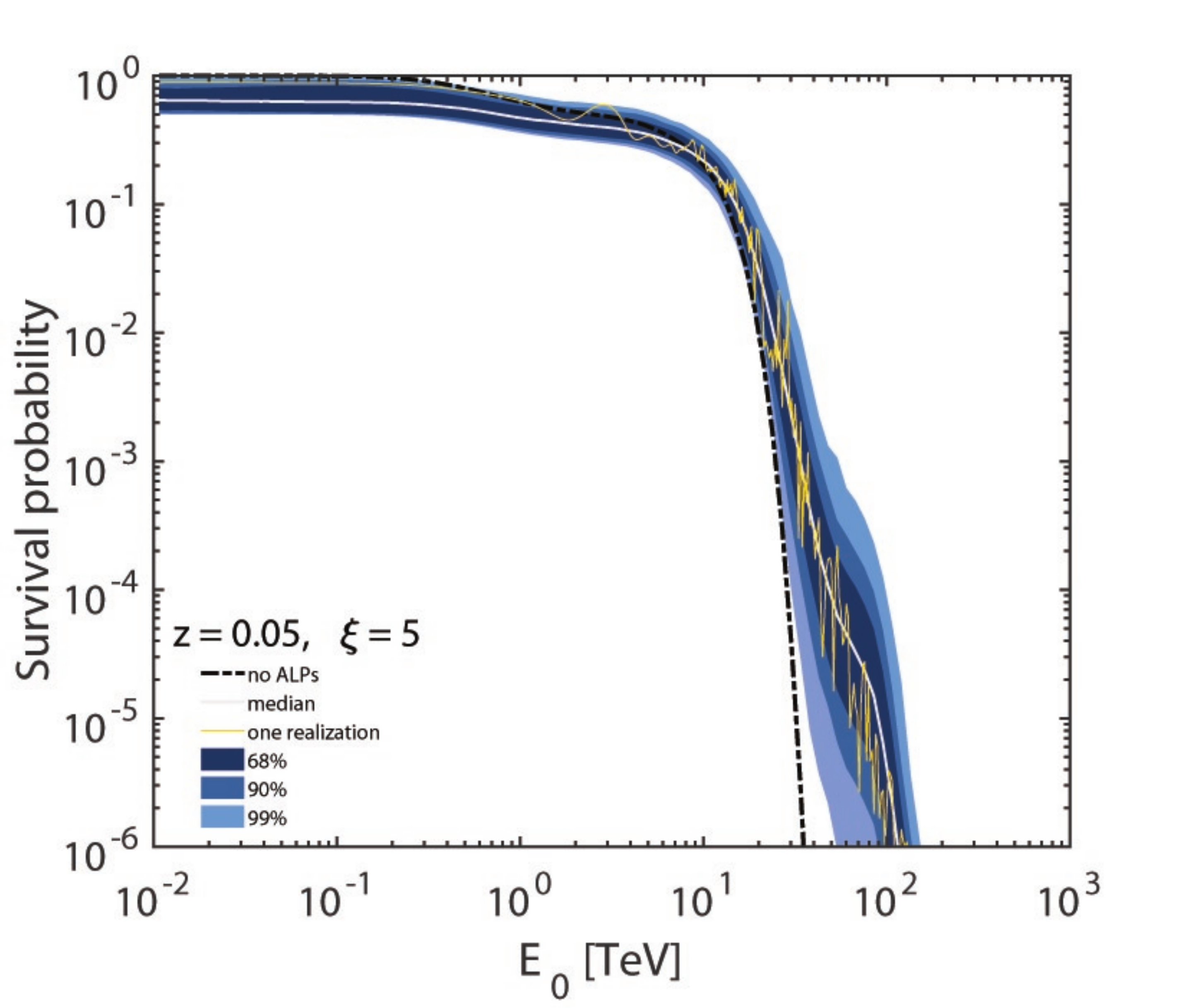}
\end{center}
\caption{\label{z005} 
Same as Figure~\ref{z002} apart from $z=0.05$.}
\end{figure}

\newpage

\section*{Figures for ${\bf z = 0.1}$}

\begin{figure}[h]       
\begin{center}
\includegraphics[width=.53\textwidth]{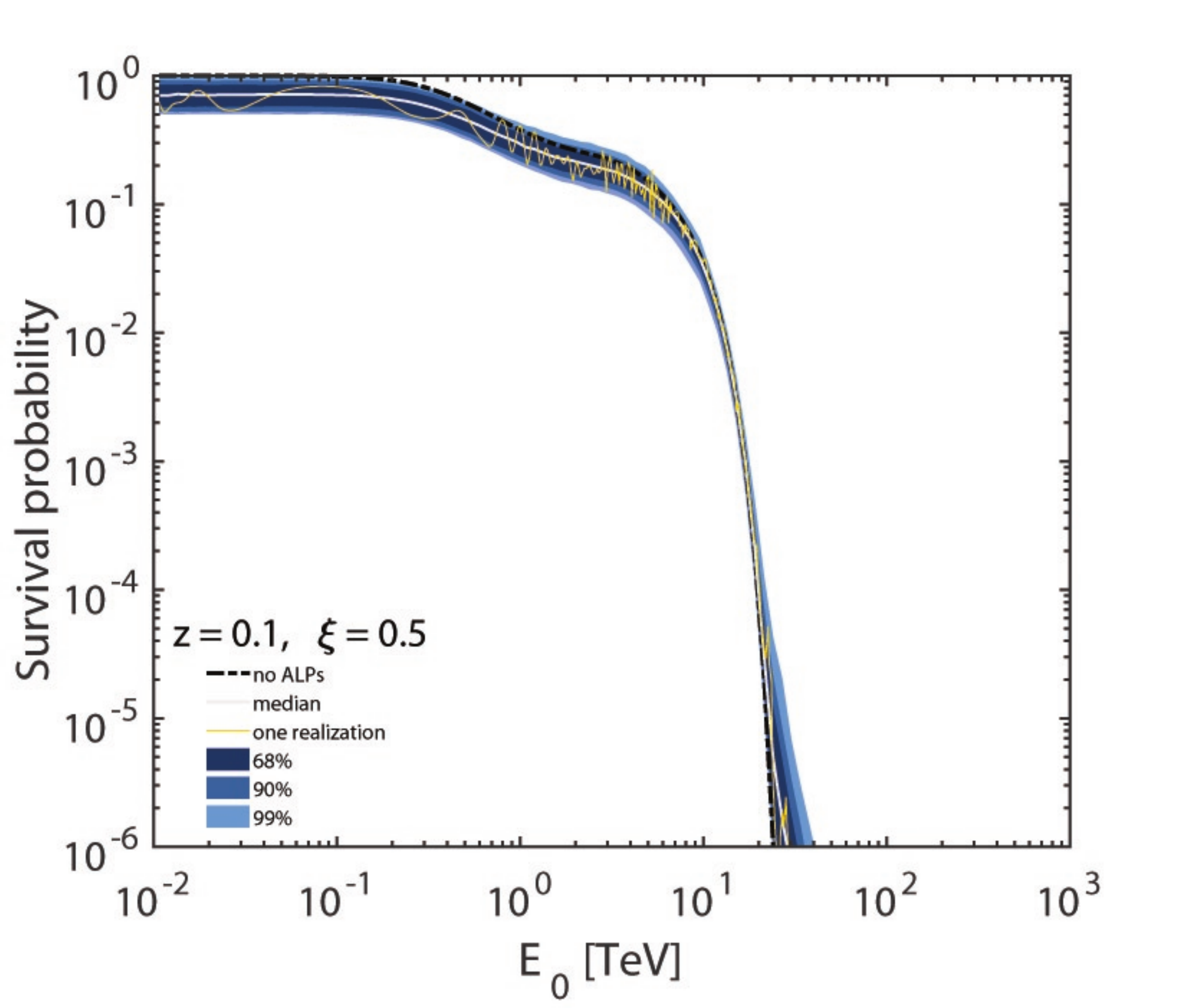}\includegraphics[width=.53\textwidth]{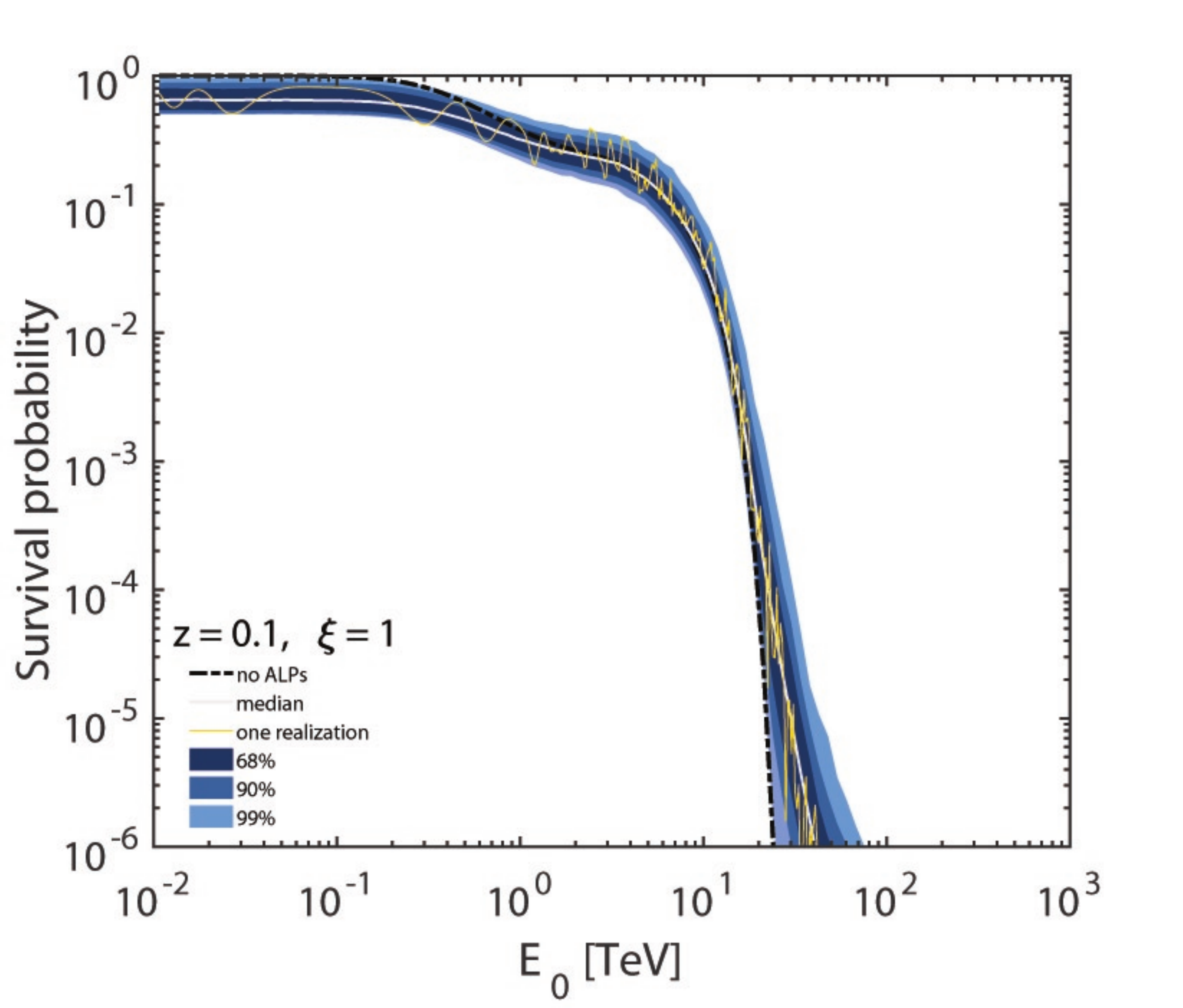}
\includegraphics[width=.53\textwidth]{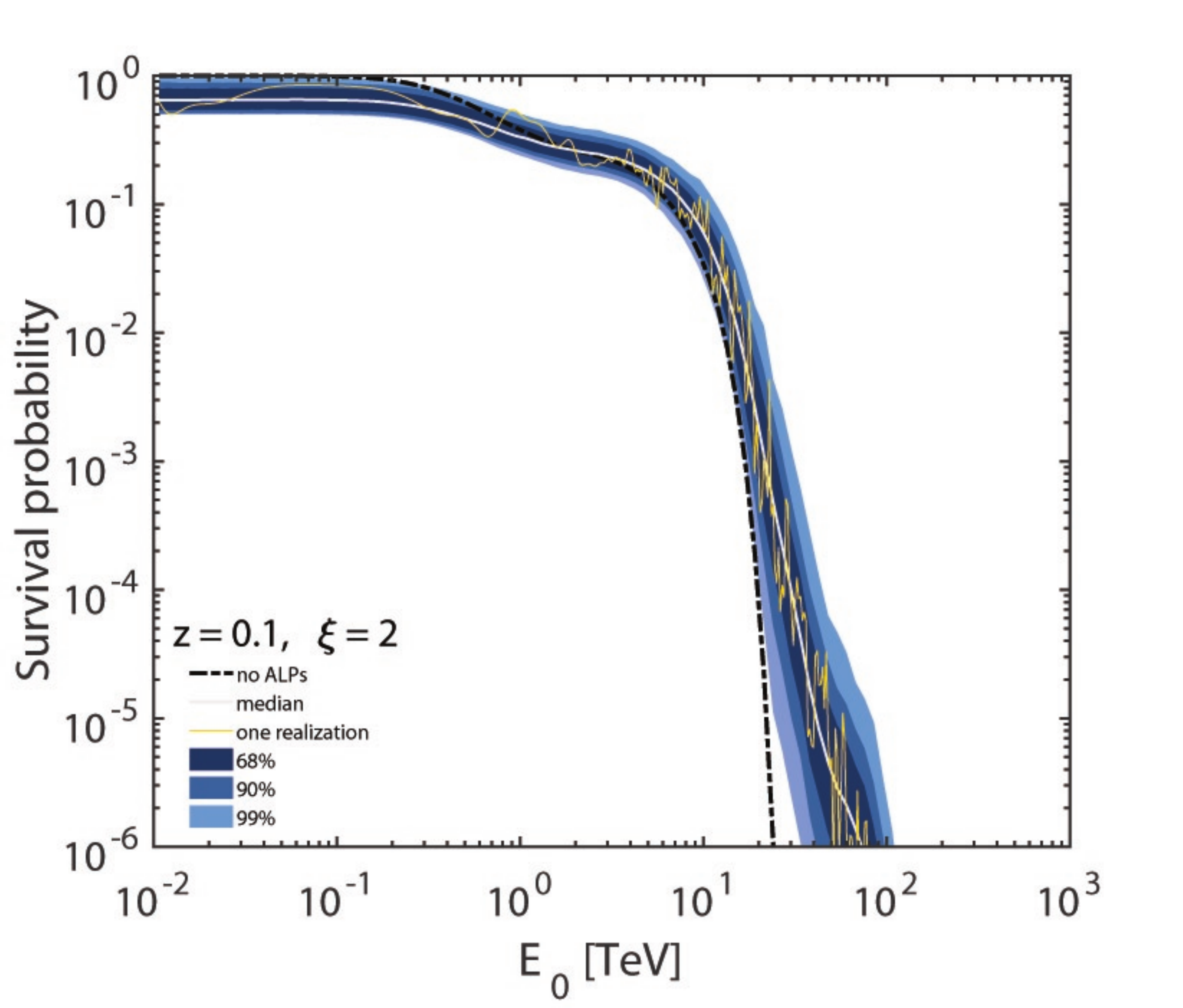}\includegraphics[width=.53\textwidth]{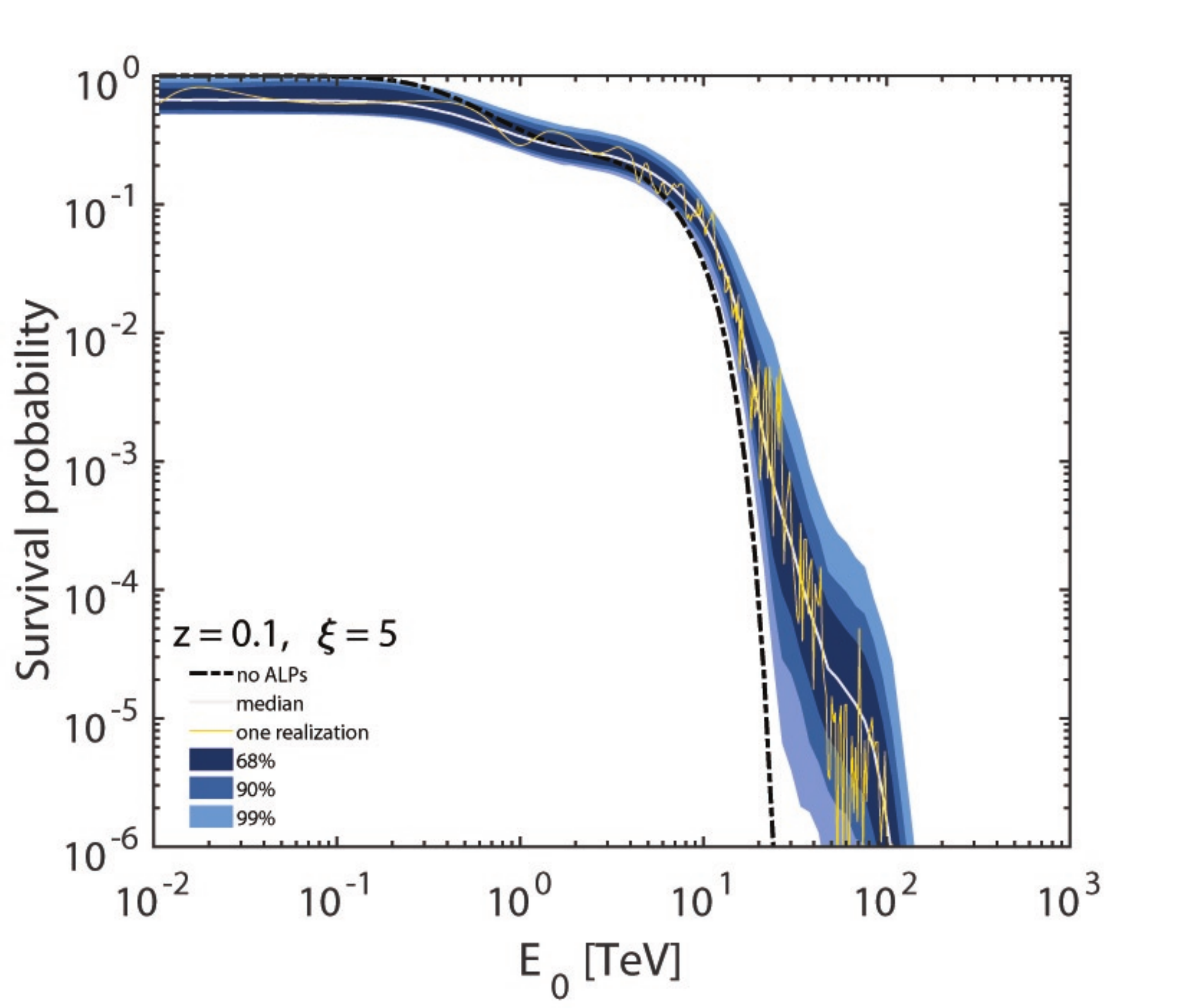}
\end{center}
\caption{\label{z01} 
Same as Figure~\ref{z002} apart from $z=0.1$.}
\end{figure}

\newpage

\section*{Figures for ${\bf z = 0.2}$}

\begin{figure}[h]       
\begin{center}
\includegraphics[width=.53\textwidth]{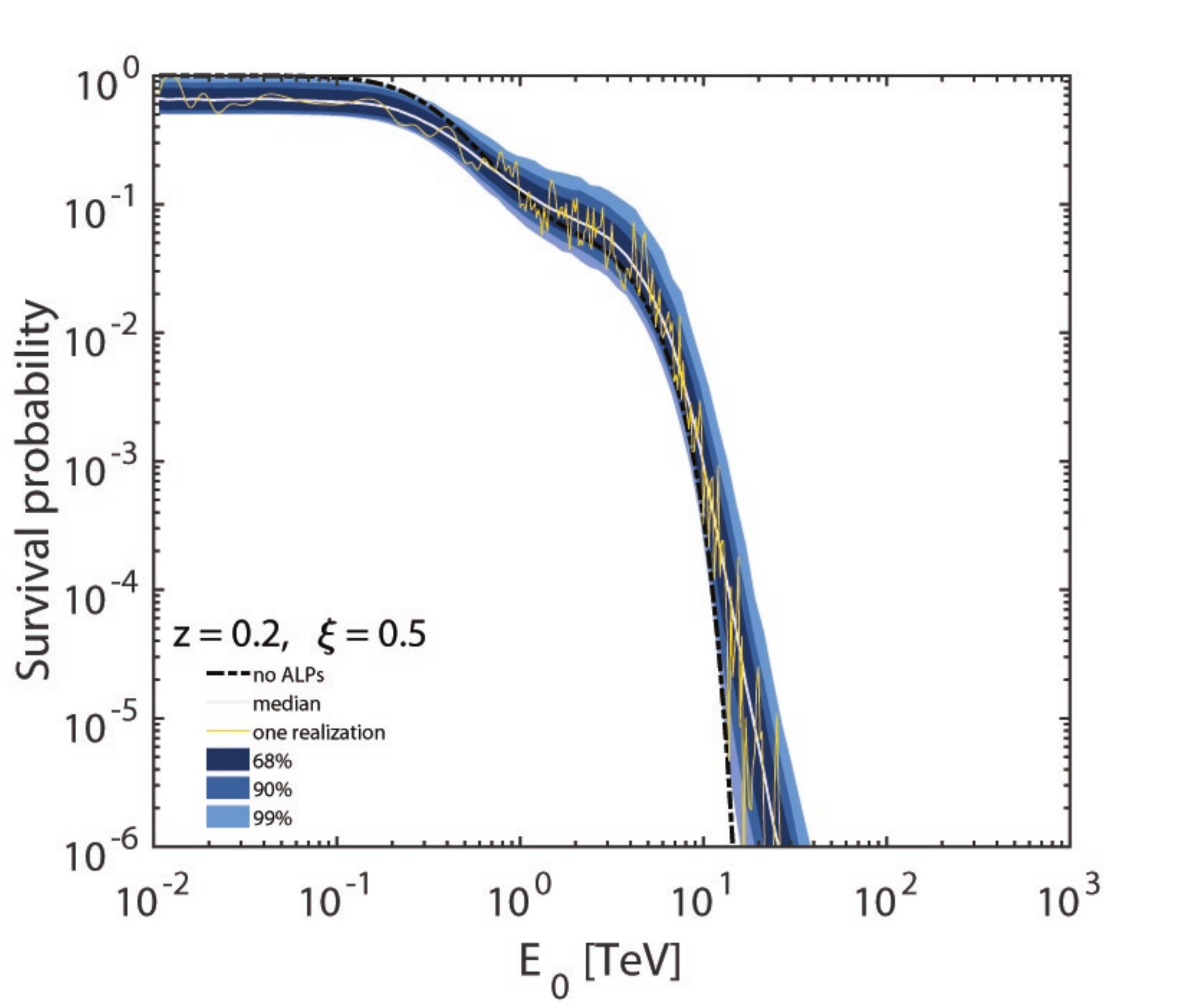}\includegraphics[width=.53\textwidth]{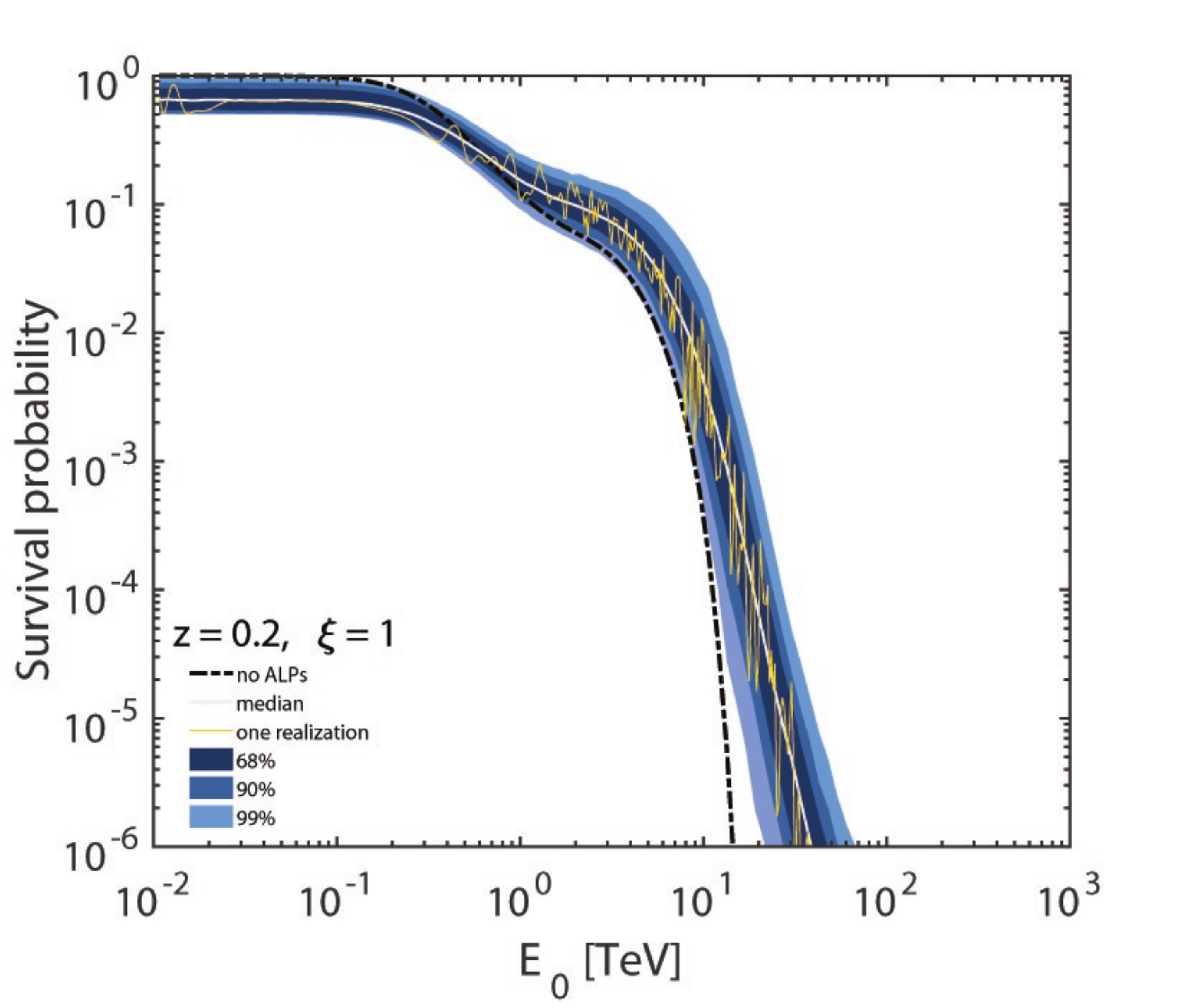}
\includegraphics[width=.53\textwidth]{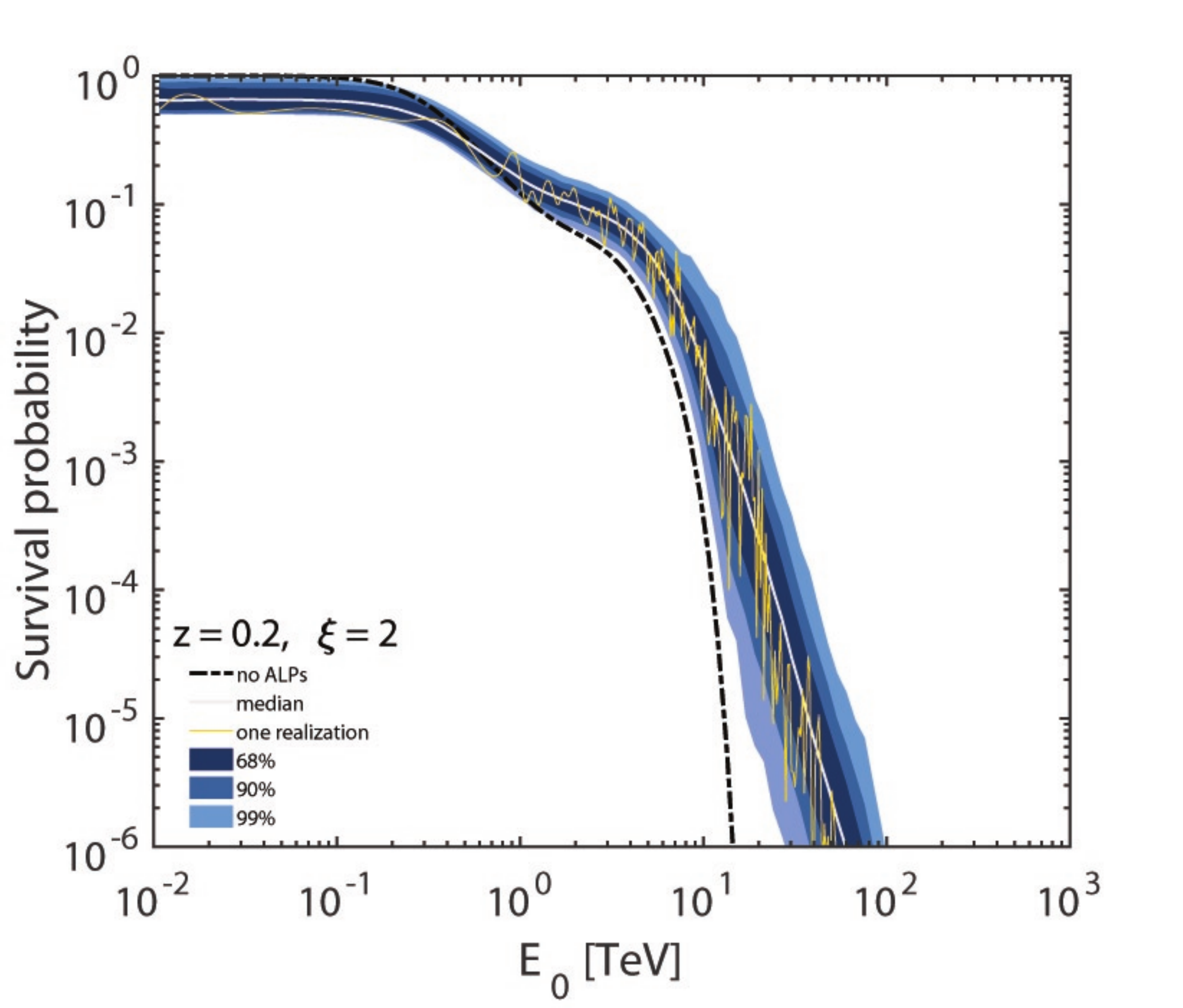}\includegraphics[width=.53\textwidth]{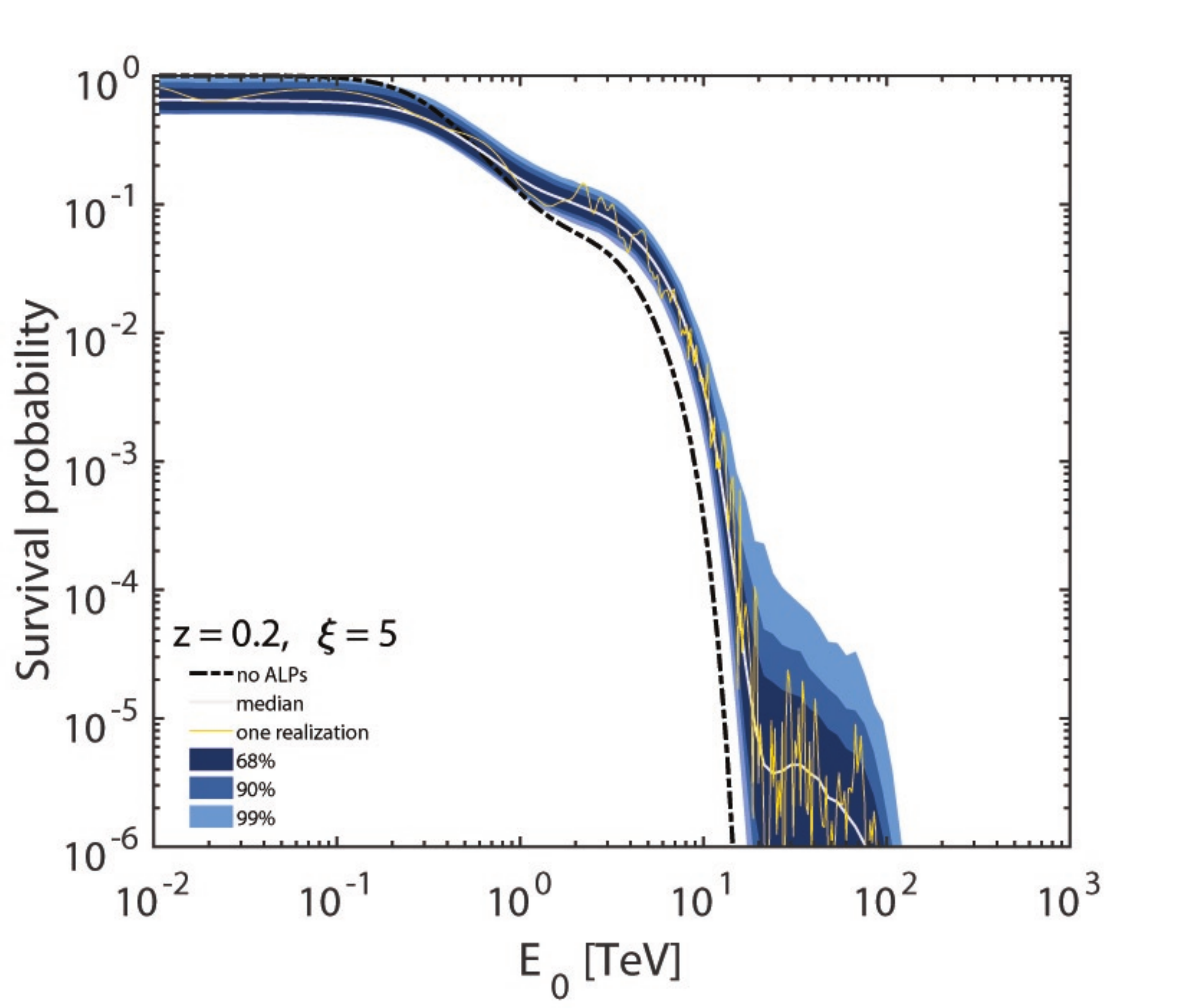}
\end{center}
\caption{\label{z02} 
Same as Figure~\ref{z002} apart from $z=0.2$.}
\end{figure}

\newpage

\section*{Figures for ${\bf z = 0.5}$}

\begin{figure}[h]       
\begin{center}
\includegraphics[width=.53\textwidth]{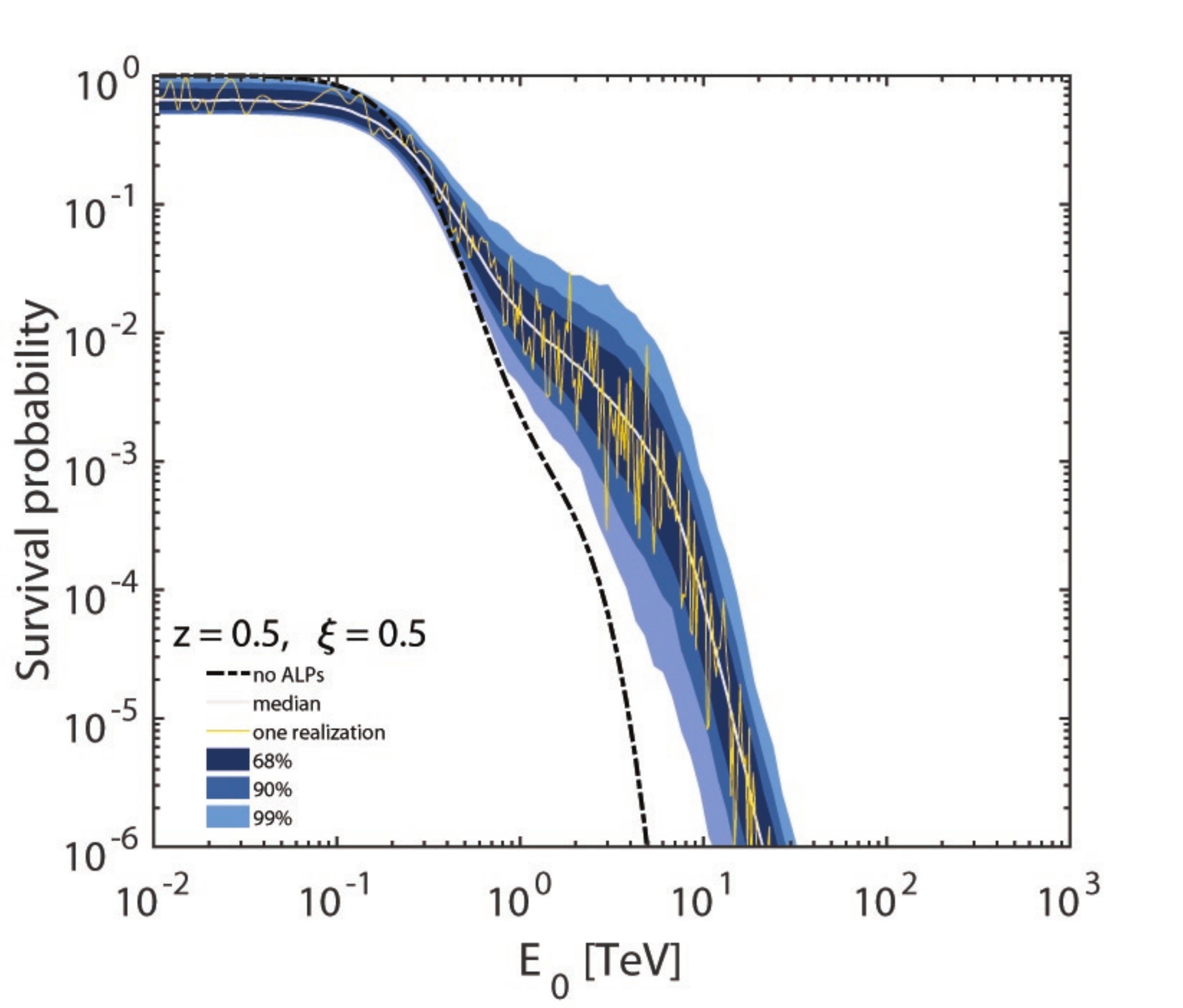}\includegraphics[width=.53\textwidth]{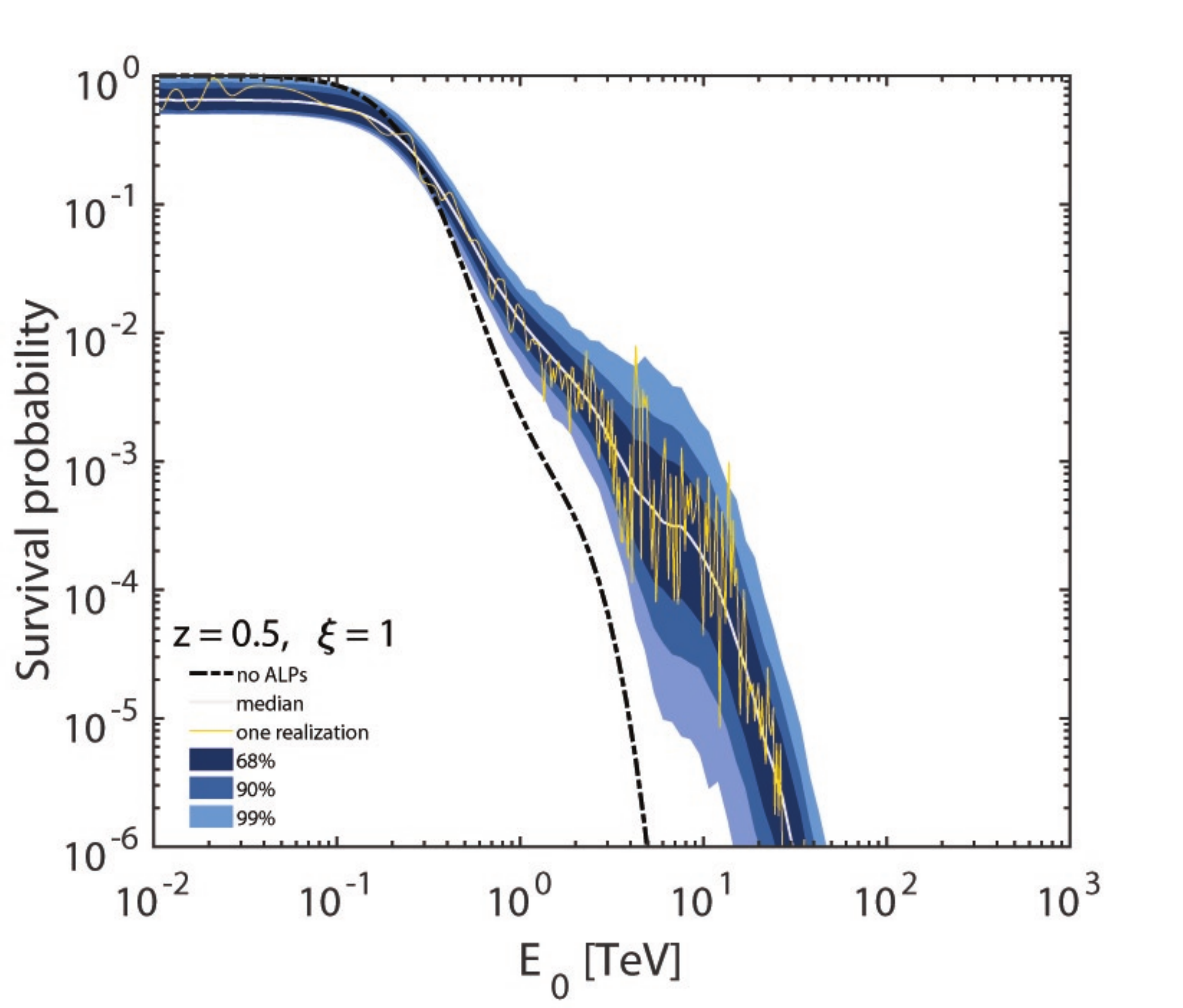}
\includegraphics[width=.53\textwidth]{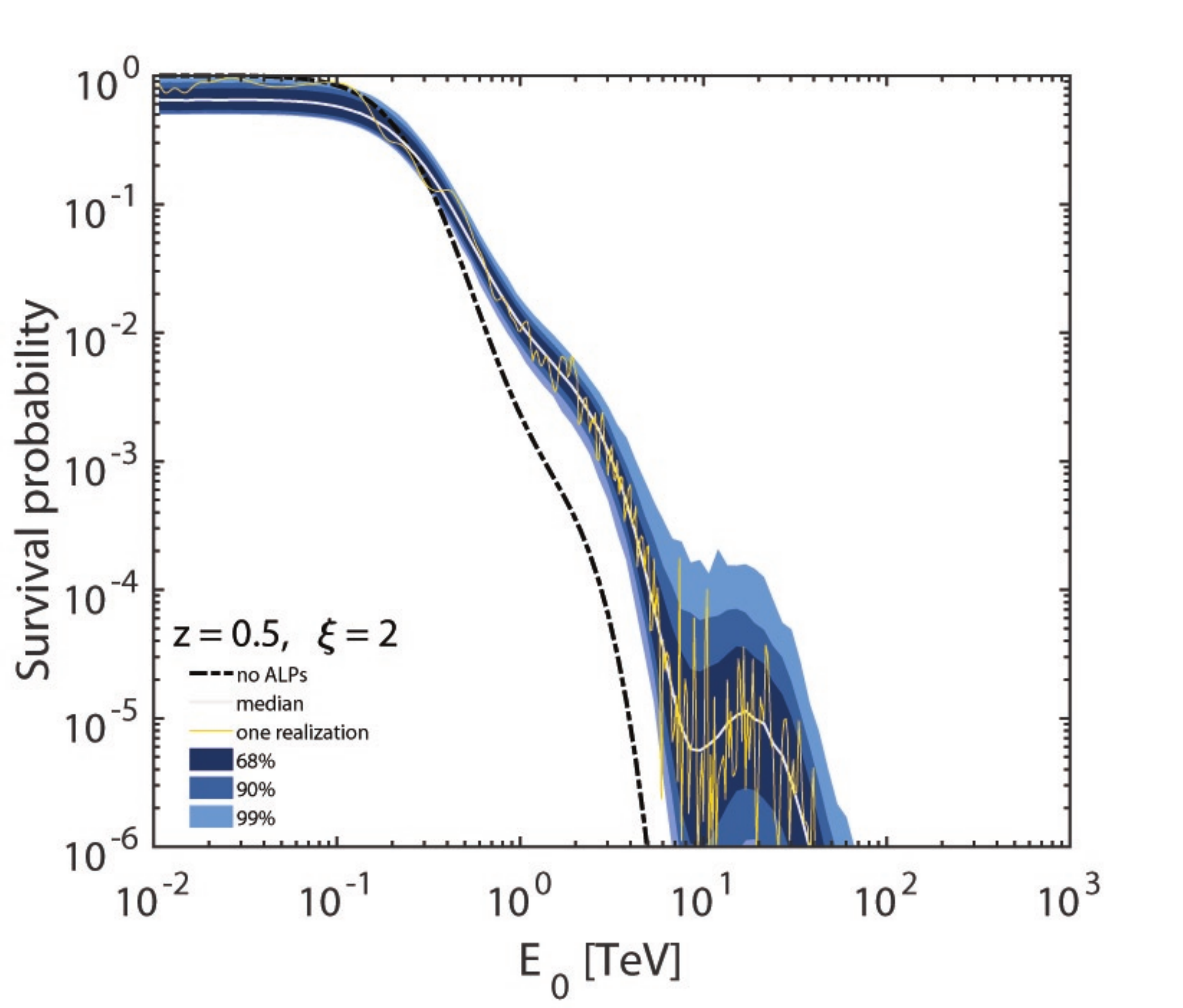}\includegraphics[width=.53\textwidth]{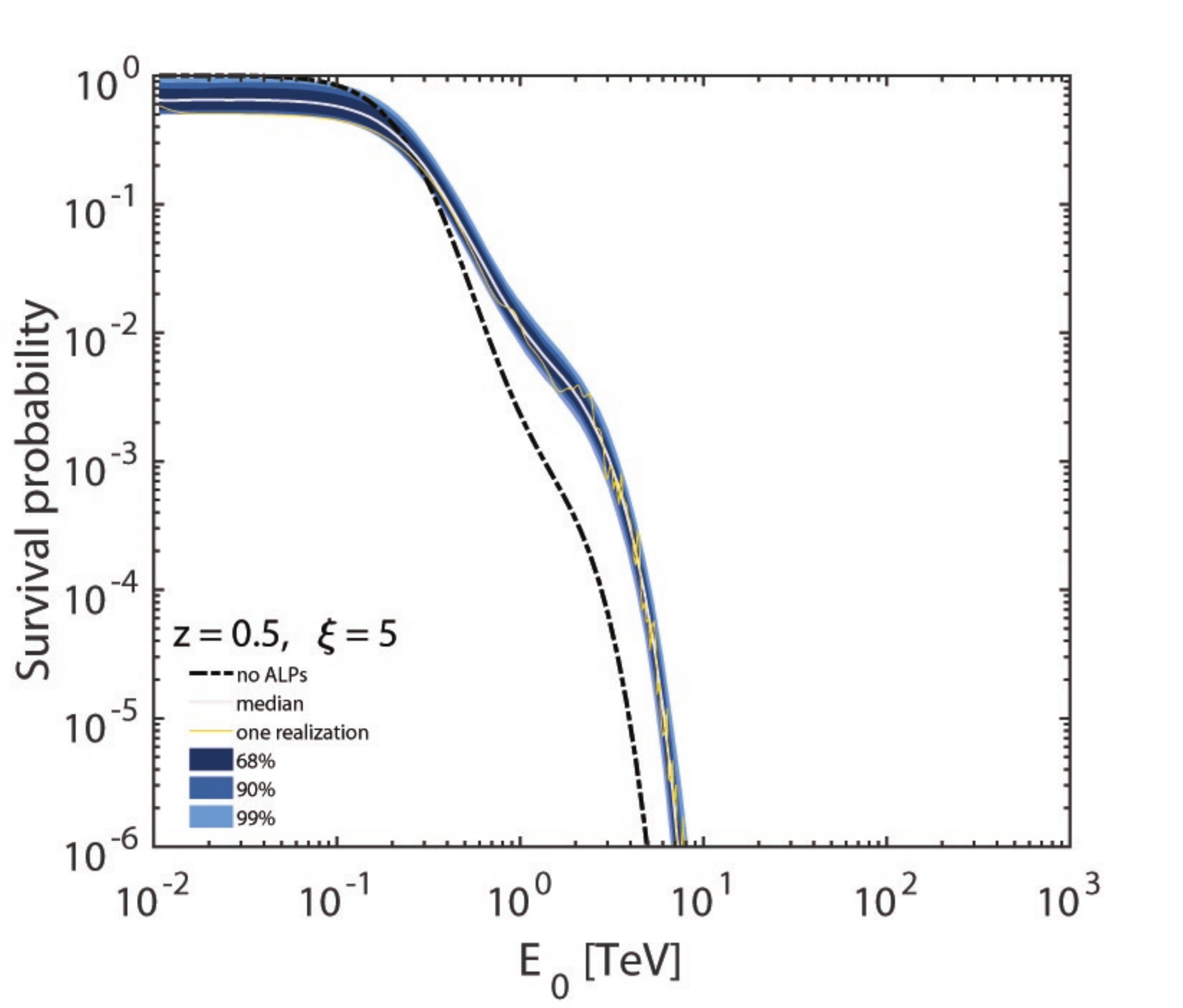}
\end{center}
\caption{\label{z05} 
Same as Figure~\ref{z002} apart from $z=0.5$.}
\end{figure}

\newpage

\section*{Figures for ${\bf z = 1}$}

\begin{figure}[h]       
\begin{center}
\includegraphics[width=.53\textwidth]{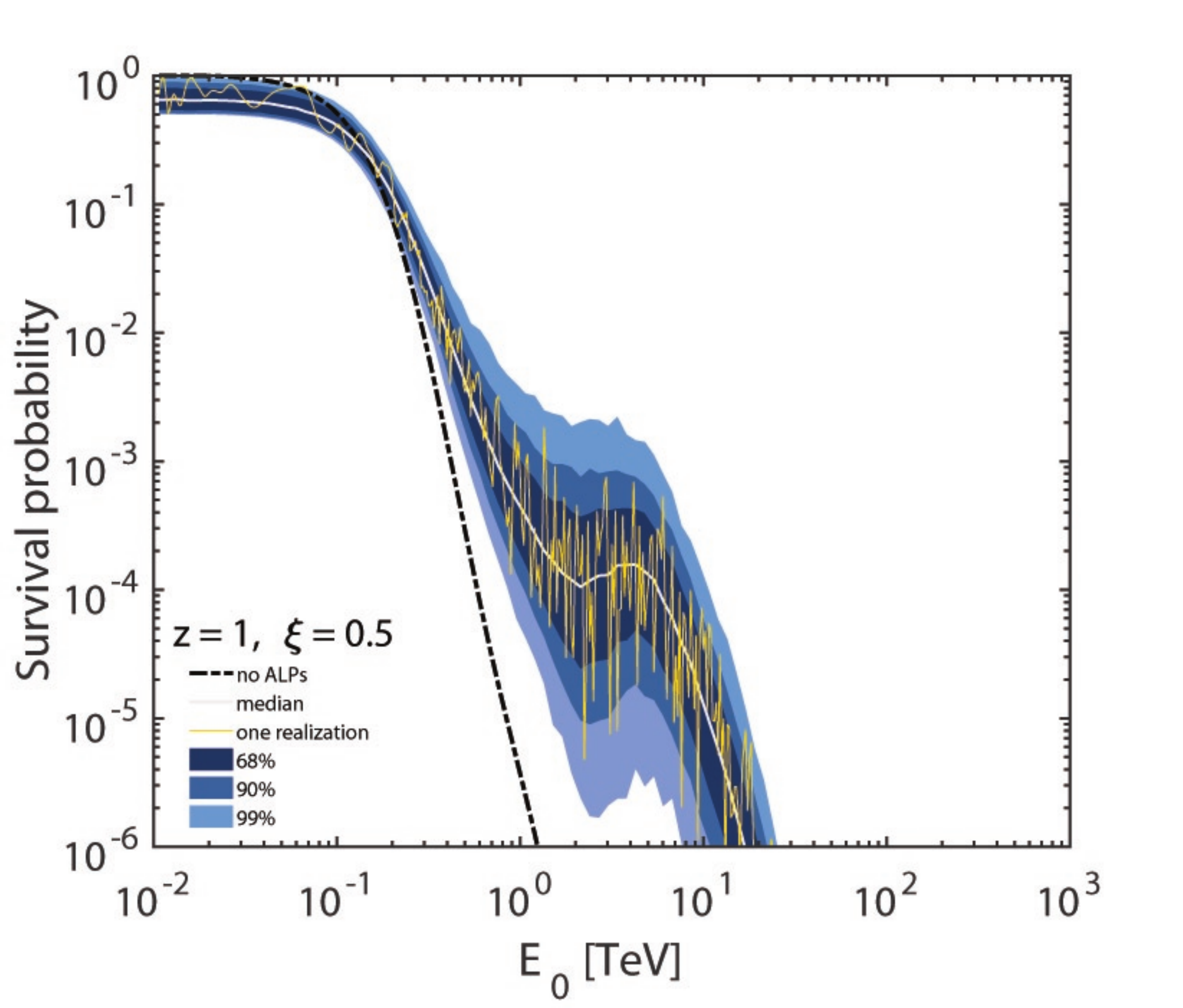}\includegraphics[width=.53\textwidth]{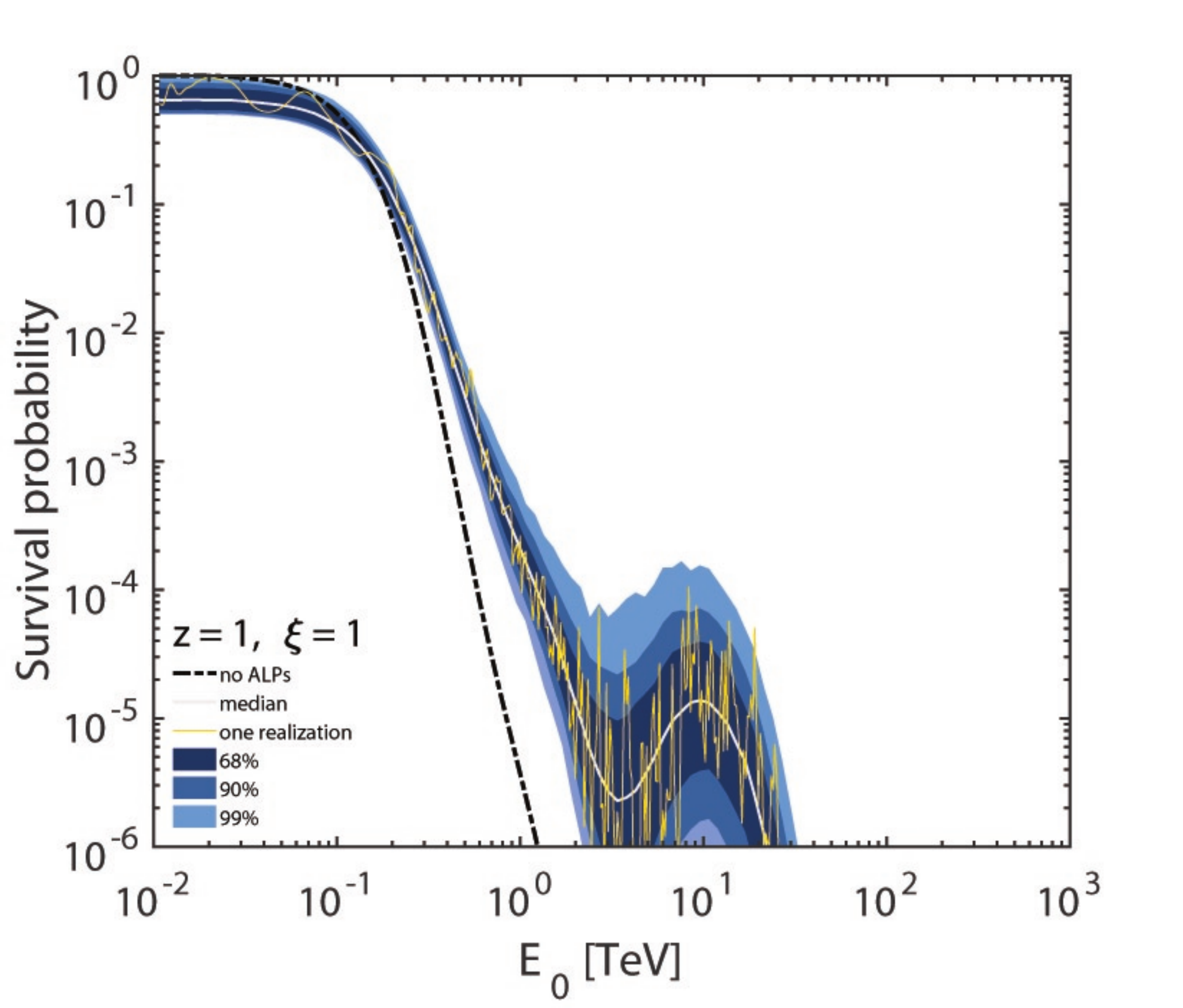}
\includegraphics[width=.53\textwidth]{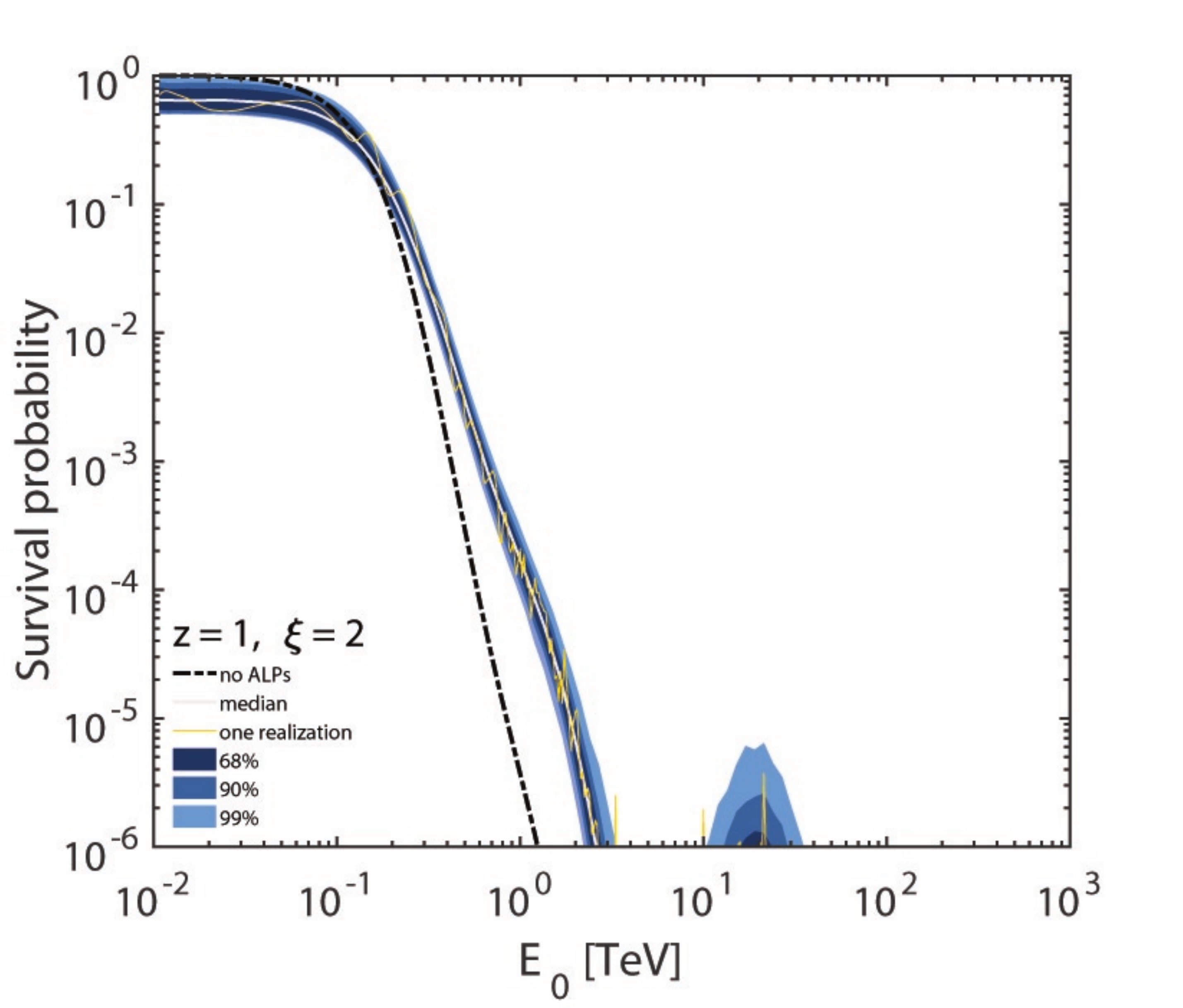}\includegraphics[width=.53\textwidth]{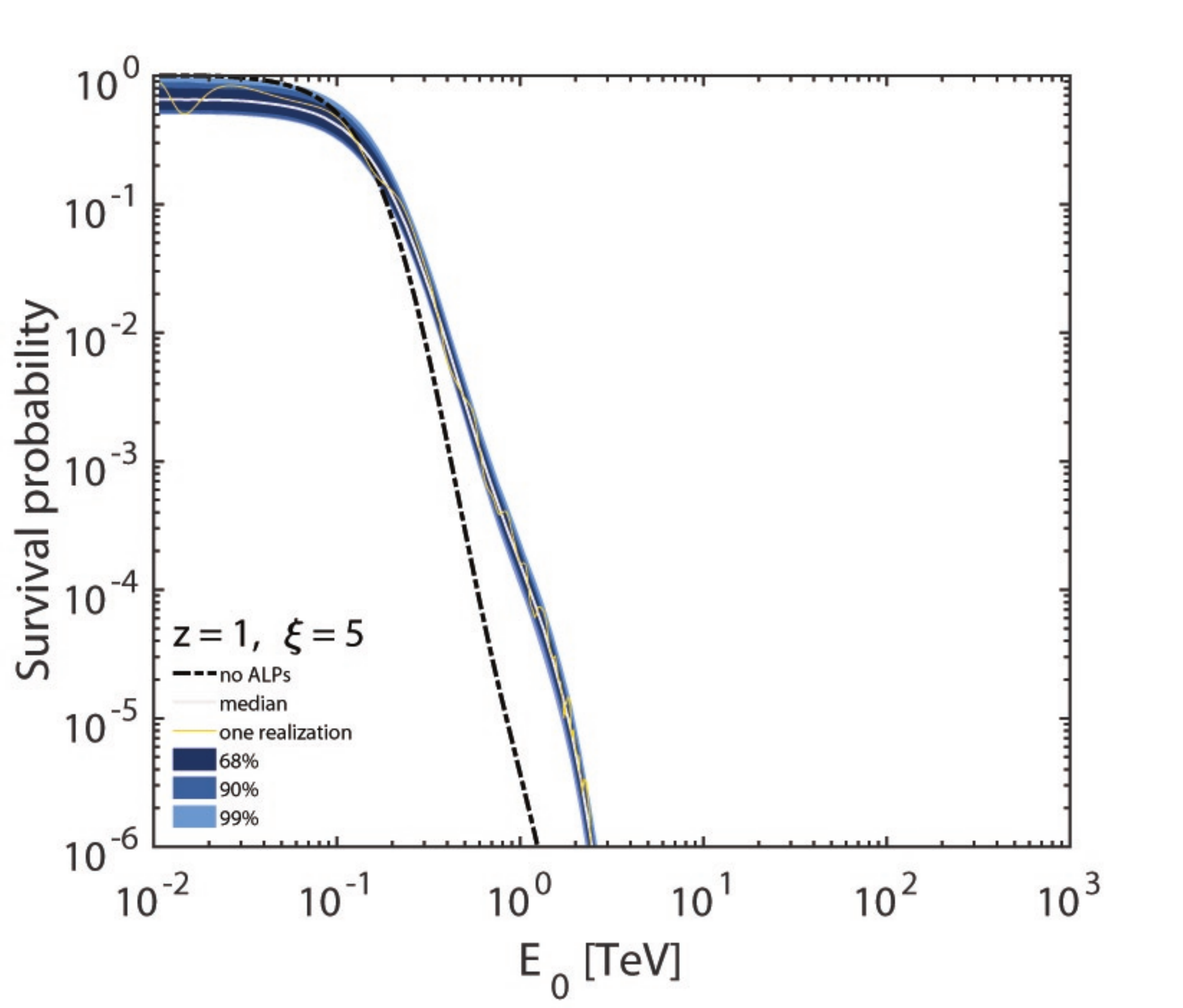}
\end{center}
\caption{\label{z1} 
Same as Figure~\ref{z002} apart from $z=1$.}
\end{figure}

\newpage

\section*{Figures for ${\bf z = 2}$}

\begin{figure}[h]       
\begin{center}
\includegraphics[width=.53\textwidth]{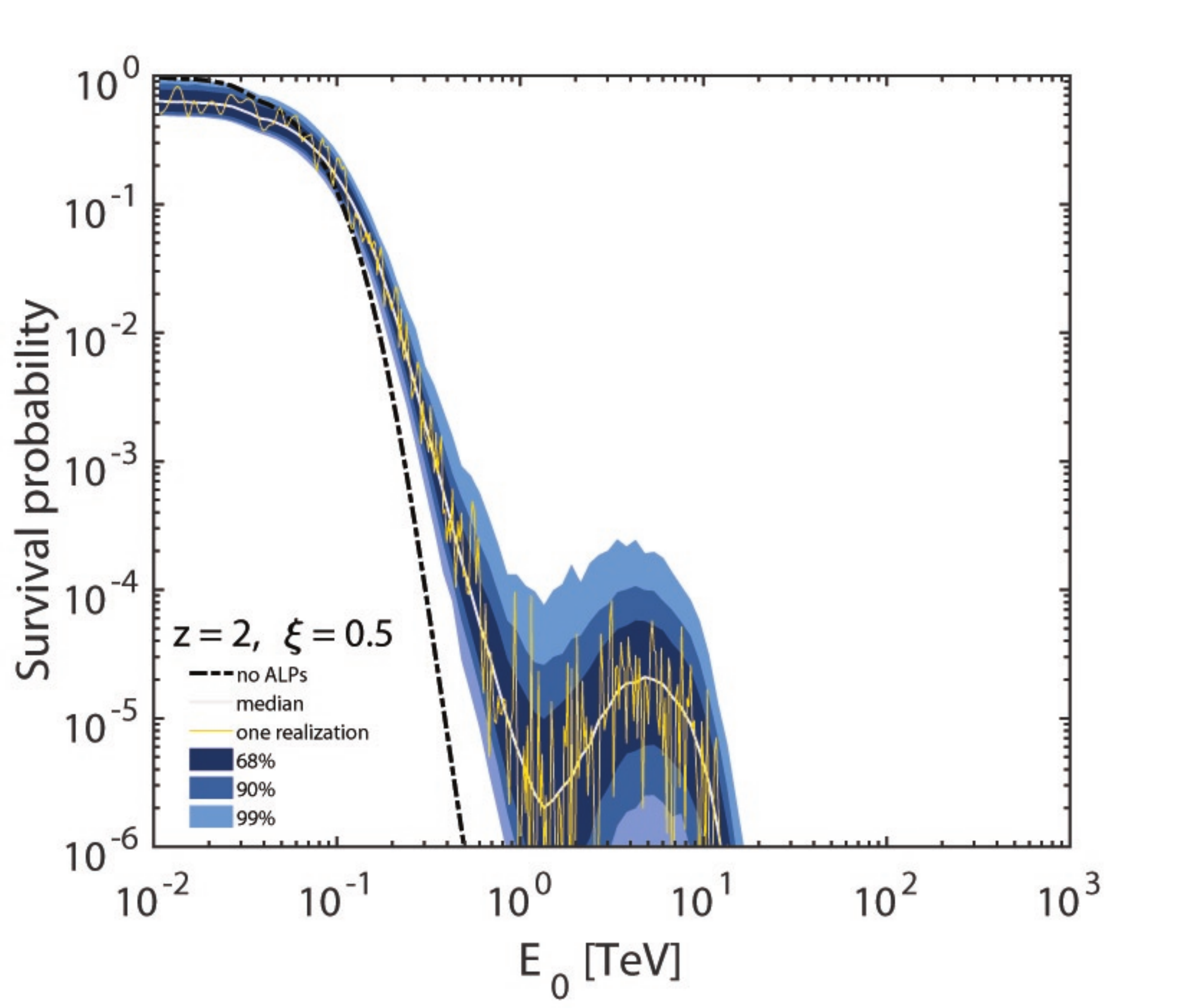}\includegraphics[width=.53\textwidth]{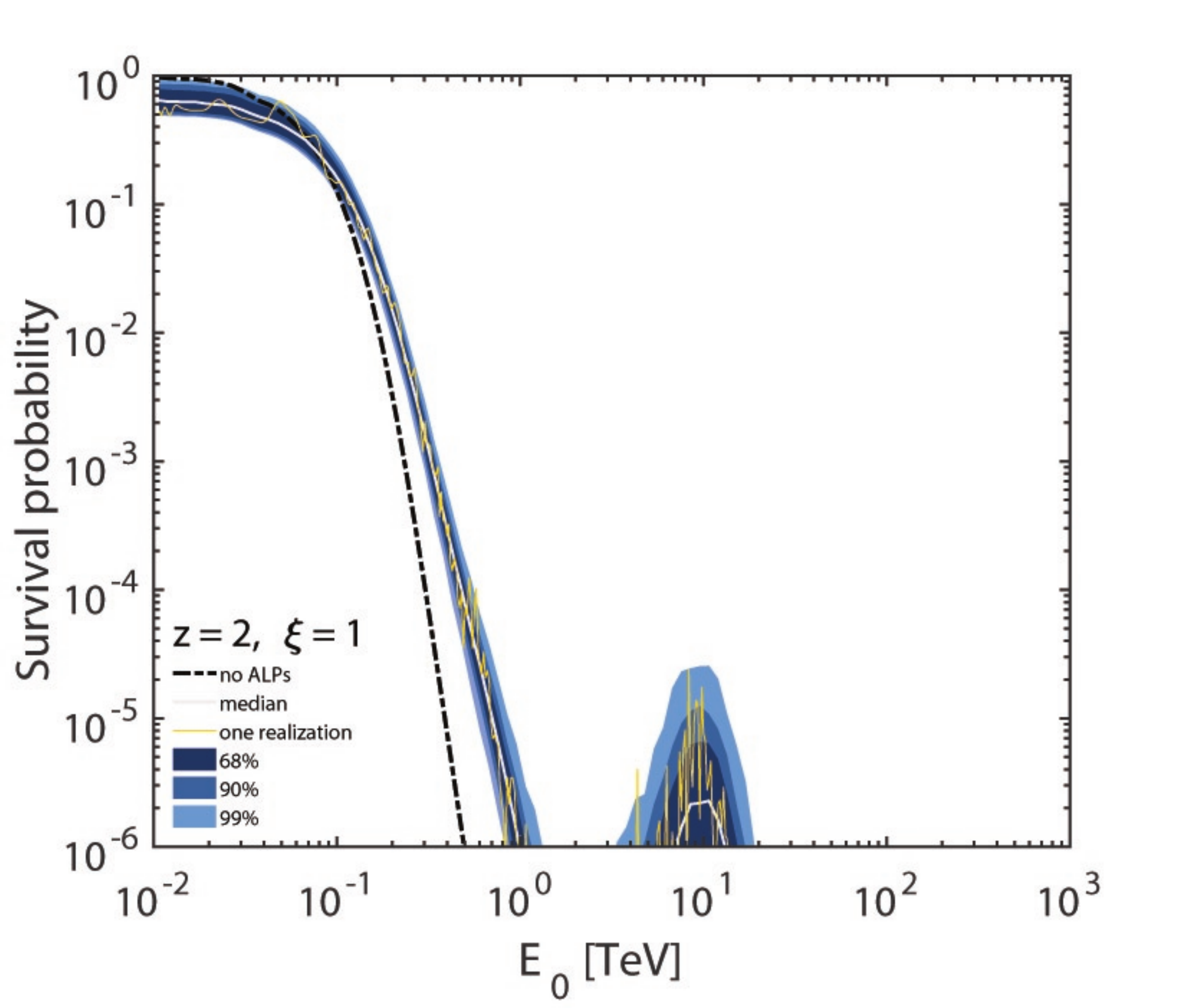}
\includegraphics[width=.53\textwidth]{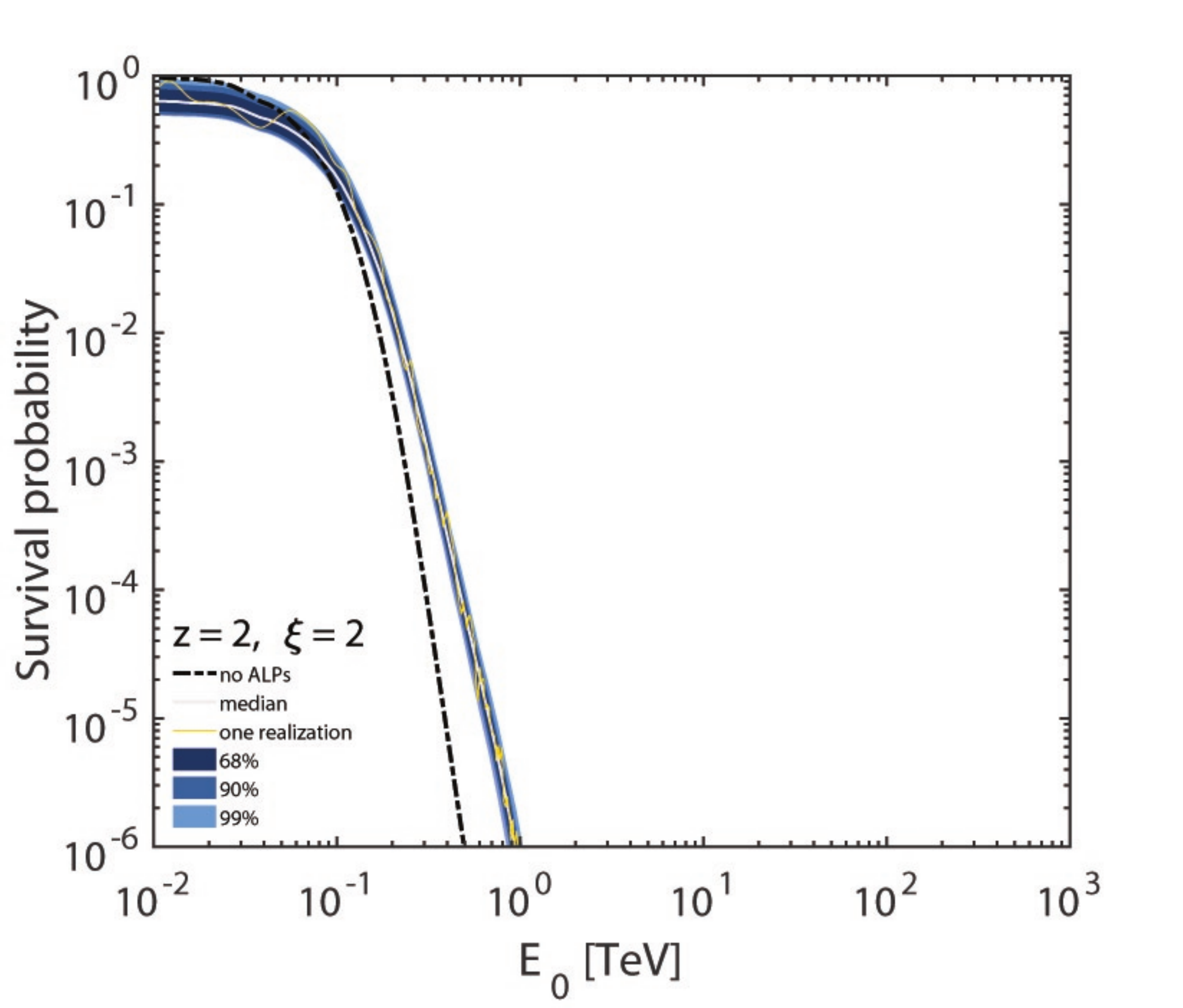}\includegraphics[width=.53\textwidth]{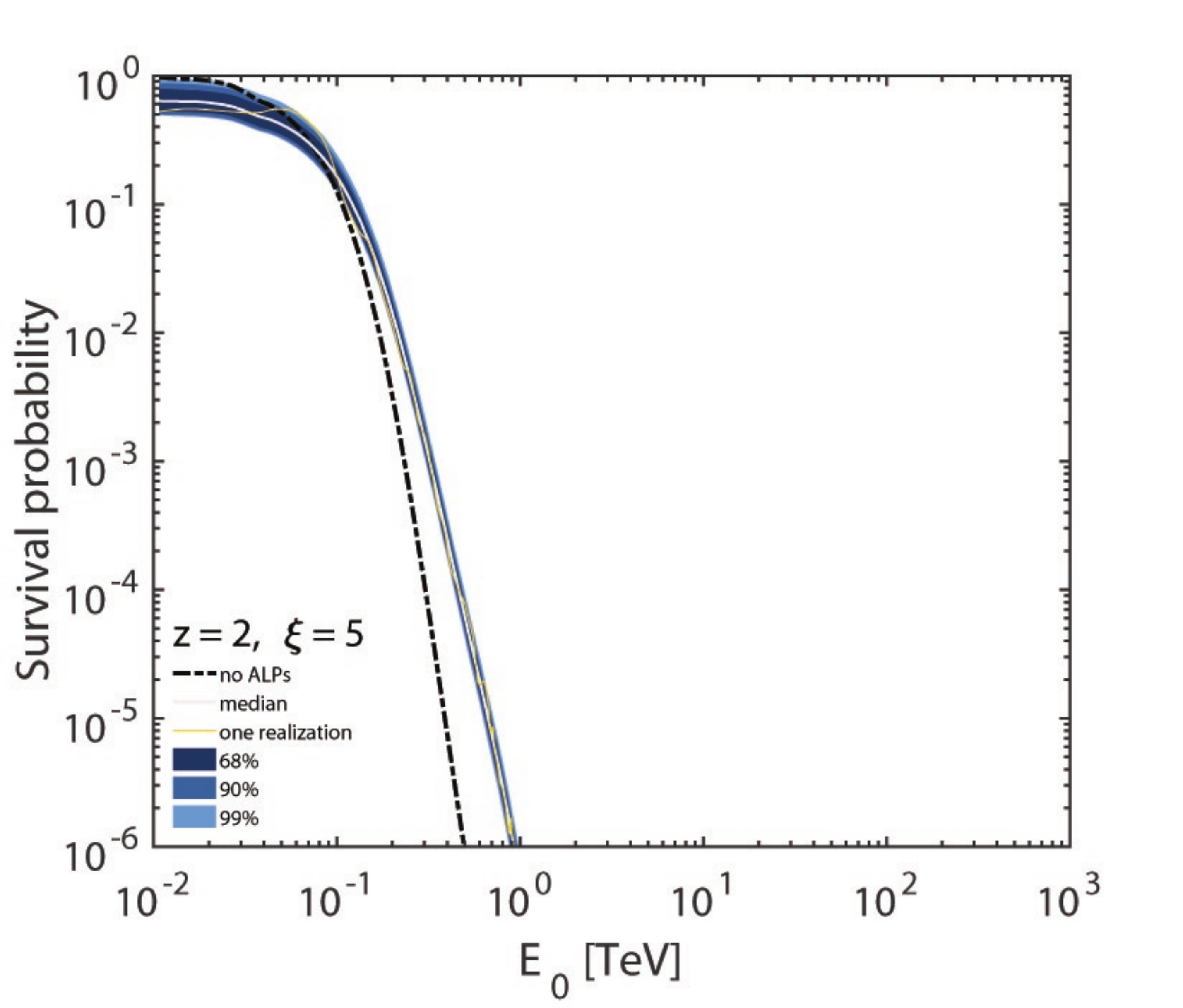}
\end{center}
\caption{\label{z2} 
Same as Figure~\ref{z002} apart from $z=2$.}
\end{figure}

\newpage

\section{Discussion}

The purpose of the present Section is to bring out as simply as possible the physical meaning of the above Figures. In order to gain in clarity, we prefer to proceed rather schematically.

\subsection{Statistical properties of the envelope of all realizations}

We start by addressing the {\it shape} of the envelope of our 1000 different individual random realizations of the considered beam propagation from the blazar to us for a given set of values of $z$, $\xi$, ${\cal E}_0$. Because both the domain lengths $\{L_{{\rm dom}, n} \}$ and the angles $\{\phi_n \}_{1 \leq n \leq N}$ are random, we are dealing with a stochastic process, which explains the oscillating nature of the single realizations.

\begin{itemize}

\item At low enough redshift -- say $z \lesssim 0.2$ -- as $\xi$ increases $P_{\gamma \to \gamma}^{\rm ALP} ({\cal E}_0, z)$ grows. This is in line with physical intuition. As a matter of fact, higher values of $\xi$ imply stronger $\gamma \to a$ conversions, and so a larger number of photons survive EBL absorption in the form of ALPs, which afterwords undergo the $a \to \gamma$ conversions. Note that this sort of behaviour occurs for any energy ${\cal E}_0$.

\item At higher redshift -- say $z > 0.2$ -- things become more complicate. A look at the corresponding Figures shows that now condition $P_{\gamma \to \gamma}^{\rm ALP} ({\cal E}_0, z) \propto \xi$ fails. Actually, above a certain energy ${\cal E}_* (z,\xi)$ the previous trend reverses: as $\xi$ increases $P_{\gamma \to \gamma}^{\rm ALP} ({\cal E}_0, z)$ decreases. Moreover, at fixed $\xi$ the energy ${\cal E}_* (z,\xi)$ slightly decreases as $z$ increases. Why such a behaviour? The answer is due to the photon absorption by the EBL. Imagine we were presently to take $\xi$ as large as possible, which would entail very efficient $\gamma \to a$ and $a \to \gamma$ conversions. As a consequence, a great number of ALPs would be converted to photons. But because at high $z$ the EBL is strong -- its level being an increasing function of $z$ -- most of the photons would be absorbed, leaving over a small number of ALPs per domain. So, in such a situation, the number of ALPs per domain increases by {\it decreasing} $\xi$. Correspondingly, for $z > 0.2$ the area of the envelope of all realizations increases as $\xi$ decreases.  
 
\item At lowest redshift and low $\xi$ -- specifically for $z = 0.02 \, (\xi = 0.5, 1.0, 2.0), 
z = 0.05 \, (\xi = 0.5, 1.0), z = 0.1 \, (\xi = 0.5)$ -- the area of the envelope of all realizations is thin, and even some ALP realizations lie below the one corresponding to conventional physics. This is due to the fact that in such a situation the EBL is small, and some photons produced by the $\gamma \to a$ transitions do not have a chance to reconvert into 
photons, and get lost. 

\item As $\xi$, $z$ and ${\cal E}_0$ increase, the available parameter space gets larger -- thereby increasing the area encompassing all realizations -- until the phenomenon discussed in the second item takes over. 

\end{itemize}

In conclusion, in order to enhance the cosmic transparency at large $z$ we better take $\xi = 0.5$, which we regard as the best of our benchmark values.

\subsection{Properties of the individual realizations}

Let us briefly consider the properties of the single realizations, which are functions of the random variables $\{L_{{\rm dom}, n} \}$ and $\{\phi_n \}_{1 \leq n \leq N}$. While within the strong mixing regime they are {\it independent} of ${\cal E}_0$ -- since $l_{\rm osc}$ and $P_{\gamma \to a} (L_{\rm dom})$ are ${\cal E}_0$-independent -- we have seen that the strong mixing dominates only the lower end of the VHE band. Next, the high-energy weak mixing regime sets in and the probability associated with the realizations becomes an oscillatory function of ${\cal E}_0$ and goes like ${\cal E}_0^{- 2}$, according to Eq. (\ref{18042018f}). We want to stress that {\it this is a crucial prediction of our model}, which can be tested with the new generation of VHE observatories provided that they have a good enough energy resolution. 

A question naturally arises. Why does the oscillatory behavior of the single realizations disappear from the boundary of their whole envelope? Consider the median, which is smooth. Then some individual oscillating realizations will lie above -- while others below -- the median. So, when they are all considered at once their oscillatory ${\cal E}_0$-dependent gets washed out.

\section{Relations with previous work} 

Since the discovery of the photon dispersion on the CMB~\cite{raffelt2015}, two paper including it have been appeared. Our aim is to discuss their analogies and differences with respect to the present work.

\begin{itemize}

\item The first analysis of the implications of the above effect for photon-ALP oscillations in the extragalactic magnetic field ${\bf B}$ has been done in 2017 by Kartavtseva, Raffelt and Vogel~\cite{krv2017}. They recognize the two main points, namely the increasing importance of the photon dispersion on the CMB as energy gets larger an larger, and the concomitant reduction of the $\gamma \leftrightarrow a$ oscillation length $l_{\rm osc} ({\cal E})$. However, they do not quantify these effects. They take for ${\bf B}$ a domain-like model somewhat similar to the one used in this paper, and plot the transfer matrix versus energy in various different regimes. Moreover, they do not deal with what is really observable -- namely the individual random realizations -- but rather with the average photon survival probability, getting approximate expressions (and not the exact and general solution for the average photon survival probability). Their paper is based on the extension of an alternative approach first developed by Mirizzi and Montanino~\cite{mm2009}. Finally, they do not consider any specific example of simulated blazar.

\item An innovative approach has been put forward in 2017 by Montanino {\it et al}.~\cite{vazzamirizzi2017}. Rather than using some sort of domain-like model for ${\bf B}$ they rely upon the magnetohydrodynamic cosmological simulations (see e.g.~\cite{vazza2014,vazza2016} and references therein). This issue has been discussed at length in Section II of GR2018a, and so we briefly recall its main aspects. One starts by assuming a cosmological magnetic field which has an arbitrary value $B_*$ at some redshift $z_* = {\cal O} (40)$ during the dark age, and investigates its evolution as driven by structure formation up to the present. The normalization condition $B_0$ is fixed by the requirements that it should reproduce the  magnetic field of regular clusters today. As a by-product, a prediction of the magnetic field ${\bf B}_{\rm fil}$ inside filaments in the present Universe arises. Unfortunately, this is not the end of the story. Within regular clusters galactic outflows are a reality, as demonstrated by the observation of strong iron lines. It has repeatedly been shown that the magnetic field ejected by a central AGN during the cluster formation can be amplified by turbulence during the cluster evolution in such a way to explain the observed cluster magnetic fields, and in addition that the strength and structure of the magnetic fields observed in clusters of galaxies are well reproduced for a wide range of the model parameters by galactic outflows~\cite{xu2009,donnert2009}. Thus -- denoting by $B_{\rm cl}$ the cluster magnetic field -- it necessarily follows that $B_0 < B_{\rm cl}$. And a realistic extreme case based on the above argument can yield $B_0 =0$! In such a situation, a cosmological magnetic field is not needed to explain the cluster magnetic field, and it would not be needed al all! What is clear at any rate is that the magnetic fields inside filaments would have values completely different from those predicted by magnetohydrodynamic cosmological simulations. Because of this uncertainty, we regard the model under consideration as unreliable.

\end{itemize}

\section{Conclusions}

We have presented a very detailed account of photon-ALP oscillations triggered by the extragalactic magnetic field $B_{\rm ext}$. The new feature of our work is to take systematically into account photon dispersion on the CMB~\cite{raffelt2015}. We have explained why this fact has required a modification of the standard domain-like model of $B_{\rm ext}$ (named DLSHE model) into a new and more complicated domain-like model (named DLSME model): this achievement has been reported in GR2018a~\cite{gr2018a} (thus previously used domain-like models of ${\bf B}_{\rm ext}$ would generally give wrong results when the whole VHE band is considered). Within this context, above an energy scale ${\cal E}_H = {\cal O} (5 \, {\rm TeV)}$ the considered effect starts to become dominant and makes the {\it single} random realizations of the beam propagation process to exhibit small energy oscillations: {\it this is a crucial prediction of our model}. 

We have been able to derive the corresponding photon survival probability along every  single random realization $P^{\rm ALP}_{\gamma \to \gamma} ({\cal E}_0, z)$ {\it analytically and exactly} up to the observed energy ${\cal E}_0 = 1000 \, {\rm TeV}$ and redshift up to $z = 2$, a fact that drastically shortens the computation time in the derivation of the Figures presented in this paper. Specifically, for 7 simulated blazars we exhibit the plots of $P^{\rm ALP}_{\gamma \to \gamma} ({\cal E}_0, z)$ versus ${\cal E}_0$, for different values of $z$ and four values of $\xi$. 

We find that -- in spite of the fact that photon dispersion on the CMB tends to decrease the photon survival probability $P_{\gamma \to \gamma}^{\rm ALP} ({\cal E}_0, z)$ -- such a quantity still remains considerably larger that the analogous one $P_{\gamma \to \gamma}^{\rm CP} ({\cal E}_0, z)$ as evaluated within conventional physics. Therefore, we do have an enhanced photon transparency in the VHE band. 

Thus, our predictions can be straightforwardly tested with the new generation of $\gamma$-ray observatories like CTA~\cite{cta}, HAWC~\cite{hawc}, GAMMA-400~\cite{g400}, LHAASO~\cite{lhaaso}, TAIGA-HiSCORE~\cite{desy} and HERD~\cite{herd}.

Quite remarkably, just the same $\gamma \leftrightarrow a$ oscillation mechanism which would lead to an increased transparency in the VHE band over cosmic distances naturally provides an excellent explanation of a vexing question in a totally different context (see Appendix): why do flat spectrum radio quasars (FSRQs) emit in the VHE band? This comes about for $m_a < 10^{- 9} \, {\rm eV}$ and two photon coupling in the range $0.83 \cdot 10^{- 11} \, {\rm GeV}^{- 1} \lesssim  g_{a \gamma \gamma} \lesssim 1.43 \cdot 10^{- 11} \, {\rm GeV}^{- 1}$. Consider now Eq. (\ref{09032018a}) with the best of our benchmark values, namely $\xi = 0.5$. Correspondingly, owing to Eq. (\ref{19052018a}) we get $0.43 \, {\rm nG} \lesssim B  \lesssim 0.74 \, {\rm nG}$: these values are well within the above quoted bound $B < 1.7 \, {\rm nG}$~\cite{pshirkov2016}. So, in a sense ``we pay one and get two''! More explicitly, whatever will happen, the existence of two so totally different  situations which are quantitatively in perfectly agreement looks to us more than a simple coincidence.

Finally, since the mass of our ALP is $m_a = {\cal O} (10^{- 10} \, {\rm eV})$ and assuming that indeed $0.83 \cdot 10^{- 11} \, {\rm GeV}^{- 1} \lesssim  g_{a \gamma \gamma} \lesssim 1.43 \cdot 10^{- 11} \, {\rm GeV}^{- 1}$, such an ALP can be detected in the laboratory within the next few years, thanks to the upgrade of ALPS II at DESY~\cite{alps2}, the planned experiments IAXO~\cite{iaxo} and STAX~\cite{stax}, as well as with other techniques developed by Avignone and collaborators~\cite{avignone1,avignone2,avignone3}. Moreover, if the bulk of the dark matter is made of ALPs they can also be detected by the planned ABRACADABRA experiment~\cite{abracadabra}.

\section*{Acknowledgments}          

We thank Alessandro De Angelis, Alessandro Mirizzi and Fabrizio Tavecchio for discussions, Luigina Feretti and Giancarlo Setti for a careful reading of the manuscript,  Georg Raffelt for discussions and for communicating to us his results about photon dispersion on the CMB  prior to publication, and especially Enrico Costa for suggestions. 
G. G. acknowledges contribution from the grant INAF CTA--SKA, `Probing particle acceleration and $\gamma$-ray propagation with CTA and its precursors', while the work of M. R. is supported by an INFN TAsP grant.

\section*{Appendix}

A totally unrelated issue concerns the VHE emission of flat spectrum radio quasars (FSRQs). They are a different kind of blazars. Apart from being more massive and luminous than BL Lac objects -- whose number density in the sky is considerably larger -- one of their properties is to have the so-called {\it broad line region} (BLR) on the jet at about $(0.1 - 0.4) \, {\rm pc}$ from the centre, which is very rich of optical/ultraviolet photons. The point is that photons start to be accelerated from the jet base, but when they enter the BLR -- with an energy ${\cal E}_{\rm HE} = (20 - 30) \, {\rm GeV}$ -- they scatter off the optical/ultraviolet photons according to the same Breit-Wheeler process considered in Section I, namely $\gamma_{\rm HE} + \gamma_{\rm UV} \to e^+ +e^-$. Because the photon density in the BLR is much larger than that of EBL photons, according to conventional physics the resulting optical depth in the BLR is $\tau_{\rm CP} ({\cal E}_{\rm HE}) \simeq 14$. As a consequence, the standard expectation is that FSRQs emit $\gamma$-ray only up to about $20 \, {\rm GeV}$~\cite{tavecchiomazin,poutanen2010}. 

Yet, several FSRQs have been detected in the VHE band up to $400 \,  {\rm GeV}$~\cite{albert2008,wagner2010,aleksic2011}. How is it possible? Various astrophysical models have been put forward to explain this fact, but so far they are all {\it ad hoc}, namely proposed just in order to explain that effect (see e.g.~\cite{stamerra2011}). 

The most impressive case is that of FSRQ PKS $1222 + 216$ at $z = 0.432$. It has been detected by Fermi/LAT in the energy range $(0.3 - 3) \, {\rm GeV}$~\cite{tanaka2011}, but also an intense VHE emission in the energy range $(70 - 400)  \, {\rm GeV}$ has been observed by the MAGIC IACT~\cite{aleksic2011}, which in addition has seen a flux 
doubling in only about 10 minutes, thereby implying that the VHE emitting region should be a very compact blob of size about $0.3 \cdot 10^{- 4} \, {\rm pc}$ at a {\it larger} central distance than the BLR. It is really a mystery how it is possible that such a small blob far-away from the centre emits like a whole BL Lac! 

Actually, just the same $\gamma \leftrightarrow a$ oscillation mechanism considered in the main text and framed within the standard blazar emission models provides an excellent explanation of the observed VHE emission for $m_a < 10^{- 9} \, {\rm eV}$ and two photon coupling in the range $0.83 \cdot 10^{- 11} \, {\rm GeV}^{- 1} \lesssim  g_{a \gamma \gamma} \lesssim  1.43 \cdot 10^{- 11} \, {\rm GeV}^{- 1}$, which are just {\it consistent} with those considered in the main text. We stress that both the Fermi/LAT and the MAGIC spectral energy distribution of $1222 + 216$ come out right~\cite{trgb2012}!

\end{document}